\documentclass[preprint,12pt]{elsarticle}
\usepackage{amssymb}
\usepackage[ruled,vlined]{algorithm2e}
\usepackage[noend]{algpseudocode}
\usepackage{amsfonts}
\usepackage{float}
\usepackage{hyperref}
\usepackage{xcolor}
\usepackage{lineno}
\usepackage{amsthm}
\newtheorem{remark}{Remark}

\usepackage{bm}
\usepackage{ragged2e}
\justifying
\usepackage{amsmath}
\usepackage{graphicx}
\journal{}
\begin{document}
\begin{frontmatter}
\title{Solving inverse problems using conditional invertible neural networks}
\author[label1]{Govinda Anantha Padmanabha}
\ead{ganantha@nd.edu}
\author[label1]{Nicholas Zabaras\corref{cor1}}
\ead{nzabaras@gmail.com}
\ead[url]{https://www.zabaras.com/}
\address[label1]{Scientific Computing and Artificial Intelligence (SCAI) Laboratory, University of Notre Dame, 311 Cushing Hall, Notre Dame, IN 46556, USA}
\cortext[cor1]{Corresponding author}
\begin{abstract}
Inverse modeling for computing  a high-dimensional spatially-varying property field from indirect sparse and noisy observations is a challenging problem. This is due to the complex physical system of interest often expressed in the form of multiscale PDEs, the high-dimensionality of the spatial property of interest, and the incomplete and noisy nature of observations. To address these challenges, we develop a model that maps the given observations to the unknown input field in the form of a surrogate model. This inverse surrogate model will then allow us to estimate the unknown input field for any given sparse and noisy output observations. Here, the inverse mapping is limited to a broad prior distribution of the input field with which the surrogate model is trained. In this work, we construct a two- and three-dimensional inverse surrogate models consisting of an invertible and a conditional neural network trained in an end-to-end fashion with limited training data. The invertible network is developed using a flow-based generative model. The developed inverse surrogate model is then applied for an inversion task of a multiphase flow problem where given the pressure and saturation observations the aim is to recover a high-dimensional non-Gaussian log-permeability field where the two facies consist of heterogeneous log-permeability and varying length-scales. For both the two- and three-dimensional surrogate models, the predicted sample realizations  of the non-Gaussian log-permeability field are diverse with the predictive mean being close to the ground truth even when the model is trained with limited data. 
\end{abstract}
\end{frontmatter}
% \linenumbers
\section{Introduction}
Inverse problems arise in many fields of engineering and science. An inverse problem aims to recover the unknown input field; however, we only have access to some indirect measurements that are the output of a physical process. For example, in reservoir engineering, it is important to find out the properties of the subsurface, such as the permeability field, given various measurements of geophysical fields. Other applications where inverse problems arise include ocean dynamics~\cite{herbei2008gyres}, medical imaging~\cite{aguilo2010inverse}, seismic inversion~\cite{russell1988introduction}, remote sensing~\cite{haario2004markov} and many others. Inverse problems are often ill-posed, i.e., one may not be able to uniquely recover the input field given the noisy and incomplete observations. This is a challenging problem, where one is interested in recovering the unknown input field given the sparse observations. To address this challenge,
most common methods used are the  Markov chain Monte Carlo methods (MCMC) method~\cite{vrugt2016markov,laloy2012high} and the ensemble-based methods~\cite{sun2009sequential,ju2018adaptive}. However, there is a challenge associated with these methods since often the forward model is computationally expensive to evaluate. Several methods have been developed in the past using inexpensive surrogate forward models such as the Gaussian process~\cite{bilionis2013solution, rasmussen2003gaussian} or polynomial chaos expansion~\cite{marzouk2009dimensionality, xiu2002wiener}. In addition to accurate surrogate models, one is also interested propagating uncertainties from the data-driven surrogate model to the inverse solution. 
\par
In recent years, enormous progress has been made in the solution of inverse problems using deep learning techniques~\cite{laloy2017inversion,canchumuni2019history,mo2020integration,hamilton2018deep}. Examples of deep learning applied for solving inverse problems in various fields of engineering and science are medical imaging~\cite{jin2017deep,hamilton2018deep,adler2017solving,fan2020solving,li2019novel}, compressed sensing~\cite{whang2020compressed,mardani2017deep},  reservoir engineering~\cite{laloy2017inversion, mo2020integration, Mo2019Deep,laloy2018training} and many more. Recently deep generative models (DGM) such as~\cite{goodfellow2014generative, kingma2013auto} have been used for learning a low-dimensional representation  of the high-dimensional input, where the input field is spatially represented within the latent space manifold and therefore making the inversion process possible using either MCMC~\cite{laloy2017inversion} or ensemble-based methods~\cite{mo2020integration}. The deep neural network (DNN) is now widely used for surrogate modeling~\cite{Densenet,res, Alom}, and it has been applied to various science and engineering fields~\cite{Ian, DNN_main}. For example, flow through porous media~\cite{Yinhao, Bilionis, Mo, zhu2019physics} or turbulence modeling~\cite{Nick, TUM}. In the deep learning community, various convolutional neural network architectures (CNN)~\cite{krizhevsky2012imagenet}  have been successfully implemented to develop various generative models that aim to learn the data generation mechanism.  Examples of deep generative models applied to various fields are fluid mechanics~\cite{xie2018tempogan}, parameterization of geological models~\cite{chan2017parametrization}, porous materials~\cite{mosser2018stochastic,mosser2017reconstruction}, and others. In recent years, generative adversarial networks (GANs)~\cite{goodfellow2014generative}, variational autoencoders (VAE)~\cite{kingma2013auto, sonderby2016ladder, hsu2018scalable}, and flow-based models~\cite{dinh2016density, kingma2018glow, dinh2014nice} have made great progress as generative models. Although there are various works in the past on parameterizing the input space for the inversion that is modeled using principal component analysis (PCA) and its variants~\cite{vo2014new, sarma2008kernel}, these methods are limited in terms of accuracy when solving a non-trivial physical problem with complex non-Gaussian fields~\cite{canchumuni2019history}.  Hence, researchers have been interested in developing efficient methods for solving the inverse problem with complex non-Gaussian fields, and one possible way is to learn a low-dimensional representation of the input space using DGM and then implement the above mentioned MCMC or the ensemble-based methods for inversion task~\cite{laloy2017inversion,canchumuni2019history,laloy2018training}. For example, Laloy et al.~\cite{laloy2017inversion} developed a Bayesian framework for the inversion of the permeability field using VAE~\cite{kingma2013auto} and build a low-dimensional representational of the input field using  encoder and decoder layers consisting of convolutional layers~\cite{krizhevsky2012imagenet}. 
\par
While the above works concentrate on non-Gaussian fields where the two facies consists of homogeneous permeability, i.e., the high-permeability channels and the low-permeability non-channels, each consists of homogeneous permeability. However, it is vital to study non-Gaussian fields where the two facies consist of heterogeneous permeability from a practical perspective. Recently, Mo et al.~\cite{mo2020integration}, developed an inversion framework by combining the convolutional adversarial autoencoder, deep residual dense convolutional network, and an iterative local updating ensemble smoother (ILUES) algorithm~\cite{zhang2018iterative} for the  inversion of a non-Gaussian log-permeability field with heterogeneous log-permeability within each facies. Here, the convolutional adversarial autoencoder is used for parameterizing the high-dimensional input space; the deep residual dense convolutional network is implemented as the forward surrogate model and the ILUES for inverse modeling.  While the above inversion framework is implemented for a high-dimensional and highly-complex forward model, the length-scale was considered to be constant for the non-Gaussian log-permeability field.
\par
 In this work, we develop an inverse surrogate model for performing an inversion task for a dynamic multiphase flow problem which aims to recover a high-dimensional non-Gaussian log-permeability field with heterogeneous log-permeability within each facies, and varying length scales given the sparse and noisy pressure observations and the saturation observations over  time. In general, inverse problems are ill-conditioned i.e., the number of data (observation data) may not be able  to uniquely recover the high dimensional input permeability field. One way to address the challenge is to learn a low-dimensional representation of the input space using a deep generative model (DGM) and then perform the inversion task.  All the above-mentioned inversion works~\cite{laloy2017inversion,canchumuni2019history,mo2020integration} which use the concept of the DGM are implemented using GANs~\cite{goodfellow2014generative} or VAEs~\cite{kingma2013auto}. However, few drawbacks are associated with these conditional deep generative models, such as training stability issues and hard to obtain samples with sharp features~\cite{ardizzone2019guided}. Recently proposed flow-based models, namely, real NVP~\cite{dinh2016density} have been successfully implemented for various scientific fields such as computer vision~\cite{kingma2018glow,ardizzone2019guided}, speech synthesis~\cite{prenger2019waveglow} and, physical systems~\cite{geneva2020multi,zhu2019physics} and have been shown to perform well without any gradient vanishing or exploding issues when training the model. These flow-based models provide an exact log-likelihood evaluation and allow more stable training  than GANs~\cite{goodfellow2014generative,ardizzone2019guided}.
 \par
Our work focuses on developing a two- and a three-dimensional inverse surrogate model that seeks to recover a high-dimensional non-Gaussian log-permeability field given the sparse and noisy observations. The model mainly consists of two networks, namely, an invertible network and a conditional network. Both   networks are trained together in an end-to-end fashion with limited training data. The invertible architecture is constructed using the concepts of real NVP~\cite{dinh2016density} that consists of several affine coupling layers at multiple scales. The conditioning network is also constructed in a  multiscale structure, so that the sparse and noisy observations are passed on to the higher-resolution features of the invertible network at multiple feature sizes. The multiscale architecture of the inverse surrogate and the exact conditional log-likelihood evaluation through the change of variables helps in capturing the ill-posedness nature that exists in an inverse problem. Here, we show the inversion task of recovering a high-dimensional non-Gaussian log-permeability field given the pressure and saturation observations at specific locations of the domain. The saturation observations are collected at   different time instances. Furthermore, in order to evaluate the efficiency of our inverse surrogate, we train our model with two different configurations that mainly depend on the number of saturation observations, i.e., for one configuration, we consider the saturation observations at the last time instant and the other configuration, we consider the saturation observations over a specified interval of time. In addition, we also train both our $2$-D and $3$-D model with different number of training data and different levels of observation noise. 
 \par
This paper is organized as follows. In Section~\ref{sec:PD}, we introduce the multiphase model and define the problem of interest. In Section~\ref{sec:methods}, we discuss the generative model, conditional invertible network with details regarding its training given in Section~\ref{sec:train}. In Section~\ref{sec:results}, the proposed model is evaluated using a synthetic example for each of the flow models. Finally, we present a summary of our work and address potential extensions in Section~\ref{sec:summary}. 
\section{Governing Equations and Problem Definition} \label{sec:PD}
In this work, we consider a two-phase flow through a heterogeneous porous media. Let $p_{o}, p_{w}, s_{o}, s_{w}$ and $\lambda_{o}, \lambda_{w}$ represent the pressure, the saturation fields, and the total mobilities of the oil (o) and water (w) phases, respectively. Here, $s_{w}+s_{o}=1$ and the capillary pressure is defined as $p_{\mathrm{c}}=p_{o}-p_{w}$. Now the governing equation for a multi-phase flow for a spatial domain $\Omega$ can be written as follows~\cite{aarnes2007introduction}:
\begin{equation}
    \nabla\cdot \bm{u} = q, \qquad \bm{u}=-\bm{K} \lambda\left(s_{w}\right) \nabla p,
\end{equation}
\begin{equation}\label{saturation_equation}
\phi \frac{\partial s_{w}}{\partial t}+\nabla \cdot\left(f_{w}\left(s_{w}\right) \bm{u}\right)={q}_{w}.
\end{equation}
Here, $p$ is the pressure field, $\bm{u}$ is the velocity field, $q$ is the source term for the water injection, ${q}_{w}$ is the source term for the saturation Eq.~(\ref{saturation_equation}) and $\bm{K}$ is the permeability. We consider no-flux boundary conditions: $\bm{u} \cdot \hat{\bm{n}}=0$ on $\partial \Omega,$ where $\hat{\bm{n}}$ is the unit normal of the boundary and zero initial saturation at time t : $s_{w}(t=0)=0$ on $\Omega$. The fractional flow $f_{w}$ is given as:
\begin{equation}
f_{w}=\frac{\lambda_{w}}{\lambda_{w}+\lambda_{o}}.
\end{equation}
The governing equations are solved using a finite-volume formulation implemented in MRST~\cite{software} with a standard two-point flux-approximation (TPFA) method for pressure and an implicit transport solver using single-point upstream mobility weighting for the  saturation~\cite{lie2019introduction}. Here, the log-permeability field $\bm{x} \in \mathbb{R}^{M}$ is represented on a two-dimensional structured, regular grids of $H \times W$ on which the simulation for the above physical system is solved, where $H$ is the height, $W$ is the width and $M = n_h \times n_w$, where $n_h$ and $n_w$ represents the number of grid points in the two axes of the spatial domain.
Similarly, for the three-dimensional problem, we consider regular grids of $H \times W \times D$, where $D$ is the depth and  $M = n_h \times n_w \times n_d$, where $n_h$, $n_w$, and $n_d$ represent the number grid points in the three axes of the spatial domain. Note that no dimensionality reduction is applied to the unknown permeability and thus  the number of unknown values that we aim to recover in this inverse problem is equal to the number of grid points.

\par
In this work, we are interested in identifying the heterogeneous log-permeability field in an oil reservoir given pressure and saturation measurements at specific locations on the domain, and therefore, we pose this problem as an inverse problem. This problem is ill-posed and considering the incomplete noisy pressure and saturation observations we may not be able  to uniquely recover the high-dimensional non-Gaussian permeability field. Our approach will be based on the development of a   surrogate model that maps noisy measurements at the sensor locations to the corresponding high-dimensional log-permeability field. This approach will be based on deep generative models~\cite{dinh2016density}. The training of the model will be based on pressure and saturation, log-permeability pairs obtained from forward simulations. The training log-permeability data will be sampled from a broad prior distribution of the input field that contains the solution to the inverse problem.
\section{Methodology} \label{sec:methods}
In this section, we start with the definition of  the inverse problem in Section~\ref{sec:IP}. Then, we present the details of the generative model in Section~\ref{sec:Generative} and finally, we provide the details of the conditional invertible neural network in Section~\ref{sec:cinn}.  
\subsection{Inverse problem}\label{sec:IP}
The inverse problem aims at recovering the log-permeability field $\bm{x} \in \mathbb{R}^{M}$  given the noisy observations (pressure and saturation) $\bm{y} \in \mathbb{R}^{D} $ at specific locations of the domain. It can be summarized as follows:
        \begin{equation}
            \bm{y}=\mathcal{\bm{S}}(\bm{x})+\bm{e}.
        \end{equation}
The mapping $\mathcal{\bm{S}}: \bm{x}\rightarrow \bm{y}$ is the forward operator and $\bm{e} \in \mathbb{R}^{D}$ is the observational noise.   With known forward operator $\mathcal{\bm{S}}$,   the regularized (deterministic) solution of this inverse problem can be written as:
\begin{equation} 
\bm{x}^{*}=\min _{\bm{x}} \mathcal{L}(\mathcal{\bm{S}}(\bm{x}), \bm{y})+\lambda \mathcal{R}(\bm{x}),
\end{equation}
where $\mathcal{R}$ is the regularizer (such as TV regularizer~\cite{rudin1992nonlinear}) and $\lambda$ is a regularization
parameter.  
\subsection{Inverse surrogate with limited forward problem evaluations}
Probabilistic solutions of inverse problems are often addressed using the Bayesian formulation, i.e., defining a prior probability measure over the unknown permeability $p(\bm{x})$ and using Bayes' rule to produce samples of the   posterior   $p(\bm{x}|\bm{y})$   using the likelihood $p(\bm{y}|\bm{x})$:
\begin{eqnarray}
p(\bm{x} \mid \bm{y}) \propto p(\bm{y} \mid \bm{x}) p(\bm{x}).
\label{eq:Bayes}
\end{eqnarray}
The likelihood $p(\bm{y}|\bm{x})$ is   computed by evaluating the forward model $\mathcal{S}$. There are two challenges associated with this approach. First,   the likelihood is computationally expensive since the physical system is non-trivial, and the approach requires multiple evaluations of the forward model. Several works have been developed in the past to address this problem that use  surrogate forward models such as Gaussian processes~\cite{bilionis2013solution} or  polynomial chaos expansion based models~\cite{marzouk2009dimensionality}.  However, it is important that one has to obtain good accuracy when evaluating the forward model using   surrogate models as well as propagate the surrogate model uncertainties to the inverse solution. Second, if the dimension of the quantity of interest is high, then it is not a trivial task to obtain the posterior samples. Altogether, it is often computationally not effective if one is directly implementing Eq.~(\ref{eq:Bayes}). 
For example, in~\cite{bilionis2013solution}, a Bayesian surrogate model was constructed for solving the inverse problem described in Section~\ref{sec:PD}. The forward model was replaced using a forward surrogate based on Gaussian processes (GP) and the model was trained using up to $10240$ forward simulations. The aim  was to recover a $12-$dimensional input (Karhunen-
Loève (KL) coefficients   of the input field) given noisy and sparse observations. The authors showed that the accuracy on the mean posterior of the input field gradually increases as the number of training data for the forward GP model increases. In general, as the dimensionality of the input field increases, the number of training data required for the forward surrogate model increases for a desired accuracy~\cite{Yinhao}.
Therefore, one requires to perform a large number of forward simulations as the dimensionality of the data increases.

In the present work, we aim to bypass Bayes' rule and directly build an inverse surrogate model for the   posterior distribution $p(\bm{x} \mid \bm{y})$ using a finite set of forward evaluations of the pressure and saturation values at the sensor locations that result from various permeability fields sampled from a prior
model $ p(\bm{x})$.  We construct the inverse surrogate as follows. First, we assume that a set of training data are available either from observations or using the forward model:
$\mathcal{D}=\left\{\bm{x}^{(i)}, \tilde{\bm{y}}^{(i)}\right\}_{i=1}^{N}$, where $N$ is the size of the training data set. The inverse surrogate will use this training data set to build a direct mapping from observations $\tilde{\bm{y}}$ to the corresponding log-permeability
$\bm{x}\in p(\bm{x})$.
Here, the conditional deep generative network consists of an invertible network and a conditioning network. The conditioning network takes in the sparse and noisy observations $\tilde{\bm{y}}$ as the input with the output being the conditioning inputs to the invertible network at different  scales. During training, the conditioning network provides smooth multiscale feature representations of the given sparse and noisy observations as if smooth observations were provided over the entire domain (or feature domain). Therefore, the multiscale nature of the network along with the regularized loss function discussed in Section~\ref{sec:loss} help in addressing the ill-posed nature of the inverse problem.

\par

In the present study, we construct  two- and three-dimensional conditional invertible networks  (cINN). The cINN consists of two components, namely, the conditioning network and the invertible network with the model parameters $\bm{\theta}_g$ and $\bm{\theta}_I$, respectively. Both the conditioning  and  invertible networks are trained in an end-to-end fashion with the regularized loss function discussed in Section~\ref{sec:loss}. Once the cINN parameters $\bm{\theta}=[\bm{\theta}_g, \bm{\theta}_I]$ are computed during the training process, we then proceed to estimate the unknown input field given the noisy and sparse observations, i.e., the conditional density $p_{\bm{\theta}}(\bm{x}|\tilde{\bm{y}})$ using the conditioning network and invertible networks. We will show that with a few forward evaluations, one can recover a high-dimensional input $\bm{x}$ field given the noisy observations $\tilde{\bm{y}}$ without constructing a forward surrogate model. In particular, we solve inverse problems with dimensionality of $4096$ and $16384$ for   two- and three-dimensional problems, respectively, and with a finite number of forward simulations up to $10000$.

\begin{remark}
\normalfont{
Contrary to standard Bayesian inference approaches to inverse problems, this work builds an inverse surrogate model that once trained can solve many inverse problems (e.g. with different observation data). The inverse surrogate models cannot be directly compared with Bayesian inverse models as e.g. variability   in the inverse solution using inverse surrogates is mainly due to the generative nature of the model whereas in Bayesian inference uncertainties are affiliated with  noisy observations and model error.}
\end{remark}

\par
Before proceeding to the details of the conditional deep generative network, we briefly discuss some of the essentials of the flow-based generative model that stand as the backbone to the conditional deep generative network.

\subsection{Flow-based generative model}
\label{sec:Generative}
\begin{figure}[H]
  \centering
\includegraphics[scale=0.3]{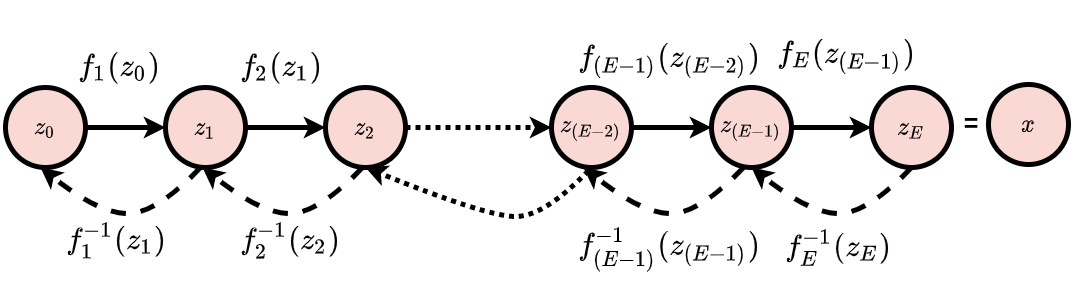}
\caption{Plate diagram of the flow-based generative model.}
\label{flow}
\end{figure}

The flow-based generative model learns the data generative mechanism using normalizing flows. Normalizing flows basically consist of a sequence of invertible transformations that can transform a simple distribution $p_{0}(\bm{z}_{0})$ to a complex distribution $p(\bm{x})$ as illustrated in Fig.~\ref{flow}.    
%
%==============================
One can obtain the complex distribution $p(\bm{x})$ from a simple distribution $p_{0}(\bm{z}_0)$ in the following steps. For simplicity, let us first derive the relationship between variable $\bm{z}_{(E-1)}$ and $\bm{z}_{(E-2)}$. First, we sample $\bm{z}_{(E-2)} \sim p_{(E-2)}(\bm{z}_{(E-2)})$. Then, we consider a bijective function $f$ that maps between the variable $\bm{z}_{(E-1)} \in \mathbb{R}^{M}$ and $\bm{z}_{(E-2)} \in \mathbb{R}^{M}$, such that $\bm{z}_{(E-1)}  = f_{(E-1)}(\bm{z}_{(E-2)})$ and $\bm{z}_{(E-2)} = f_{(E-1)}^{-1}(\bm{z}_{(E-1)})$. Next, using change of variables, we can compute $p_{(E-1)}(\bm{z}_{(E-1)})$:
        \begin{equation} 
            p_{(E-1)}\left(\bm{z}_{(E-1)}\right)=p_{(E-2)}\left(\bm{z}_{(E-2)}\right)\left|\operatorname{det} \frac{d f_{(E-1)}^{-1}\left(\bm{z}_{(E-1)}\right)}{d \bm{z}_{(E-1)}}\right|.
        \end{equation}
Now taking log on both sides and further simplifying the above equation leads to the following:
\begin{equation}
     \log p_{(E-1)}\left(\bm{z}_{(E-1)}\right)= \log p_{(E-2)}\left(\bm{z}_{(E-2)}\right)-\log \left|\operatorname{det} \frac{d f_{(E-1)}\left(\bm{z}_{(E-2)}\right)}{d \bm{z}_{(E-2)}}\right|.
 \end{equation}
 Finally, the above equation can be  extended to obtain the complex distribution $p(\bm{x})$ where the transformation function should satisfy the following properties. First, the transformation must be invertible. Next, the Jacobian determinant should be easy to compute, and lastly, the input and the output dimensions must be the same. Therefore, the complex distribution $p(\bm{x})$ given $\{\bm{z}_{E-1},\ldots,\bm{z}_1,\bm{z}_0\}$ and the bijective function $\{f_{j}\}_{j=1}^{E}$ can be computed as follows:
     \begin{equation}\label{eq:NF}
        \log p(\bm{x}) = \log p_0(\bm{z}_0)-\sum_{j=1}^{E} \log \left|\operatorname{det} \frac{d f_{j}\left(\bm{z}_{j-1}\right)}{d \bm{z}_{j-1}}\right|.
    \end{equation}
%==============================
\par
\begin{figure}[H]  
  \label{test}
  \begin{minipage}[b]{0.48\linewidth}
    \centering
    \includegraphics[scale=0.3]{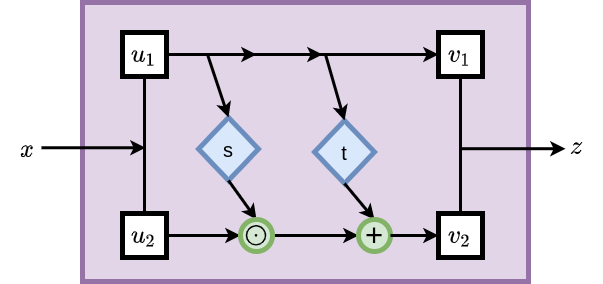}\\{(a)}
  \end{minipage}
  \hspace{0.5cm}
  \begin{minipage}[b]{0.48\linewidth}
    \centering
    \includegraphics[scale=0.3]{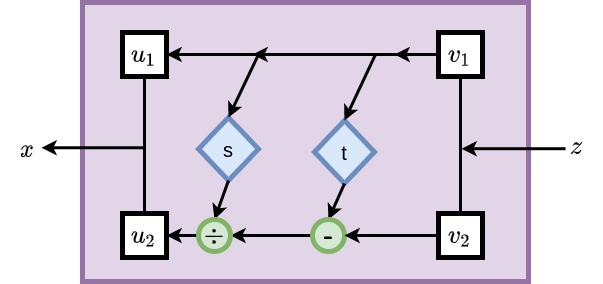}\\{(b)}
  \end{minipage}
 \caption{RealNVP: Illustration of the (a) forward propagation and (b) inverse propagation, where $\bm{u}_1 = \bm{x}_{1: m}$, $\bm{u}_2 = \bm{x}_{m+1: M}$, $\bm{v}_1 = \bm{z}_{1: m}$ and $\bm{v}_2 = \bm{z}_{m+1: M}$.}
    \label{fig:RealNVP}
\end{figure}

Notably, an enormous effort has been made in developing models for normalizing flows~\cite{dinh2016density, kingma2018glow, dinh2014nice}. The real-valued Non-Volume Preserving (RealNVP)~\cite{dinh2016density} model is developed by constructing a sequence of   normalizing flows by invertible bijective transformations, where each transformation    is referred as the \textit{affine coupling layer} that basically satisfies the aforementioned properties. The real NVP model consists of forward and inverse propagation as illustrated in Fig.~\ref{fig:RealNVP}. 

To enumerate the details of the real NVP, let us consider the forward propagation $f: \bm{x} \rightarrow \bm{z}$. Here, $\bm{x}\in\mathbb{R}^{M}$ is considered as the input for the forward propagation.
In the forward propagation, the input is split into two parts. The first part, $\bm{x}_{1:m},~1<m<M$ remains the same whereas the second part,  $\bm{x}_{m+1:M}$, undergoes transformation as follows:
\begin{eqnarray}
&\bm{z}_{1: m} = \bm{x}_{1: m},\nonumber \\
&\bm{z}_{m+1: M} =\bm{x}_{m+1: M} \odot \exp \left(s\left(\bm{x}_{1: m}\right)\right)+t\left(\bm{x}_{1: m}\right),
\end{eqnarray}
where, $s(\cdot)$ and $t(\cdot)$ are scale and translation functions, respectively, that are modeled by  neural networks   trained  in the forward pass (Fig.~\ref{fig:RealNVP}(a)). Here, $\odot$ is the element-wise product. The above transformation is easy to invert and therefore the inverse  propagation (Fig.~\ref{fig:RealNVP}(b)) with   input $\bm{z} \in \mathbb{R}^{M}$ is given as follows:
\begin{eqnarray}
&\bm{x}_{1: m} = \bm{z}_{1: m},\nonumber \\
&\bm{x}_{m+1: M}=\left(\bm{z}_{m+1: M}-t\left(\bm{z}_{1: m}\right)\right) \odot \exp \left(-s\left(\bm{z}_{1: m}\right)\right).
\end{eqnarray}

The Jacobian determinant for the above transformation is given by~\cite{dinh2016density}:
\begin{equation}\label{Jacobian}
\bm{J}=\left[\begin{array}{cc}
{\mathbb{I}_{m}} & {\bm{0}_{m \times(M-m)}} \\
{\frac{\partial \bm{z}_{m+1: M}}{\partial \bm{x}_{1: m}}} & {\operatorname{diag}\left(\exp \left(s\left(\bm{x}_{1: m}\right)\right)\right)}
\end{array}\right].
\end{equation}
Therefore, the Jacobian determinant is easy to compute. This invertible transformation together with Eq.~(\ref{eq:NF})   can be used to transform   a simple distribution $p(\bm{z})$, say, for example, standard Gaussian distribution to a complex distribution $p(\bm{x})$.
\subsection{Conditional invertible neural network}\label{sec:cinn}
In this section, we present the two- and three-dimensional inverse surrogate models by extending the concept of real NVP (as discussed in the previous section) by concatenating the observations ($\tilde{\bm{y}}$) at multiple scales with the aid of a conditioning network to the affine coupling layer. Here, the observations ($\tilde{\bm{y}}$) are not directly concatenated to the affine coupling layer, but rather we construct a conditioning network that will output the conditioning inputs to the invertible network at multiple scales. In this work, we construct the invertible network with several affine coupling layers at multiple scales. Therefore, our conditional invertible neural network (or inverse surrogate) consists of an invertible network and a conditional network that is trained in an end-to-end fashion.
\par
Note that in recent years  several conditional deep generative models have been proposed~\cite{isola2017image, mirza2014conditional, sohn2015learning}. However, there are few drawbacks associated with   existing conditional deep generative models, such as training stability issues and hard to obtain samples with sharp features~\cite{ardizzone2019guided}. Therefore, in this work, we use a flow-based generative model as it provides an exact log-likelihood evaluation~\cite{dinh2016density, zhu2019physics}. The concept of concatenating the conditioning inputs to the scale, $s(\cdot)$, and shift, $t(\cdot)$, networks is similar to the work of Geneva and Zabaras~\cite{geneva2020multi}, Zhu et al.~\cite{zhu2019physics} and Ardizzone et al.~\cite{ardizzone2019guided}. However, in this work, the conditioning and the invertible architecture are modified to incorporate the sparse observations present in the inverse problem.

\subsubsection{Two-dimensional conditional invertible neural network}
In this section, we enumerate the details regarding the construction of a two-dimensional conditional invertible neural network. Aforementioned, in this work, the conditional invertible neural network (inverse surrogate) consists of two networks, namely, the conditioning network and the invertible network. 
\par
\begin{figure}[H]
    \centering
    \includegraphics[scale=0.165]{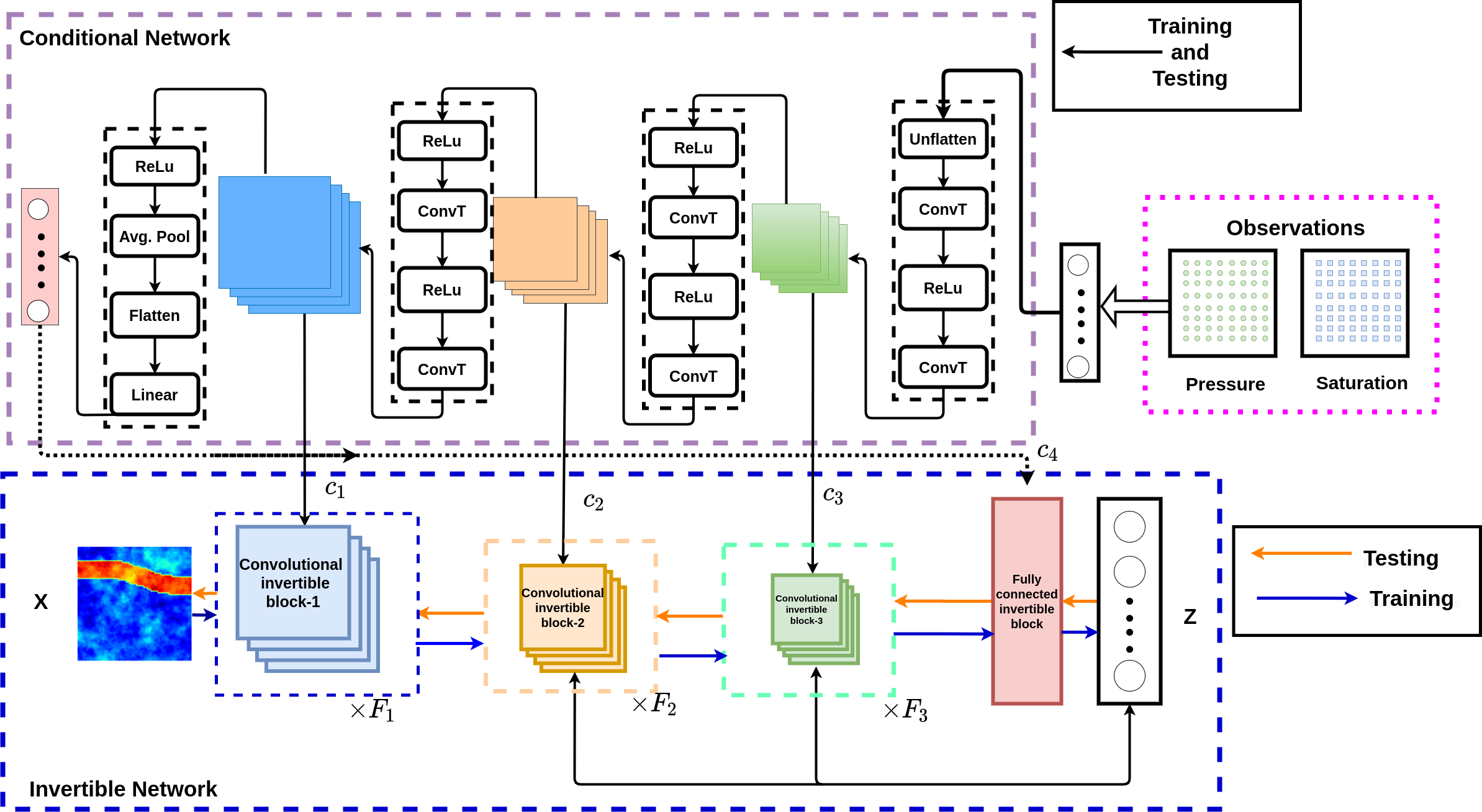}
    \caption{Multiscale $2$-D conditional invertible neural network. Here, the blue dashed box is the invertible network with $\bm{x}$ as the input field and $\{\bm{c}_l\}_{l=1}^{4}$ as the output from the conditional network (indicated in brown dashed box) considered as the input to the invertible network. The figure shows the direction of the training (indicated in blue line) and testing processes (indicated in yellow line).}
    \label{fig:main_fig}
\end{figure}

The conditional network consists of $L$ conditional blocks (in our implementation and the following schematics, $L=4$, see Fig.~\ref{fig:main_fig}). The input to the conditioning network are the observations $\tilde{\bm{y}}$. Here the output features $\{\bm{c}_l\}_{l=1}^{L}$ at each conditional block as they derive from the observations $\tilde{\bm{y}}$ are provided as the conditional input to the invertible network. In the first conditional block, we unflatten the given observations channel-wise, i.e., in this work, the observation to the network is a multi-channel input of size $C_o \times T$, where $C_o$ is the number of observation channels, and $T$ is the number of observations. If we consider the multiphase flow problem of interest, then the number of observation channels are the saturation at different time instances and the pressure. The unflattening operation involves converting the one-dimensional data into two-dimensional data. The resulting unflattened feature maps are then passed to a transposed convolution layer, a $ReLU$ (Rectified Linear Unit), and then a transposed convolution layer to increase the feature size. The $ReLU$ is a non-linear activation function that is defined as $ReLU(\bm{x}) = \max(0,\bm{x})$. The conditional blocks $2$ and $3$ perform the following operations: a $ReLU$ followed by a transposed convolution layer and then again a $ReLU$ followed by a transposed convolution layer, thereby doubling the feature size of the input to the respective conditional blocks. The last conditional block ($L=4$) consists of the following operations: a $ReLU$, average pooling that performs downsampling along the spatial dimensions by dividing the input feature into small patches, and evaluating the average values of each patch and finally  flattening the average pooled feature. The flattened data is then passed through fully-connected layers. Here, the features extracted after each conditional block are considered as the conditional input features to the invertible network. Note that the conditional network has only a forward pass.

\begin{figure}[H]
   \begin{minipage}[b]{0.85\linewidth}
    \centering
    \includegraphics[scale=0.14]{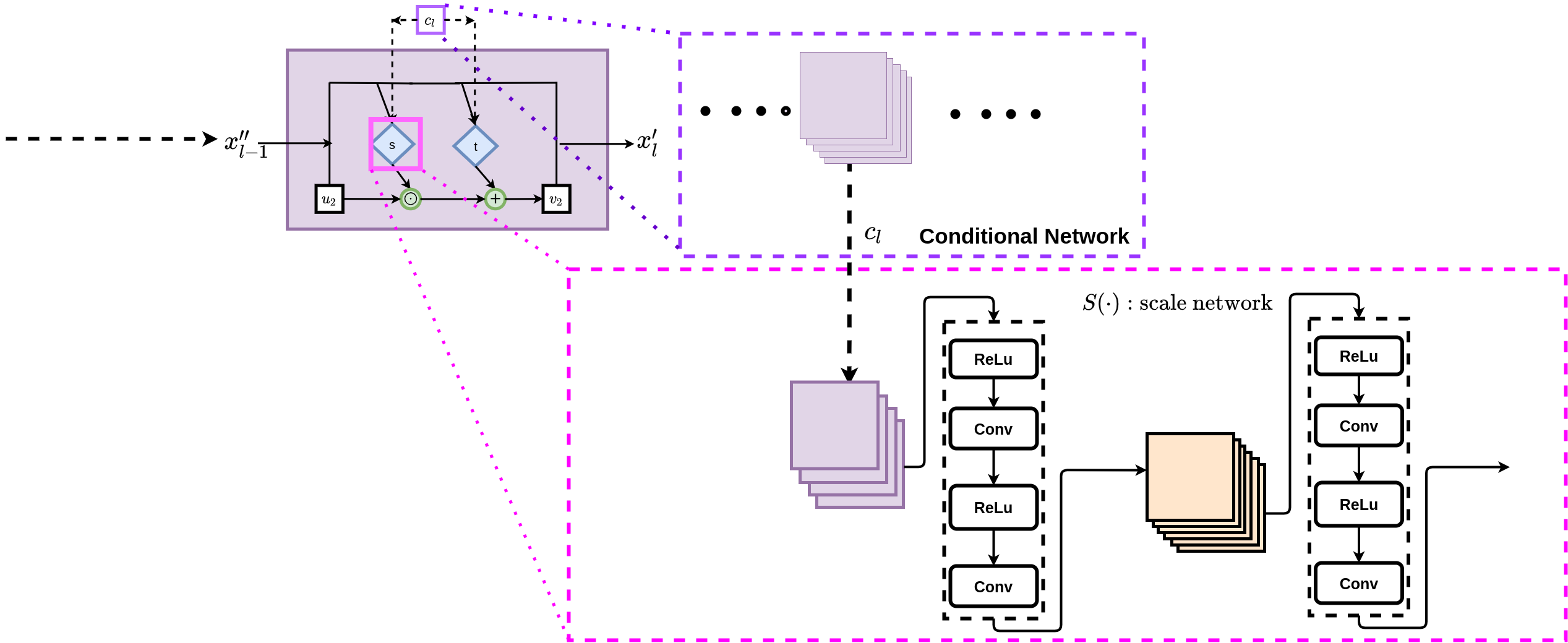}\\
    {(a)} 
  \end{minipage} \\
  \begin{minipage}[b]{0.85\linewidth}
    \centering
    \includegraphics[scale=0.14]{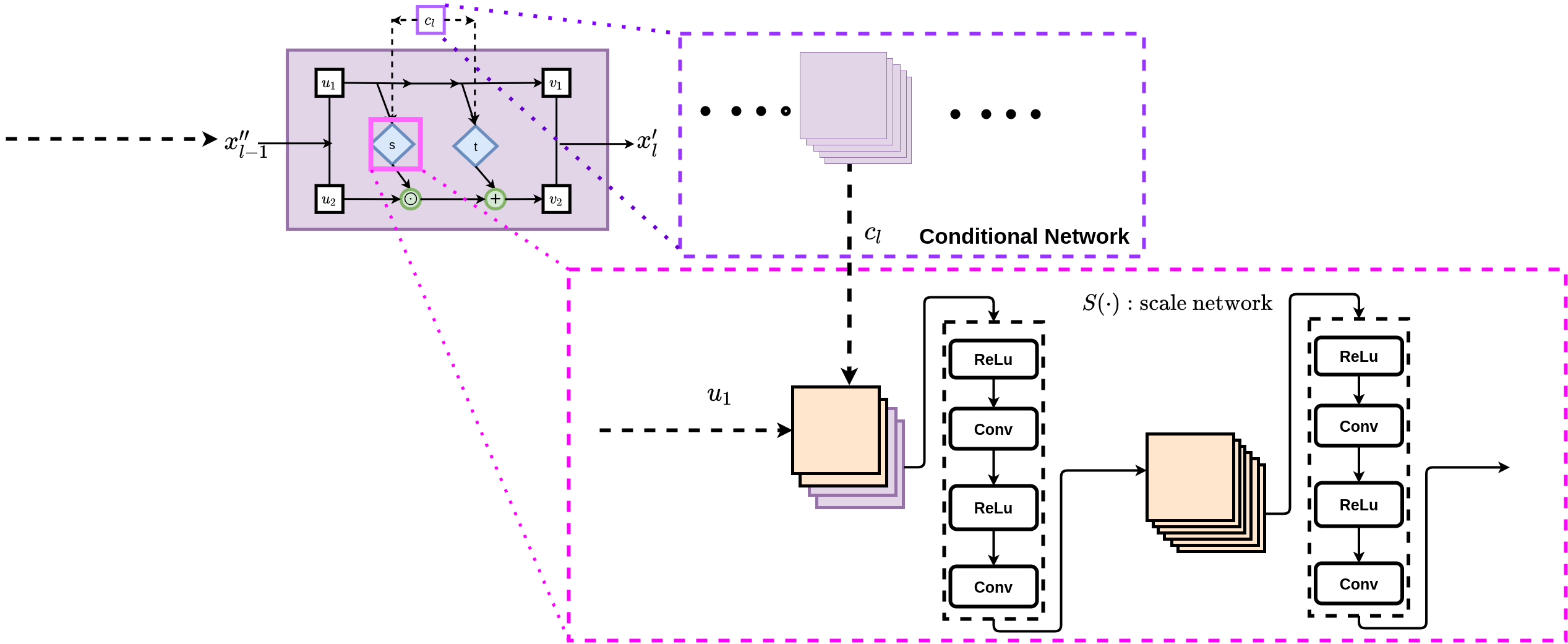}
    \\
    {(b)} 
    \end{minipage}
 \caption{Scale network, $s(\cdot)$, for (a) the convolutional invertible block-$1$. Here, $\bm{u}_2 = \bm{x}$, and $\bm{v}_2 = \bm{x}^{\prime}$, (b) the convolutional invertible block-$2$. In each sub-figure, the box with dashed lines (magenta) illustrates the details of the scale network that consists of two convolutional blocks. Here, $\bm{u}_1 = \bm{x}^{\prime\prime}_{1: m}$, $\bm{u}_2 = \bm{x}^{\prime\prime}_{m+1: M}$, $\bm{v}_1 = \bm{x}^{\prime}_{1: m}$ and $\bm{v}_2 = \bm{x}^{\prime}_{m+1: M}$}
 \label{fig:scale_net}
\end{figure}

\begin{figure}[H]
    \centering
    \includegraphics[scale=0.22]{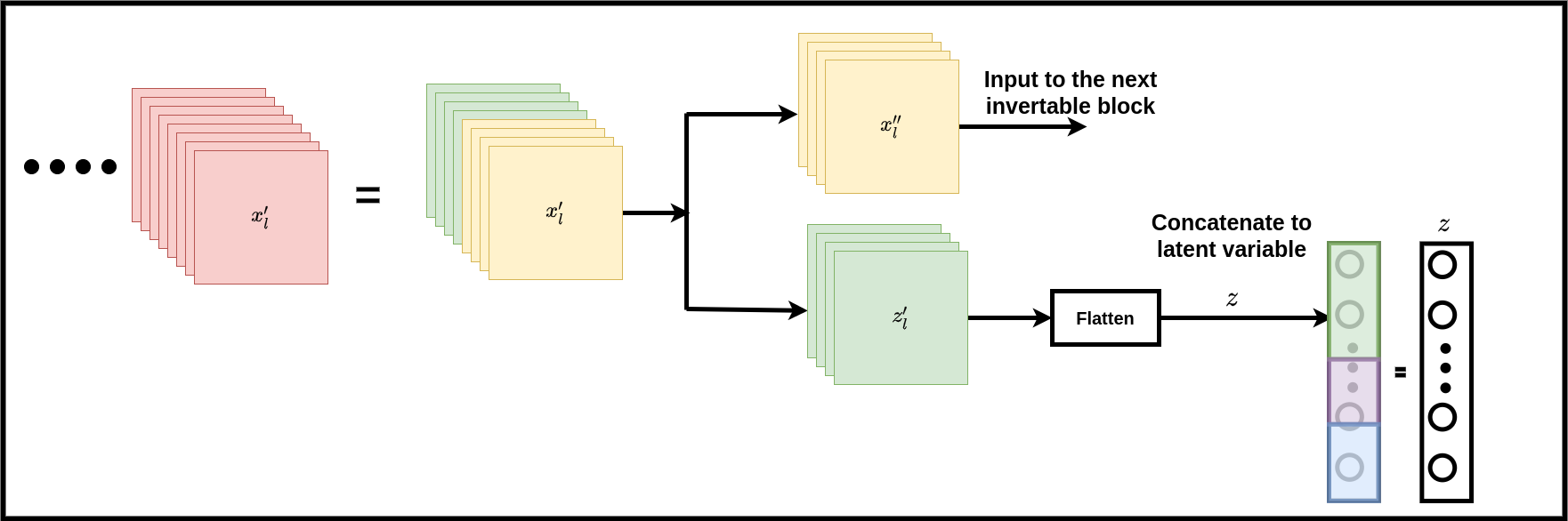}
    \caption{Split forward operation during training, where the output from the previous invertible block $\bm{x}^{\prime}_l$ is divided channel-wise  and the first half $\bm{x}^{{\prime\prime}_l}$ is the input to the next invertible block. The other half of the data $\bm{x}^{\prime}_l$ is concatenated to the latent variable $\bm{z}$.}
    \label{fig:forward_split}
\end{figure}

\begin{figure}[H]
    \centering
    \includegraphics[scale=0.22]{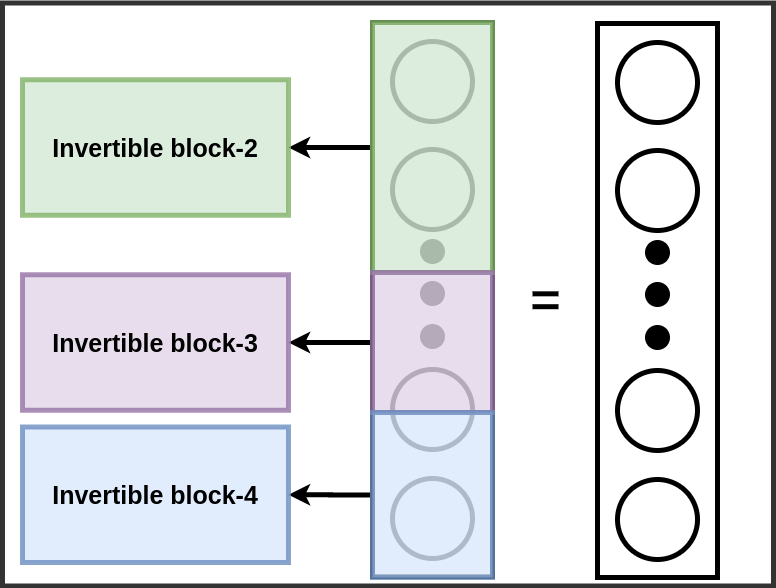}
    \caption{Split inverse operation during inference, where the latent space $\bm{z} \in \mathbb{R}^{M}$ is divided into three parts $1<m_1<m_2<M$ such that the first part $(\bm{z}_{(1:m_1)})$ is concatenated to the invertible block-$2$, the second part $(\bm{z}_{(m_1:m_2)})$ is concatenated to the invertible block-$3$ and the last part $(\bm{z}_{(m_2:M)})$ is provided as an input to the fully-connected invertible block-$4$.}
    \label{fig:inverse_split}
\end{figure}

The invertible architecture transforms the latent space $\bm{z}$ that is concatenated at different scales to the input $\bm{x}$ conditioned on $\{\bm{c}_l\}_{l=1}^{L}$, where $L$ is number of invertible blocks. In this work, we construct $4$ invertible blocks with the first three  blocks being convolutional invertible blocks and the last invertible block being a fully-connected block. Here, we denote $\{{\bm{x}^{\prime}\}_{l=1}^{L}}$ as the output of each invertible blocks. The multiscale architecture is constructed such that the output $\{{\bm{x}^{\prime}\}_{l}}$ of each convolutional invertible block is half the feature map size of its input $\{{\bm{x}^{\prime}\}_{l-1}}$.  Figure~\ref{fig:main_fig} illustrates the invertible and conditional multiscale architectures. The output features $\{\bm{c}_l\}$ from the conditional network are provided as conditional input to the invertible network at different scales. More specifically, the output features $\{\bm{c}_l\}$ are concatenated with first half of the input features ($\bm{x}_{1:m}$ (during forward propagation) or $z_{1:m}$ (during inverse propagation)) before they are passed to the scale ${s}(\cdot)$ and shift $t(\cdot)$ networks. Figure~\ref{fig:scale_net}(b) shows the output features $\{\bm{c}_l\}$ are concatenated to the first half of the input data, $x_{1:m}$,  before passing to the scale ${s}(\cdot)$. The details regarding the forward and reverse of each affine coupling layer with the conditioning of the output features $\{\bm{c}_l\}$ are provided in  Table~\ref{tab:affine_layer}.  
\begin{table}[H]
\caption{Forward and inverse propagation of each affine coupling layer  conditioned on the output features $\{\bm{c}_{l}\}$. During forward propagation, the input is $\bm{x}^{\prime\prime}_{l-1} \in \mathbb{R}^{M}$ and the output is $\bm{x}^{\prime}_l \in \mathbb{R}^{M}$. During inverse propagation, the input is the latent space $\bm{z}^{\prime}_{l-1}$ and the output is $\bm{z}^{\prime}_{l}$. Here, the split operation divides the data into two parts: $1<m<M$, such that $\bar{\bm{x}}_{1,{l-1}}$ = $\bm{x}^{\prime\prime}_{(1:m),\{{l-1}\}}$ and $\bar{\bm{x}}_{2,{l-1}}$ = $\bm{x}^{\prime\prime}_{(m:M),\{{l-1}\}}$. Similarly, $\bar{\bm{z}}_{1,{l-1}}$ = $\bm{z}^{\prime}_{(1:m),\{{l-1}\}}$ and $\bar{\bm{z}}_{2,{l-1}}$ = $\bm{z}^{\prime}_{(m:M),\{{l-1}\}}$.} 
\centering
\begin{tabular}{l|l}
 Forward & Inverse \\ \hline
 $\overline{\bm{x}}_{1,{l-1}}, \overline{\bm{x}}_{2,{l-1}}$={split}$\left(\bm{x}^{\prime\prime}_{{l-1}}\right)$&  $\overline{\bm{z}}_{1,,{l-1}}, \overline{\bm{z}}_{2,{l-1}}$={split}$\left(\bm{z}^{\prime}_{l-1}\right)$\\
$\hat{\bm{x}}_{1,{l-1}}$={concat}$\left(\overline{\bm{x}}_{1,{l-1}}, \bm{c}_{l}\right)$ &  $\hat{\bm{z}}_{1,{l-1}}$={concat}$\left(\overline{\bm{z}}_{1,{l-1}}, \bm{c}_{l}\right)$\\
 $({\bm{s}}, \bm{t})$={AffineCouplingNetwork}$\left(\hat{\bm{x}}_{1,{l-1}}\right)$& $({\bm{s}}, \bm{t})$={AffineCouplingNetwork}$\left(\hat{\bm{z}}_{1,{l-1}}\right)$ \\
$\overline{\bm{x}}_{2,{l}}=\exp(\bm{s}) \odot \overline{\bm{x}}_{2,{l-1}}+\bm{t}$ & $\overline{\bm{z}}_{2,l}=\left(\overline{\bm{z}}_{2,{l-1}}-\bm{t}\right) / \exp(\bm{s})$ \\
 $\overline{\bm{x}}_{1,{l}}=\overline{\bm{x}}_{1,{l-1}}$& $\overline{\bm{z}}_{1,{l}}=\overline{\bm{z}}_{1,{l-1}}$  \\
 $\bm{x}^{\prime}_{l}$={concat}$\left(\overline{\bm{x}}_{1,{l}}, \overline{\bm{x}}_{2,{l}}\right)$& $\bm{z}^{\prime}_{l}$={concat}$\left(\overline{\bm{z}}_{1,{l}}, \overline{\bm{z}}_{2,{l}}\right)$
\end{tabular}
\label{tab:affine_layer}
\end{table}
\noindent \textit{Single-side affine conditional coupling layer:} 
As mentioned earlier in Section~\ref{sec:methods}, in a conventional affine coupling layer, the data is split into two parts such that the first part remains unchanged, whereas the second part goes through a transformation. In~\cite{dinh2016density}, the data is divided into two parts using the masking schemes, where the data is split spatially in a checkerboard pattern. However, in the problem considered in this work, we construct a single-side affine conditional coupling layer for the convolutional invertible block-$1$, such that we process the given input as a single entity without dividing it into two components. In this manner,  we retain the high-resolution information of the input during both the forward and inverse propagation processes. Here, the input field, i.e., the log-permeability field is a single channel input feature of size $C \times H \times W$, where $C$ is the number of channels (here, $C= 1$), $H$ is the height, and $W$ is the width of the feature maps. Fig.~\ref{fig:scale_net}(a) illustrates the single affine conditional layer where the output features $\bm{c}_3$ from the conditioning network are considered as the input to the scale ${s}(\cdot)$ and the input field $\bm{x}$ undergoes transformation along with the outputs from the scale ${s}(\cdot)$ and shift network ${t}(\cdot)$.
Other convolutional invertible blocks consist of the regular affine coupling layer as enumerated in Section~\ref{sec:Generative} and shown in~\ref{fig:RealNVP}. During the forward propagation, we split the output $\{{\bm{x}^{\prime}\}_{l=1}^{L}}$ of each convolutional invertible block channel-wise into two halves, where the first half $\bm{x}^{\prime\prime}_l$ is considered as the input to the next invertible block and the other half of the data $\bm{z}^{\prime}_l$ is concatenated to the latent variable $\bm{z}$ as shown in Fig.~\ref{fig:forward_split}. During inverse propagation, we divide  the latent dimension $\bm{z}$ into three parts $1<m_{1}<m_{2}<M$ such that the first  convolutional part  $\bm{z}^{(j)}_{(1:m_1)}$ is concatenated to the invertible block-$2$, where we first unflatten the data and then concatenate. Similarly, the second part,  $\bm{z}^{(j)}_{(m_1:m_2)}$, is first unflattened and then concatenated to the invertible block-$3$ and the last part $(\bm{z}^{(j)}_{(m_2:M)})$ is provided as the input to the fully-connected invertible block-$4$ as shown in Fig.~\ref{fig:inverse_split}. We also perform permutation that reverses the ordering of the channels~\cite{dinh2016density} between two invertible blocks. The reasons for constructing the conditional network and the invertible network with multiscale features are the following. First, training the network with multiscale features helps in reducing the computational cost rather than training the model~\cite{ardizzone2018analyzing} with constant feature size. Second, conditioning the observations at multiscale features helps in the inference stage, i.e., the samples generated in the inference stage will be diverse and, at the same time, dependent on the observations. Lastly,  the conditioning network is constructed in a multiscale structure, so that sparse information (or observation) is passed on to the higher-resolution features of the invertible network at multiple feature sizes.
\subsubsection{Three-dimensional conditional invertible neural network }
\begin{figure}[H]
    \centering
    \includegraphics[scale=0.14]{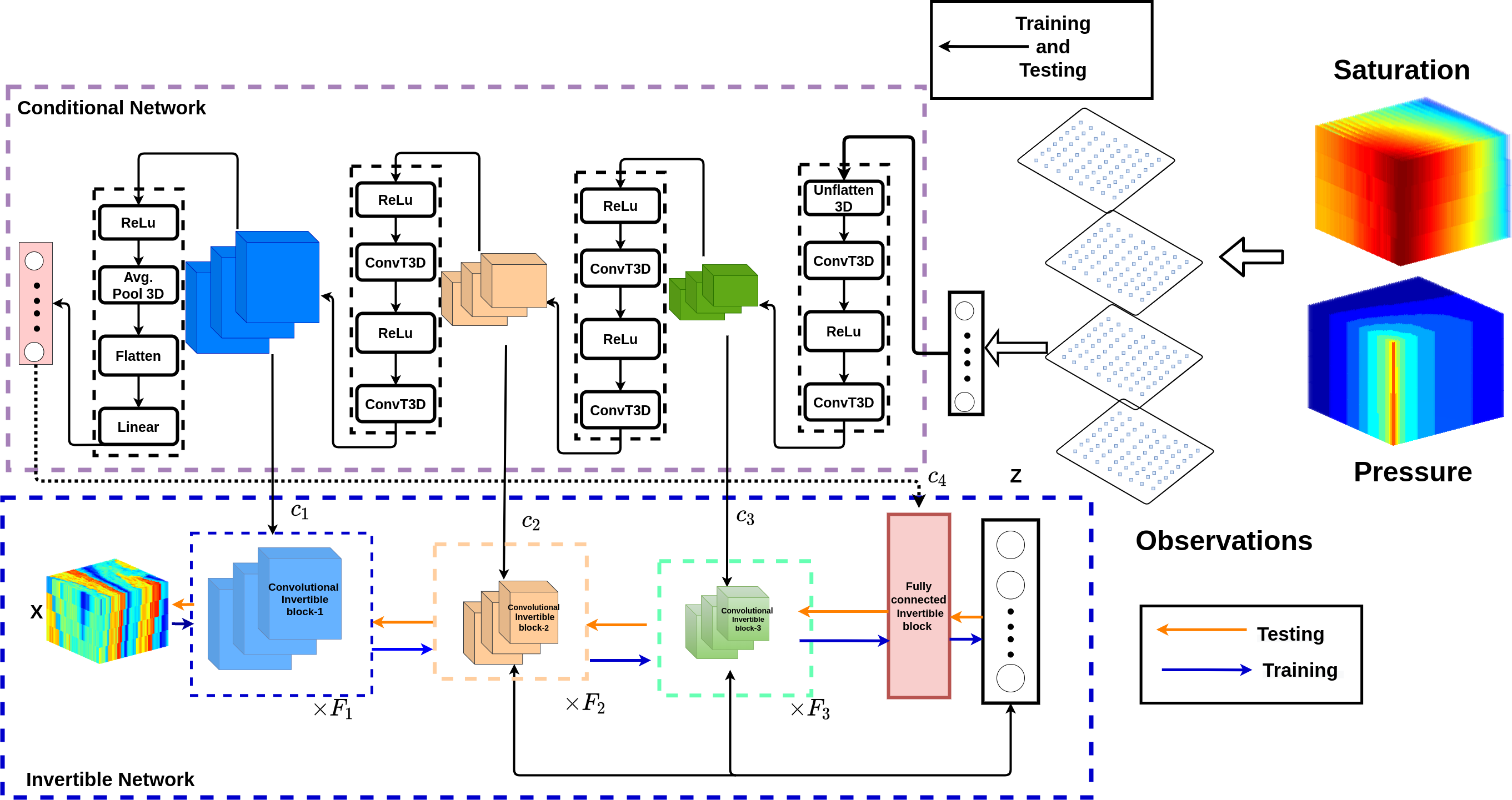}
    \caption{Multiscale $3$-D conditional invertible neural network. Here, the blue dashed box is the invertible network with $\bm{x}$ as the input field and $\{\bm{c}_l\}_{l=1}^{4}$ as the output from the conditional network (indicated in brown dashed box) considered as the input to the invertible network. The figure shows the direction of the training (indicated in blue line) and testing processes (indicated in yellow line).}
    \label{fig:three_D_model}
\end{figure}

In this section, we enumerate the details of the three-dimensional conditional invertible neural network. The concept of conditioning inputs to the scale $s(\cdot)$ and shift $t(\cdot)$ networks is similar to the two-dimensional conditional INN. However, the network architecture is modified to incorporate the data with depth $D$, height $H$, and width $W$ of the input features. Here, the three-dimensional conditional INN consists of two networks, namely, conditional network and invertible network as illustrated in Fig.~\ref{fig:three_D_model}.
In this work, the input to the model is a three-dimensional single-channel data, i.e., the input feature is of the size $C\times D \times H \times W$, where here, $C=1$. Similar to the two-dimensional cINN, we implement the single-side affine coupling layer in the convolutional invertible block-$1$ and the conventional affine coupling layer for the other invertible blocks. Here, the scale $s(\cdot)$ and shift $t(\cdot)$ networks consist of three-dimensional convolutional blocks and $ReLU$. The three-dimensional convolutional blocks are constructed such that the output of each invertible block has the same feature size as that of the conditioning input at multiple feature sizes. Additionally, one can also add dropout~\cite{srivastava2014dropout} and also initialize the model parameters using Xavier weight initialization technique~\cite{glorot2010understanding}. 
\par
The conditional network consists of four conditional blocks, and the output of each block is considered as the conditional input to the invertible network. The first conditional block unflattens the observations  channel-wise and depth-wise. Given a multi-channel observation of size $C_o$ $\times$ $D_o$ $\times$ $T$, where $C_o$ is the number of observation channels, $D_o$ is the depth-wise information and $T$ is the number of observations, the data is unflattened to obtain a three-dimensional representation. Then the data is passed to a   $3$-D convolution transpose and $ReLU$ to increase the feature size. In order to maintain the conditioning input to the same feature size of the output of each invertible blocks, a $3$-D convolution transpose is performed at the end of each conditioning block. Conditional blocks-$2$ and $3$ perform the following operations:  a ReLU followed by a $3$-D transposed convolution layer, and then again a ReLU followed by a $3$-D transposed convolution layer, thereby doubling the feature size of the input to the respective conditional blocks. The last conditional block consists of average pooling the three-dimensional information and then flattening the data. The resulting flattened data is then passed through the fully-connected layer.
\subsubsection{Network architecture and hyperparameter tuning}
Designing the network architecture and hyperparameter tuning are some of the most challenging problem-specific tasks. In this work, we address these challenges by empirically performing an extensive hyperparamater search and in particular focusing on the number of affine coupling layers in each invertible block and the number of conditioning input scales. Here, we also perform an extensive hyperparameter search to find the  learning-rate, weight decay, and the number of epochs that work well across different datasets and observation noise for the problems discussed in this work. 
\par 
As mentioned earlier, we consider three convolutional invertible blocks at multiple scales for both the two- and three-dimensional conditional invertible neural network. First, we test the model by changing the number of affine coupling layers in each convolutional invertible block. To assess the accuracy of each model, we consider the mean NLL loss (as discussed in Section~\ref{sec:loss}), and the results for each model are provided in~\ref{sec:Appendix-A}. Second, we test the model by adjusting the conditioning network for various conditioning scales. We provide the details for different conditioning scales in~\ref{sec:Appendix-B}.
The details regarding the training and testing procedures for both the $2$-D and $3$-D models are provided in Section~\ref{sec:loss}. 
\section{Application}\label{sec:train}
We consider here the PyTorch implementation of the  methodology in Section~\ref{sec:methods}.
For full-reproducibility of the results, the computer code and data can be found in \href{https://github.com/zabaras/inn-surrogate}{https://github.com/zabaras/inn-surrogate}. 

\subsection{Identification of the permeability field of an oil reservoir}
\begin{figure}[H] 
  \begin{minipage}[b]{0.3\linewidth}
    \centering
    \includegraphics[width=1\linewidth]{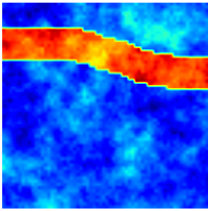} \\{(a)} 
    \vspace{4ex}
  \end{minipage}
  \begin{minipage}[b]{0.5\linewidth}
    \centering
    \includegraphics[width=0.9\linewidth]{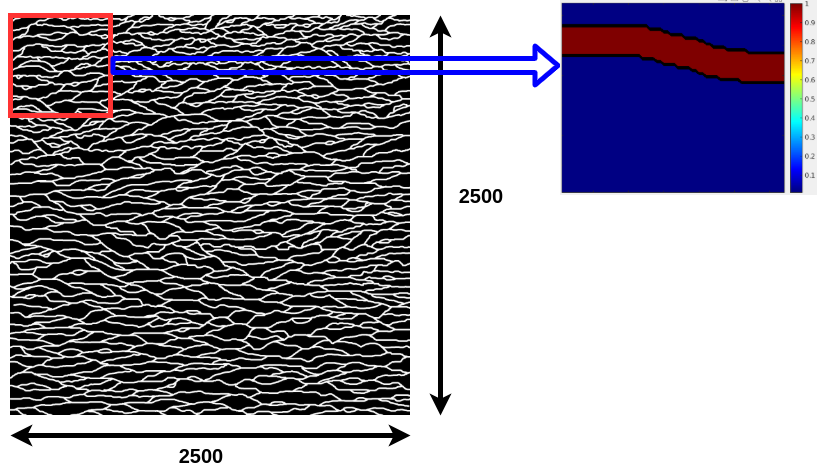} 
    \\{(b)} 
    \vspace{4ex}
  \end{minipage} 
  \begin{minipage}[b]{0.48\linewidth}
    \centering
    \includegraphics[width=1.05\linewidth]{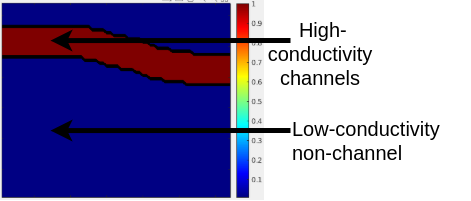}\\{(c)} 
    \vspace{4ex}
  \end{minipage}
  \begin{minipage}[b]{0.48\linewidth}
    \centering
    \includegraphics[width=1\linewidth]{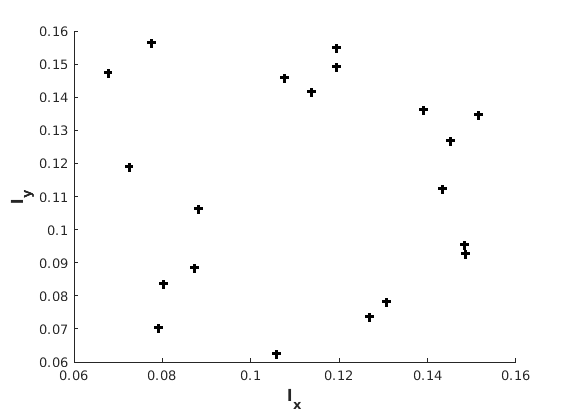}\\{(d)} 
    \vspace{4ex}
  \end{minipage} 
\centering
\includegraphics[scale=0.35]{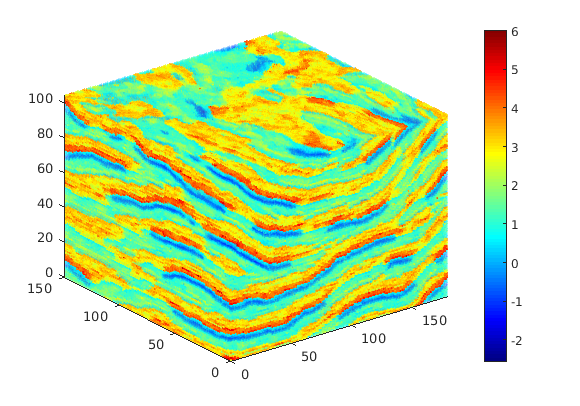}\\{(e)}
 \caption{(a) $2$-D sample log-permeability, (b) a channelized field where the samples of size $64 \times 64$ are cropped from a large training image, (c) a channelized field consists of two facies:  high-permeability channels and low-permeability non-channel, (d) sampled pair of length-scales, and (e) $3$-D sample log-permeability used for training the inverse surrogate.}
 \label{fig:domain_fig}
\end{figure}

In this work,  we consider a non-Gaussian log-permeability field with a $64 \times  64$ grid, as illustrated in Fig.~\ref{fig:domain_fig}(a) for the $2$-D case and $4 \times 64 \times  64$ grid for the $3$-D case. For the $2$-D case, the left and right boundaries are Dirichlet, with pressure values $1$ and $0$, respectively. The boundary conditions for the upper and lower walls of the domain are Neumann, with zero flux. For the $3$-D case, there is an injection well at the left bottom of the reservoir and a production well at the top right corner. For both the $2$-D and $3$-D cases, we assume a constant porosity $\phi = 0.25$. The log-permeability field for both cases is generated in the following steps. First, we consider a channelized field where the samples of size $64 \times 64$ for $2$-D are cropped from a large training image~\cite{laloy2018training}. Similarly, for the $3$-D case, the samples of size $4 \times 64 \times  64$ are cropped from a large training data (available at http://www.trainingimages.org/training-images-library.html). As illustrated in Fig.~\ref{fig:domain_fig}[c], this field consists of two facies: high-permeability channels and low-permeability non-channel, which consists of  $0$'s and $1$'s, respectively. Finally, we assign each facies with log-permeability values independently generated Gaussian random fields with constant mean $\mu$, and covariance function \textit{k}  specified in the following form~\cite{mo2020integration}:
\begin{equation}
k\left(\bm{r}, \bm{r}^{\prime}\right)=\sigma_{\log (K)}^{2} \exp\left(-\sqrt{\left(\frac{r_{1}-r_{1}^{\prime}}{l_{1}}\right)^{2}+\left(\frac{r_{2}-r_{2}^{\prime}}{l_{2}}\right)^{2}}\right),
\end{equation}
where $\bm{r}=\left(r_{1}, r_{2}\right)$ and $\bm{r}^{\prime}=\left(r_{1}^{\prime}, r_{2}^{\prime}\right)$ denote two arbitrary spatial locations, $\sigma_{\log (K)}^{2}$ is the variance, and $\ell_{1}$ and $\ell_{2}$ are the correlation lengths along the $x$ and $y$ axes, respectively. The length scales used in the training dataset were generate using  Algorithm \ref{algo_length} and 
\begin{equation}\label{GRF}
    K(r) = exp(G(r)), \hspace{2cm} G(\cdot) \sim \mathcal{N}(\mu,k(\cdot,\cdot)).
\end{equation}
In this work, we assign each facies with log-permeability values independently generated Gaussian random fields with a mean value of $4$ for high-permeability channels and $0$ for low-permeability non-channel, i.e., in the imported channelized data, we replace $1$ in the channelized data with one Gaussian random field and $0$ with another Gaussian random field. For both the $2$-D and $3$-D cases, the means for the two facies, namely, low-permeability non-channel and high-permeability channel, are $0$ and $4$, respectively.  Here, we train the cINN model with varying length-scales instead of training the model with a constant length scale (this will allow us to solve inverse problems with true log-permeabilities sampled from different length scales than those used for training). In this case, we obtain the length-scale samples using Algorithm~\ref{algo_length} rather than sampling the length-scales uniformly~\cite{tripathy2018deep}. This is due to the fact that small length-scales have more variability in the solution rather than larger length-scales, and therefore, we generate the length-scales samples which are more concentrated in the smaller length-scales regions as illustrated in Fig.~\ref{fig:domain_fig}[d]. 
As illustrated in Algorithm~\ref{algo_length}, the number of length scales pairs $n$, and the lower bound $h$ are pre-specified. In this work, we define $n=20$ and $h=0.016$. However, one can choose any number of length scale pairs $n$ depending upon any prior information available. In this work, we have $20$ pairs of length scales for both the $2$-D and $3$-D cases.
\begin{algorithm}[H]
    {\textbf{Input:} Number of length-scale pairs $n$; lower bound $h$}
    \SetAlgoLined
     initialization $\mathcal{H}$ and $\epsilon = 1$\\
     \While{$\epsilon \leq n$}{
      Sample $u = [u_1,u_2,u_3]\sim \mathcal{U}[0,1]^3$\\
      \If{$\exp(-u_1-u_2)<u_3$}{
       $l_c = (l_x,l_y)=(h+u_1(1-h),u_2(1-h))$\\
       $\mathcal{H} \leftarrow l_c$\\
       Increment $\epsilon$ $\leftarrow$ $\epsilon+1$
       }
     }
     \caption{Sampling length scale for generating training data.}
     \label{algo_length}
\end{algorithm}

\subsection{Observations}\label{sec:Observations}
In this section, we enumerate the details of the synthetic observations that are used for the multiphase flow. Here, the synthetic observations are generated by running the forward model with the permeability field as the input. 
\subsubsection{$2$-D model}
In this work, we obtain a set of noisy measurements by corrupting the observations with $1$\%, $3$\%, and $5$\% independent Gaussian noise. For the multiphase flow, we construct two configurations based on the saturation at $3$-time instances (i.e., $100$, $200$, and $300$ days). In configuration $1$, the pressure is collected at  $8 \times 8$, i.e., $64$ observation locations, and the saturation is collected at the last time instant (i.e., $300$ days) with $64$ number of observations. In configuration-$2$, the pressure is collected at $8 \times 8$, i.e., $64$ observation locations and the saturation is collected at  all  time instants (i.e., $100$, $200$, and $300$ days)  with $8 \times 8$  observations at each time instant and therefore resulting in  $192$ observation locations.

\subsubsection{$3$-D Model}
In the present study for the $3$-D model, we consider a set of noisy measurements by corrupting the observations with $1$\%, $2$\%, and $5$\% independent Gaussian noise. Here, we consider two configurations based on the saturation time instances (i.e., $100$, $200$, and $300$ days). For configuration-$1$, we consider the pressure at $8 \times 8$ observation locations at each depth, and similarly, the saturation is collected at the last time step (i.e., $300^{th}$ day) with $8 \times 8$ observation locations at each depth. In configuration-$2$, we consider saturation at all the time instances (i.e., $100$, $200$, and $300$ days) with $8 \times 8$ observations at each depth, and also, the pressure is collected at $8 \times 8$ observations at each depth.

\subsection{Network design and training} \label{sec:train_model}
In this work, we construct two networks, namely, the invertible and conditioning networks. The details regarding both  networks are given in Section~\ref{sec:cinn}.

\subsubsection{$2$-D Model}
The invertible network consists of four invertible blocks, as illustrated in Fig.~\ref{fig:main_fig}. Each invertible block consists of the scale $s(\cdot)$ and shift $t(\cdot)$ networks.
Here, the first three blocks are fully-convolutional networks, and the last block is a fully-connected network. In the convolutional invertible blocks, the scale $s(\cdot)$ and shift $t(\cdot)$ networks are fully-convolutional networks with $4$ convolutional layers. The output of each convolutional invertible blocks is half the feature map size of its input feature map size, i.e., $32 \times 32$ , $16 \times 16$, and $8 \times 8$ are the output feature map size of the convolution invertible blocks-- $1,\ 2$, and $3$, respectively. The kernel size for all the convolutional layers is $3$, and the stride is $1$ for maintaining the same feature size. Here, the number of steps that each convolutional invertible block is evaluated are $F_1=6, F_2=5$ and $F_3=4$, and for the fully-connected invertible block $F_4=1$.
\par
The conditional network consists of four conditional blocks, as illustrated in Fig.~\ref{fig:main_fig}. Given the observations, the conditional blocks $1-3$  double the feature sizes (ex. $16 \times 16$, $32 \times 32$, and $64 \times 64$) using the transposed convolution layers and each conditional blocks consist of two transposed convolution layers with $2$ and $3$ kernel size, respectively. The stride $1$ and $2$ are used to maintain the feature size and to decrease the feature size, respectively. The last conditional block contains an average pool that performs down-sampling along the spatial dimensions and then flattens the downsampled image. Here, the feature extracted after each conditional block is considered as the conditional input to the respective invertible blocks. 
\subsubsection{$3$-D Model}
In the $3$-D model, we consider the observations at each depth and the number of observation locations are $8\times8$ at each level and therefore resulting in $256$ observation locations. Similarly to the $2$-D, we consider four invertible blocks that consist of the scale $s(\cdot)$ and shift networks $t(\cdot)$. The first three invertible blocks are fully $3$-D convolutional networks and each block consists of four $3$-D convolutional layers. Also, the output of each convolutional invertible blocks is half the feature map size of its input feature map size. The kernel size and the stride are $3$ and $1$, respectively.  
\par
Similarly to the $2$-D framework, we consider four conditional blocks. Given the observations, the first three conditional blocks double the feature size of their input feature size using the $3$-D transposed convolution layers. Here, the kernel size for the first three conditional blocks are $2$ and $3$, respectively. Lastly, we perform the average pooling in the last conditional block and flatten the data. 
\subsubsection{Training} \label{sec:loss}
In this section, we provide the training procedure for the cINN model. Given a set of training data $\mathcal{D} = \{\bm{x}^i,\bm{\tilde{y}}^i\}_{i=1}^{N}$, where $\bm{x}^{(i)}$ is the non-Gaussian log-permeability field (input) and $\bm{\tilde{y}}^{(i)}$ is the corresponding pressure and saturation observations at specific locations on the domain, the objective is to learn a \textit{probabilistic inverse surrogate model}   that specifies a conditional density $p_{\bm{\theta}}(\bm{x}|\tilde{\bm{y}})$, where $\bm{\theta}$ are the model parameters.
\par 
In this work, the $2$-D and the $3$-D surrogate models are trained with $N = 6000$, $8000$, and $10000$ training data 
($\mathcal{D}$). Aforementioned, we develop a surrogate model that maps the noisy and the sparse observations $\tilde{\bm{y}}^{(i)}$ to the non-Gaussian field $\bm{x}^{(i)}$. This inverse mapping is confined to a broad prior distribution of the log-permeability field with which we train our cINN model. Here, the set of training data is obtained by evaluating the forward (well-posed) model. The cINN model simultaneously trains to compute the invertible network parameters, $\bm{\theta}_I$, and the conditional network parameters, $\bm{\theta}_g$, in an end-to-end fashion on a NVIDIA Tesla V$100$ GPU card. As illustrated in Fig.~\ref{fig:main_fig} for $2$-D and Fig.~\ref{fig:three_D_model} for $3$-D (indicated in blue arrow), we provide the observations, $\tilde{\bm{y}}$, as the input to the conditioning network and the non-Gaussian field, $\bm{x}$,  as the input to the invertible network during the training process. The mini-batch size considered for both the $2$-D and $3$-D cases is $Q = 16$ for all the   training cases considered. Here, we train the $2$-D model for $100$ epochs and the $3$-D model for $200$ epochs. For both models, the Adam optimizer~\cite{kingma2014adam} was considered with a learning rate of $1 \times 10^{-4}$ for the $2$-D case and $5 \times 10^{-5}$ for the $3$-D case. For the $2$-D case, the weight decay value is $1e-5$, and for the $3$-D case, the weight decay value is $5e-6$. In this work, we consider the number of invertible blocks to be $4$ ($L=4$). At the end of the training process, the invertible network parameters, $\bm{\theta}_I$, and the conditional network parameters, $\bm{\theta}_g$,  are computed such that the trained surrogate model   will then allow us to estimate the unknown log-permeability field for  given  sparse and noisy observations. As discussed earlier, even though deep generative models have shown a great potential to generalize, the unknown underlying true log-permeability is assumed here to be well represented by the assumed broad prior model. The training procedure is illustrated in Algorithm~\ref{algo_train}.
\par
During training, the input to the invertible network is the log-permeability field ${\bm{x}^{(i)}}$. Let us consider the output of each invertible block to be $\{{\bm{x}^{\prime(i)}\}_{l=1}^{L}}$, where $L$ is number of invertible blocks that includes both the convolutional invertible blocks ($L=1,2,3$) and the fully-connected  invertible block ($L=4$). Here, the output $\{{\bm{x}^{\prime(i)}\}_{l}}$ of each convolutional invertible block will be half the feature map size of its input $\{{\bm{x}^{\prime(i)}\}_{l-1}}$ feature map size. For example, in the $2$-D case, if the feature size of the input field ${\bm{x}^{(i)}}$ is $C \times H \times W$, where $C$ is the number of channels, $H$ is the height, and $W$ is the width of the feature size, then, the output $\bm{x}^{\prime(i)}_1$ for the first invertible block ($L=1$) will be $4C \times H/2 \times W/2$. The last fully-connected invertible block consists of the same input $\bm{x}^{\prime(i)}_3$ and output $\bm{x}^{\prime(i)}_4$ dimensions. As illustrated in Fig.~\ref{fig:forward_split}, we split the output of each convolutional invertible block channel-wise into two halves, where the first half is considered as the input $\bm{x}^{\prime\prime(i)}_l$ to the next invertible block and the other half of the data $\bm{z}^{\prime(i)}_l$ is concatenated to the latent variable $\bm{z}$. As enumerated in Section~\ref{sec:Generative}, the dimension of the input $\bm{x}^{(i)}$ and the latent space $\bm{z}^{(i)}$ should be the same for the transformation function to be bijective. Therefore, the dimension of the latent space $\bm{z}^{(i)}$ which includes the output of the fully-connected invertible block $\bm{x}^{\prime(i)}_4$ and the concatenation from the other invertible blocks ($\bm{z}^{\prime(i)}_2$ and $\bm{z}^{\prime(i)}_3$) should be same as the dimension of the input $\bm{x}^{(i)}$. Also, the output of each conditional blocks $\{\bm{c}_l\}$ from the conditioning network is concatenated to the scale $s(\cdot)$ and shift $t(\cdot)$ networks.
\par
\noindent \textit{Loss function.} 
Let us define the loss function that is used for training our cINN model. Given a set of training data  $\mathcal{D} = \{\bm{x}^i,\bm{\tilde{y}}^i\}_{i=1}^{N}$, let us consider the \textit{maximum a posteriori} (MAP) estimate of the model parameters $\bm{\theta}$:
\begin{equation}\label{MAP_1}
\bm{\theta}^{*} = \arg \max _{\bm{\theta}} \prod_{i=1}^{N}  p(\bm{\theta} | \bm{x}^{(i)}, \tilde{\bm{y}}^{(i)}).
\end{equation}
Here, the model parameters, $\bm{\theta}$, include the invertible network parameters, $\bm{\theta}_I$, and the conditional network parameters ($\bm{\theta}_g$), i.e., $\bm{\theta} = [\bm{\theta}_g,\bm{\theta}_I]$. Now, Eq.~(\ref{MAP_1}) can further be simplified as follows:
\begin{equation}\label{MAP_2}
\bm{\theta}^{*} = \arg \max _{\bm{\theta}} p(\bm{\theta}) \prod_{i=1}^{N} p(\bm{x}^{(i)}|\tilde{\bm{y}}^{(i)}, \bm{\theta}),
\end{equation}
where $p(\bm{\theta})$ is the prior on the model parameters and the evaluation of the conditional likelihood $p(\bm{x}|\tilde{\bm{y}}, \theta)$ is discussed next.
\par
\begin{figure}[H]
    \begin{minipage}[b]{0.48\linewidth}
    \centering
    \includegraphics[width=0.85\linewidth]{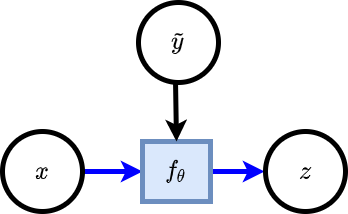} \\
    {(a)}  
  \end{minipage}
      \begin{minipage}[b]{0.48\linewidth}
    \centering
    \includegraphics[width=0.85\linewidth]{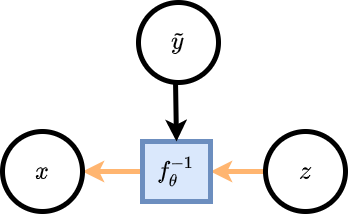} \\
    {(b) }  
    \end{minipage}
 \caption{(a) Training of the cINN model and (b) testing of the cINN model. During training, apart from $\bm{z}$, we also obtain the Jacobian (not shown in this Figure) from $f_{\theta}(\cdot)$. Details regarding the Jacobian can be found in Section~\ref{sec:Generative} and its implementation in Algorithm~\ref{algo_train}.}
 \label{fig:loss_fig}
\end{figure}

In this work, the invertible network is constructed using the concepts of the flow based generative model (as discussed in Section~\ref{sec:Generative}). Therefore, given a latent variable $\bm{z}$ and its known probability density function $\bm{z}\sim p(\bm{z})$, a bijection ${f}_{\bm{\theta}}(\cdot)$ (see Fig.~\ref{fig:loss_fig}) is introduced, that is parametrized by  the model parameters $\bm{\theta}$ and conditioned on the     data $\bm{y}$, that maps  the input $\bm{x}$ to $\bm{z}\sim p(\bm{z})$. Through the change of variables formula, we can write the conditional likelihood as follows:
\begin{equation}
 p_{\bm{\theta}}(\bm{X} \mid \tilde{\bm{Y}})= \prod_{i=1}^{N} p(f_{\theta}(\bm{x}^{(i)},\tilde{\bm{y}}^{(i)})) \cdot \left|\operatorname{det}\left(\frac{\partial( f_{\theta}(\bm{x}^{(i)},\tilde{\bm{y}}^{(i)}))}{\partial \bm{x}^{(i)}}\right)\right|,
\end{equation}
where, $\bm{X}=\left\{\bm{x}^{1}, \bm{x}^{2}, \ldots, \bm{x}^{N}\right\}$, $\tilde{\bm{Y}}=\left\{\tilde{\bm{y}}^{1}, \tilde{\bm{y}}^{2}, \ldots, \tilde{\bm{y}}^{N}\right\}$ and the latent variable $\bm{z}^{(i)} = f_{\theta}(\bm{x}^{(i)},\tilde{\bm{y}}^{(i)})$.  Therefore, the conditional log-likelihood $\log p_{\bm{\theta}}({\bm{X}} \mid \tilde{\bm{Y}})$ can be exactly evaluated as the following:
\begin{equation} \label{cond_LL}
 \log p_{\bm{\theta}}(\bm{X} \mid \tilde{\bm{Y}})= \sum_{i=1}^{N} \log p(f_{\theta}(\bm{x}^{(i)},\tilde{\bm{y}}^{(i)})) + \log \left|\operatorname{det}\left(\frac{\partial( f_{\theta}(\bm{x}^{(i)},\tilde{\bm{y}}^{(i)}))}{\partial \bm{x}^{(i)}}\right)\right|.
\end{equation}   
The MAP estimate in Eq.~(\ref{MAP_2}) can then be re-written as the minimization of the following loss function: 
\begin{equation}
\mathcal{L}= - \sum_{i=1}^{N} \left[\log p(f_{\theta}(\bm{x}^{(i)},\tilde{\bm{y}}^{(i)})) + \log \left|\operatorname{det}\left(\frac{\partial( f_{\theta}(\bm{x}^{(i)},\tilde{\bm{y}}^{(i)}))}{\partial \bm{x}^{(i)}}\right)\right|\right] - \log p(\bm{\theta}).
\end{equation}   
In this work, we consider a standard normal distribution for the latent variable $p(\bm{z})$.
%and we know that $\bm{z}^{(i)} = %f_{\theta}(\bm{x}^{(i)},\tilde{\bm{y}}^{(i)})$. 
Therefore, $\log p(f_{\theta}(\bm{x}^{(i)},\tilde{\bm{y}}^{(i)}))$ can be further simplified as $\frac{\left\|f_{\theta}(\bm{x}^{i},\tilde{\bm{y}}^{i})\right\|_{2}^{2}}{2}$. Therefore, 
\begin{equation}
\mathcal{L}=  \sum_{i=1}^{N} \left[\frac{\left\|f_{\theta}(\bm{x}^{(i)},\tilde{\bm{y}}^{(i)})\right\|_{2}^{2}}{2} - \log \left|\operatorname{det}\left(\frac{\partial( f_{\theta}(\bm{x}^{(i)},\tilde{\bm{y}}^{(i)}))}{\partial \bm{x}^{(i)}}\right)\right|\right] - \log p(\bm{\theta}).
\end{equation}   
Here, we consider a Gaussian prior on the model parameters $p(\bm{\theta})$ with mean $0$ and variance $\sigma^{2}_{\bm{\theta}}$. Therefore, 
\begin{equation} \label{loss_eq}
\mathcal{L}=\frac{1}{N}\sum_{i=1}^{N}\left[\frac{\left\|f_{\theta}(\bm{x}^{(i)},\tilde{\bm{y}}^{{(i)}})\right\|_{2}^{2}}{2}-\log \left|J_{i}\right|\right]+\tau\|\theta\|_{2}^{2},
\end{equation}
where, $\tau = 1/2\sigma^2_{\bm{\theta}}$ and the Jacobian determinant $J_{i} = \operatorname{det}\left(\frac{\partial( f_{\theta}(\bm{x}^{(i)},\tilde{\bm{y}}^{(i)}))}{\partial \bm{x}^{(i)}}\right)$. The training process is shown in Fig.~\ref{fig:loss_fig}(a). Here, the $L_2$ regularization is implemented using PyTorch optimizers~\cite{paszke2017automatic} by providing the weight decay value ($\tau$). As illustrated in Algorithm~\ref{algo_train}, the Jacobian for each invertible block is calculated separately using Eq.~(\ref{Jacobian}) and then summed over all the blocks. Therefore, the determinant of the Jacobian in Eq.~(\ref{loss_eq}) is computed as the summation of  all the individual determinants of the Jacobians in the flow model. 
\par
The minimization of the above regularized loss function makes it feasible to compute the solution of the inverse problem of interest. Regularization is introduced in multiple ways both explicitly (in terms of the prior parameter model) but also implicitly via the conditioning process.
Recall that the sparse observations using the conditioning network are mapped hierarchically during training to  higher-size features all the way up to the problem domain size. This conditions the training process with smooth observations over the domain of each feature. Similarly, once the model is trained, the given observation data in the inverse problem are lifted in a continuous fashion to the size of each feature and eventually over the whole domain. This    has implicitly a regularizing effect allowing us in computing  the conditional density $p(\bm{x}|\bm{y})$ and recovering the unknown high-dimensional input field. For example, in the $2$-D case, given the $8 \times 8$ pressure/saturation observations at  specific locations of the domain, we upscale this incomplete information to $16 \times 16$, $ 32 \times 32 $ and $64 \times 64$, respectively.    
\subsubsection{Testing}
Once the invertible network parameters, $\bm{\theta}_I$, and the conditional network parameters, $\bm{\theta}_g$ are computed using Algorithm~\ref{algo_train}, we can proceed to the solution of the inverse problem and estimate the unknown log-permeability field, $\hat{\bm{x}}$, for an unseen (not used during training) pressure and saturation observation data, $\tilde{\bm{y}}$. 
As discussed earlier, 
%the unseen observation data ($\tilde{\bm{y}}$) is obtained by evaluating the forward model with 
we are assuming that the unknown  (ground truth) $\bm{x}$, lies within the space specified by our broad prior model $p(\bm{x})$. 
%which is from the same broad prior distribution with which we trained our surrogate model. 
The inversion process is as follows. First, we sample $\bm{z}$ from a standard normal distribution with $S = 1000$ samples: $\mathcal{\bm{Z}} = \{\bm{z^{(j)}}\}_{j=1}^{S}$, where $\bm{z^{(j)}}\sim \mathcal{N}(0,1)$ and provide these samples as the input to the invertible network that consists of several affine coupling layers. Simultaneously, we also provide the observations, $\tilde{\bm{y}}$, as the input to the conditioning network as shown in Fig.~\ref{fig:main_fig} for $2$-D and Fig.~\ref{fig:three_D_model} for $3$-D (indicated in yellow arrow). As enumerated in Section~\ref{sec:Generative}, these affine coupling layers are referred as the invertible bijective transformation function. Therefore, taking the advantage of the bijective nature of the affine coupling layers one can obtain the input field $\{\hat{\bm{x}}^{(j)}\}$ given the latent variable $\bm{z}^{(j)}$ conditioned on the observations, $\tilde{\bm{y}}$, such that for $j=\{1,\ldots,S\}$ (see Fig.~\ref{fig:loss_fig}(b)): 
\begin{equation}\label{inverse_map}
    \hat{\bm{x}}^{(j)} = f_{\bm{\theta}}^{-1} (\bm{z}^{(j)},\tilde{\bm{y}}).
\end{equation}
The above inversion operation of transforming the latent variable  $\bm{z}^{(j)}$ to the input field $\hat{\bm{x}}^{(j)}$ conditioned on the observations $\tilde{\bm{y}}$ using the transformation function $f^{-1}(\cdot)$ is performed as follows. Consider $\bm{z}^{(j)} \in \mathbb{R}^{M}$  to be the   latent variable sampled from a standard Gaussian distribution. Here, we first split the latent dimension into two parts $1<m<M$ such that the first part $(\bm{z}^{(j)}_{1:m})$ remains the same, therefore, $\bar{\bm{x}}_1^{(j)} = \bm{z}^{(j)}_{1:m}$  and the second part  $\bm{z}^{(j)}_{m+1:M}$ undergoing through the transformation with the concatenation of the observation $\tilde{\bm{y}}$ resulting in $\bar{\bm{x}}_2^{(j)}$ as illustrated in Table~\ref{tab:affine_layer}. Now we concatenate $\bar{\bm{x}}_1^{(j)}$ and $\bar{\bm{x}}_2^{(j)}$ to obtain the input field $\hat{\bm{x}}^{(j)}$. Note that this inverse mapping is different from the traditional forward deterministic approach where the output is just a mere point estimate. However, in this work, during testing, the cINN model provides us samples of the input field for given observation data. This is due to the nature of the design of our neural network that hinges on the concept of generative models~\cite{dinh2016density}, where the output will be the samples of the input field conditioned on the observations $\tilde{\bm{y}}$. 
The latent variable that is the input to Eq.~(\ref{inverse_map}) is sampled from a standard Gaussian distribution. Once the input field for all the samples $S$ is predicted, we then estimate the predictive mean and $95$\% confidence interval for all the predicted input fields. In this work, we also present a one-dimensional cut along the diagonal of the domain.
\par
Algorithm~\ref{algo_test} illustrates the inversion process using the cINN model for a given pressure and saturation observation data, $\tilde{\bm{y}}$. Aforementioned in Section~\ref{sec:cinn} and the network design for the $2$-D and $3$-D models described in this section, the invertible network is designed with four invertible blocks. During the inference process, we first sample $\bm{z}^{(j)} \in \mathbb{R}^{M}$ from a standard Gaussian distribution. As shown in Fig.~\ref{fig:inverse_split},  we split the latent dimension into three parts $1<m_1<m_2<M$ such that the first part $\bm{z}^{(j)}_{(1:m_1)}$ will be  concatenated to the invertible block-$2$, the second part $\bm{z}^{(j)}_{(m_1:m_2)}$ will be concatenated to the invertible block-$3$ and the last part $\bm{z}^{(j)}_{(m_2:M)}$ will be provided as the input to the fully-connected invertible block-$4$. In this work, $m_1 = M/2$, and $m_2 = 3M/4$. During the inversion process, let us consider the output of each invertible block to be $\{{\bm{z}_{l}^{\prime(j)}\}_{l=1}^{L}}$, where $L$ is number of invertible blocks that includes both the convolutional invertible blocks ($L=1,2,3$) and the fully-connected  invertible block $L=4$. Here, in the inversion process, the output of each convolutional invertible block will be twice the feature map size of its input feature map size. For example, in the $2$-D case, if the input $\bm{z}^{\prime(j)}_2$ feature size for the invertible block-2 ($L=2$) is  $4C \times H/2 \times W/2$ then, the output $\bm{z}^{\prime(j)}_1$ feature size will be $C \times H \times W$. Similar to the training, in the inference stage, the output of each conditional blocks $\{\bm{c}_l\}$ from the conditional network is concatenated to the scale $s(\cdot)$ and shift $t(\cdot)$ networks. 
In this work, other than mean NLL the relative $L_2$ error is considered as the evaluation metric:
\begin{eqnarray}
L_{2}=\frac{1}{N} \sum_{i=1}^{N} \frac{\left\|\bar{\bm{x}}^{(i)}-\bm{x}^{(i)}\right\|_{2}}{\left\|\bm{x}^{(i)}\right\|_{2}},
\end{eqnarray}
where $\bar{\bm{x}}^{(i)}$ is the mean of the predicted log-permeability samples $\bar{\bm{x}} = \frac{1}{S}\sum_{j=1}^{S}\left(p(\bm{x}|\tilde{\bm{y}},\bm{\theta})\right)$ and $\bm{x}^{(i)}$ is the corresponding ground truth log-permeability field. For both  models, $128$ test data was considered and $S = 1000$ samples were generated using the invertible network. 

\begin{algorithm}
\caption{Training procedure for the inverse surrogate model.}
\label{algo_train}
\KwIn{Training data: $\{\bm{x}^{(i)}, \tilde{\bm{y}}^{(i)}\}_{i=1}^{N}$,  number of epochs: $E_{\text{train}}$, learning rate: $\eta$, number of steps in each flow model (each scale): $F$ and number of scales: $L$.}
\For{epoch =  $1$ to $E_{\text{train}}$}{
Sample a minibatch from the the training data: $\{\bm{x}^{(i)}, \tilde{\bm{y}}^{(i)}\}_{i=1}^{Q}$ and pass the observations $\{\tilde{\bm{y}}^{(i)}\}_{i=1}^{Q}$ to the conditioning network to obtain the conditioning inputs $\{\bm{c}_{l}^{(i)}\}_{i=1,l=1}^{Q,L}$: $\bm{c}_{l}^{(i)} = \bm{g}_{\bm{\theta_g}}(\tilde{\bm{y}}^{(i)})$. \\
Pass the input $\{\bm{x}^{(i)}\}_{i=1}^{Q}$ to the invertible network: \\
\For{$l$ = $1$ to $L$}{
$\bm{x}^{\prime\prime(i)}_{0} = \bm{x}^{(i)}$ \\
\If{
$1<l<L$}{$\bm{x}^{\prime(i)}_l, {\text{J}}^{\prime(i)}_l = {f}_{l,\bm{\theta_{I}}}(\bm{x}^{\prime \prime(i)}_{(l-1)}, \bm{c}_{(l)}^{(i)})$ \Comment{${f}$ is the invertible network that includes permutation and $F$ steps.\\
$ \bm{x}^{\prime\prime(i)}_l, \bm{z}^{\prime(i)}_l$ = Split($\bm{x}^{\prime(i)}_l$)}}
\ElseIf{$l=L$}{$\bm{x}^{\prime(i)}_l, {\text{J}}^{\prime(i)}_l = {f}_{l,\bm{\theta_{I}}}(\bm{x}^{\prime \prime(i)}_{(l-1)}, \bm{c}_{(l)}^{(i)})$}
\ElseIf{$l=1$}{$\bm{x}^{\prime \prime(i)}_l, {\text{J}}^{\prime(i)}_l = {f}_{l,\bm{\theta_{I}}}(\bm{x}^{\prime \prime(i)}_{(l-1)},\bm{c}_{(l)}^{(i)})$}
$\hat{\bm{z}}^{(i)}$ = {concatenate}$\{\bm{z}^{\prime(i)}_l,\bm{x}^{\prime(i)}_{L}\}_{l=2}^{L}$ \\
Compute $J_{\text{final}}^{(i)} = \sum_{l=1}^{L}({\text{J}}^{\prime(i)}_{l})$
}
$\mathcal{L}$ = Loss($\hat{\bm{z}}^{(i)},J_{\text{final}}^{(i)}$) \Comment{Loss 
is evaluated using Eq.~(\ref{loss_eq})
}\\
$\nabla \bm{\theta}\leftarrow$Backprop$\left(\mathcal{L}\right)$ \\
$\bm{\theta} \leftarrow \bm{\theta}-\eta \nabla \bm{\theta}$
}
\KwOut{Simultaneously trained invertible and conditional network.}
\end{algorithm}
\begin{algorithm}
\caption{Inverse solution: Conditional invertible neural networks.}
\label{algo_test}
\KwIn{Trained invertible and conditional network and $\bm{\tilde{y}}$ (observations) 
  \SetAlgoLined \\
$\bm{z^{(j)}}\sim \mathcal{N}(0,1)$; $\mathcal{\bm{Z}} = \{\bm{z}^{(j)}\}_{j=1}^{S}$ and $\bm{z}^{(j)} \in \mathbb{R}^{M}$. {Here, S is number of samples} \\
Split the latent dimension into $(L-1)$ parts ($m_{0}<m_{1}<\dots<m_{L-1}<m_{L}$): $\{\bm{z}^{(j)}_{(m_{l-1}:m_{l})}\}_{l=1}^{L}$.
Here. $m_0=1$  and $m_L=M$.
{Fig.~\ref{fig:inverse_split} shows the latent dimension split into three parts.} \\
${{\bar{\bm{y}}}} = \text{tile}(\bm{\tilde{y}})$; ${\bm{\bar{y}}} = {\{\tilde{\bm{y}}^{(j)}\}_{j=1}^{S}}$   {tile: constructs a new array by repeating $\tilde{\bm{y}}$} \\  
Pass the observations ${\{\tilde{\bm{y}}^{(j)}\}_{j=1}^{S}}$ to the conditioning network to obtain the conditioning inputs 
$\{\bm{c}_{l}^{(j)}\}_{j=1,l=1}^{S,L}$:\\ $\bm{c}_{l}^{(j)} = {g}_{\bm{\theta_g}}(\tilde{\bm{y}}^{(j)})$.\\
Pass the samples ${\{\bm{z}^{(j)}\}}_{j=1}^{S}$ to the invertible network:}
\For{$l$ = $1$ to $L$}{
$\bm{z}^{(j)}_l = \bm{z}^{(j)}_{(m_{(l-1)}:m_{l})}$ \\
%=============
\If{
$l=L$}{
$\bm{z}^{\prime(j)}_l
= {f}^{-1}_{l,\bm{\theta_{I}}}(\bm{z}^{(j)}_{(L-1)}, \bm{c}_{(l)}^{(j)})$ \Comment{${f}^{-1}$ is the inverse mapping that includes permutation,  $F$ steps and split
}}
\ElseIf{$1<l<L$}{
$\tilde{\bm{z}}^{(j)}$ = {concatenate}$\{\bm{z}^{\prime(j)}_{(l+1)},\bm{z}^{(j)}_{(l-2)})\}_{l=2}^{L}$ \\
$\bm{z}^{\prime(j)}_l = {f}^{-1}_{l,\bm{\theta_{I}}}(\tilde{\bm{z}}^{(j)}, \bm{c}_{(l)}^{(j)})$}
\ElseIf{$l=1$}{
$\hat{\bm{x}}^{(j)} = {f}^{-1}_{l,\bm{\theta_{I}}}(\bm{z}^{\prime(j)}_{(l+1)},\bm{c}_{(l)}^{(j)})$}
}%for loop
\KwOut{Samples:$\{\hat{\bm{x}}^{(j)}\}_{j=1}^{S}$}
\end{algorithm}
\section{Results and Discussion}\label{sec:results}
In this section, the results for the two-dimensional and three-dimensional inverse multiphase flow problem are presented. In the present study, the inverse surrogate model is trained with $6000$, $8000$, and $10000$ training data  and tested with $128$ data. Here each training and the test data consists of a pair the log-permeability field and the corresponding pressure and saturation observations. Both the training and test data are generated using the MRST software~\cite{software}. Aforementioned, in this work, for both the $2$-D and $3$-D cases, we consider two different configurations, three different noise levels, and three different numbers of training data for a multiphase flow problem as described in Section~\ref{sec:PD}. For the present problem, the number of observations and the location of the observations for a given $2$-D or $3$-D domain is fixed, i.e., we consider $8 \times 8$ observations for a given two-dimensional domain and for the $3$-D domain, we consider $8 \times 8$ observations at each depth. In the present study, we investigate the effect of the number of saturation data (considered at regular time instants), the effect of the number of training data, and, finally, the effect of observation noise for both the two- and three-dimensional data.
\subsection{Varying the number of training data}
\begin{figure}[H]
  \begin{minipage}[b]{0.98\linewidth}
    \centering
    \includegraphics[width=0.6\linewidth, height=0.55\linewidth]{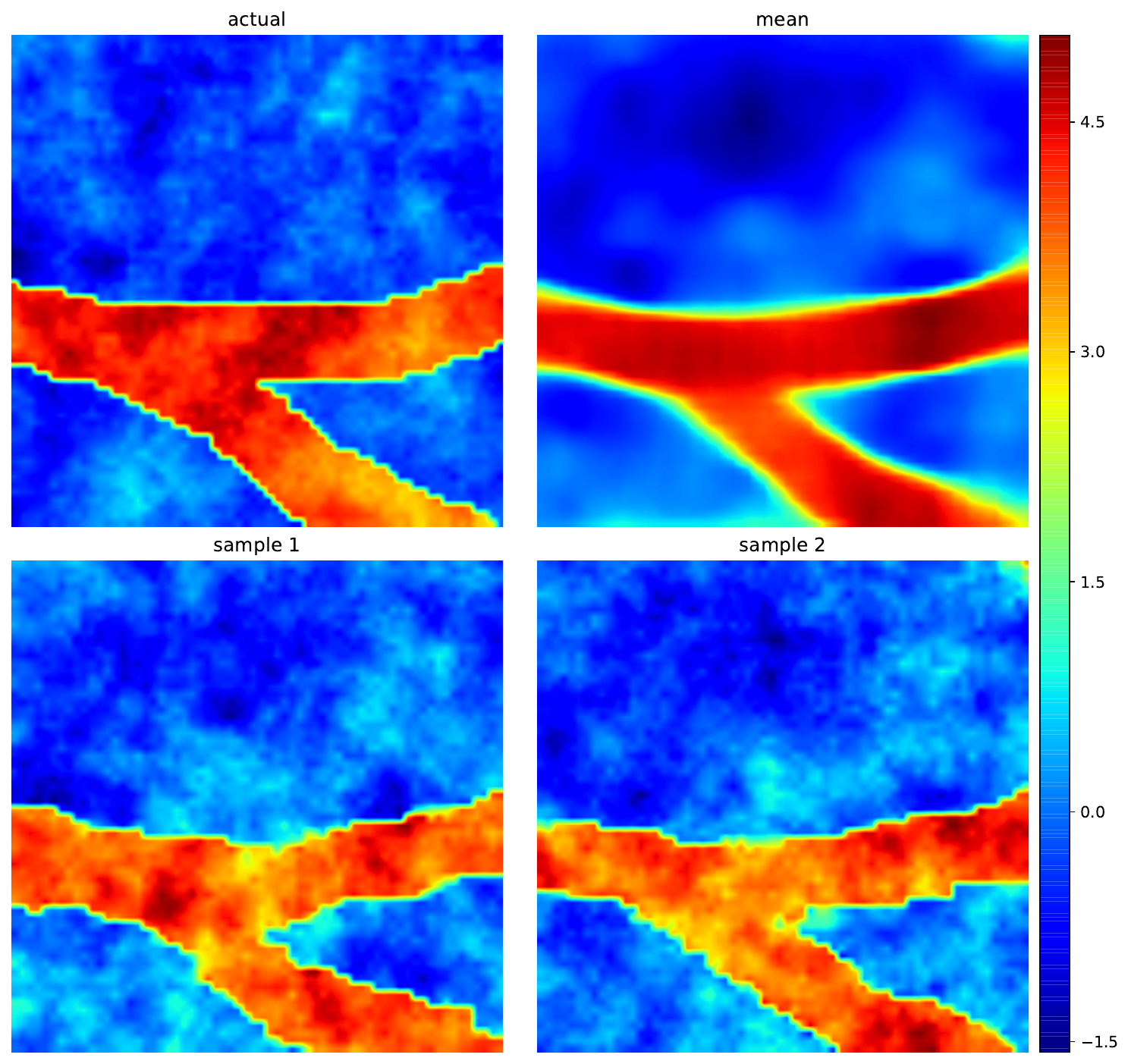}\\
    {(a) 2D-case configuration-1}  
  \end{minipage}
\\
  \begin{minipage}[b]{0.98\linewidth}
    \centering
    \includegraphics[width=0.6\linewidth,height=0.55\linewidth]{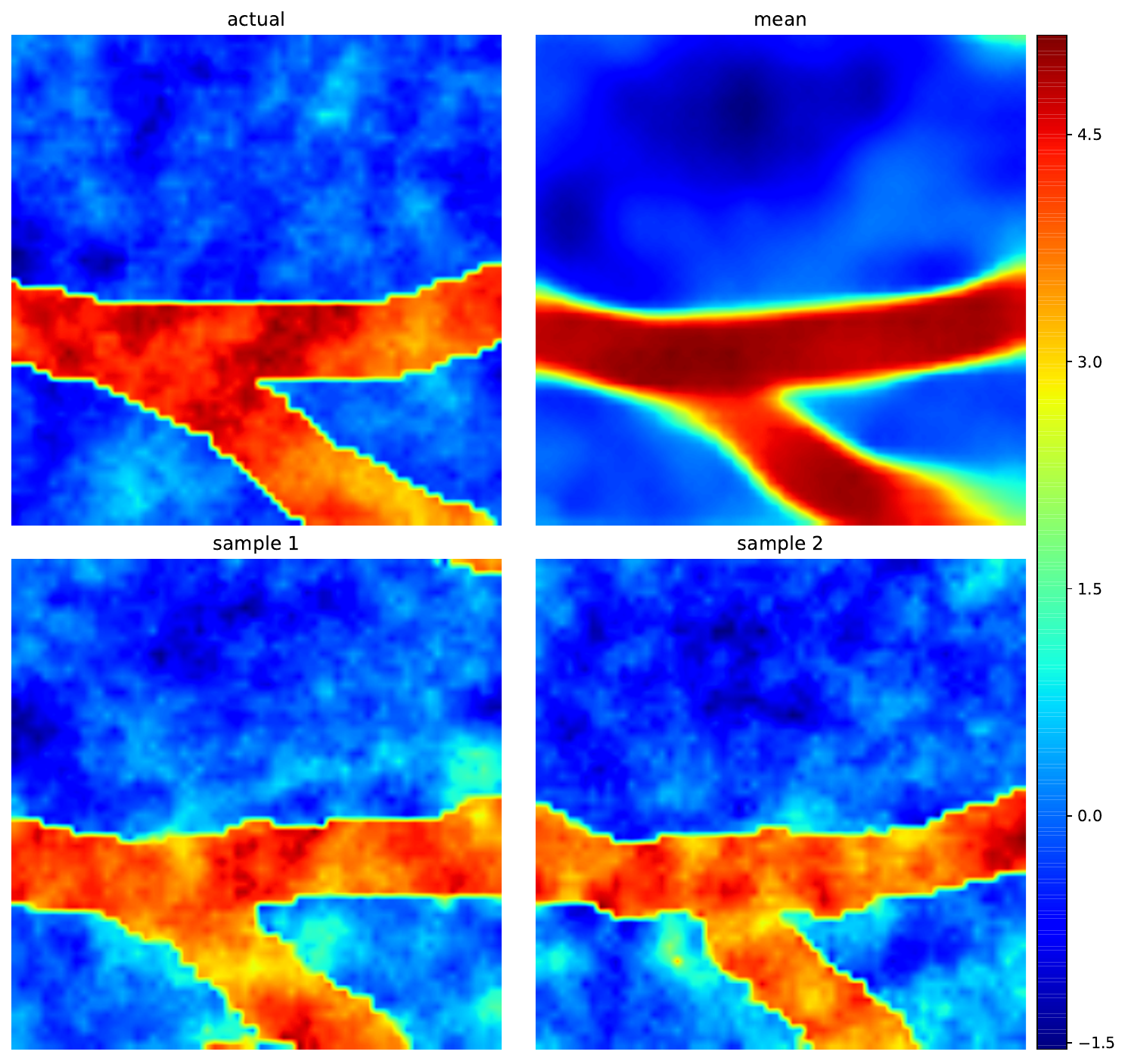} \\
    {(b) 2D-case configuration-2}  
  \end{minipage}
 \caption{In each subfigure for the $2$-D problem: (a) configuration-$1$ and (b) configuration-$2$: first image (first row and first column) shows the actual log-permeability field, second image (first row and second column) shows the mean for all the samples and the other images (second row) are samples.}
 \label{fig:mean_figure}
\end{figure}

\begin{figure}[H]
    \begin{minipage}[b]{0.48\linewidth}
    \centering
    \includegraphics[width=0.88\linewidth]{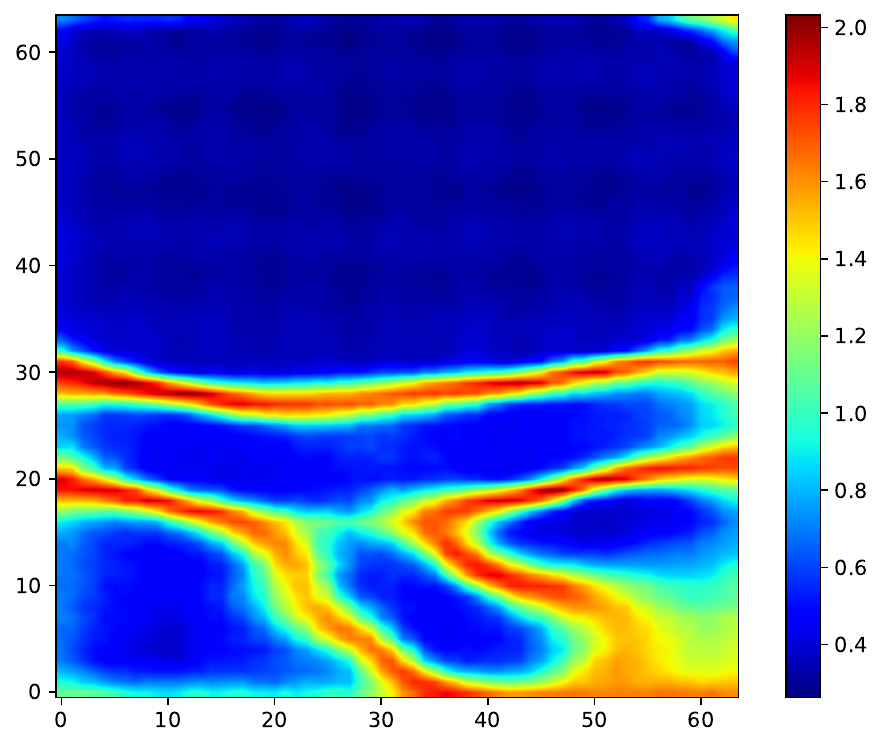} \\
    {(a)}  
  \end{minipage}
      \begin{minipage}[b]{0.48\linewidth}
    \centering
    \includegraphics[width=0.88\linewidth]{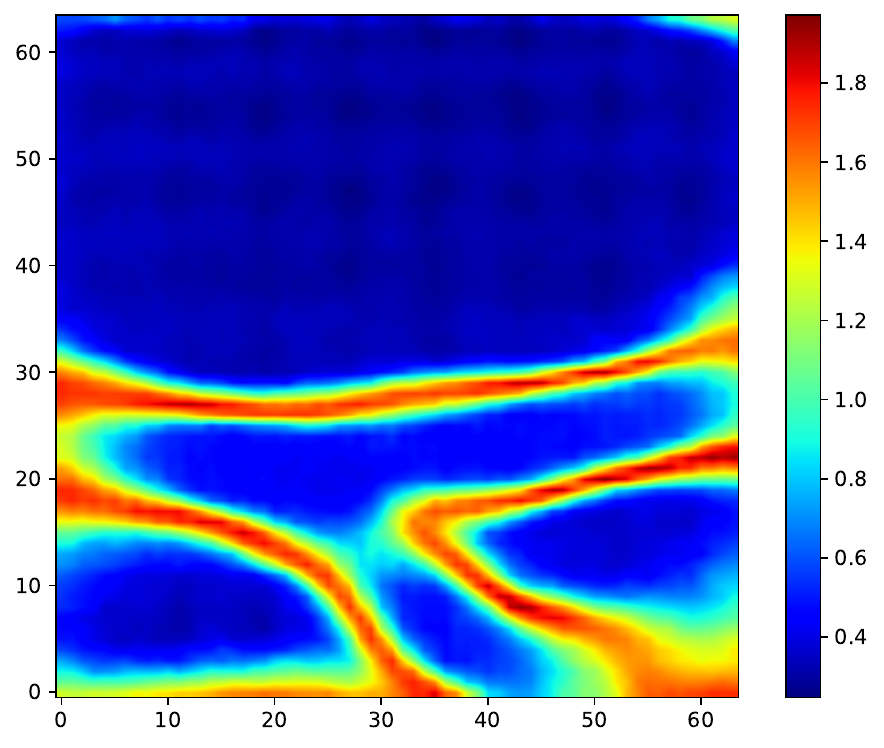} \\
    {(b) }  
    \end{minipage}
 \caption{Standard deviation of the predicted log-permeability based on the 
 generated samples (a) configuration-$1$ and (b) configuration-$2$ for the $2$-D problem.}
 \label{fig:std_2D}
\end{figure}

\begin{figure}[H]
    \centering
    \includegraphics[scale=0.4]{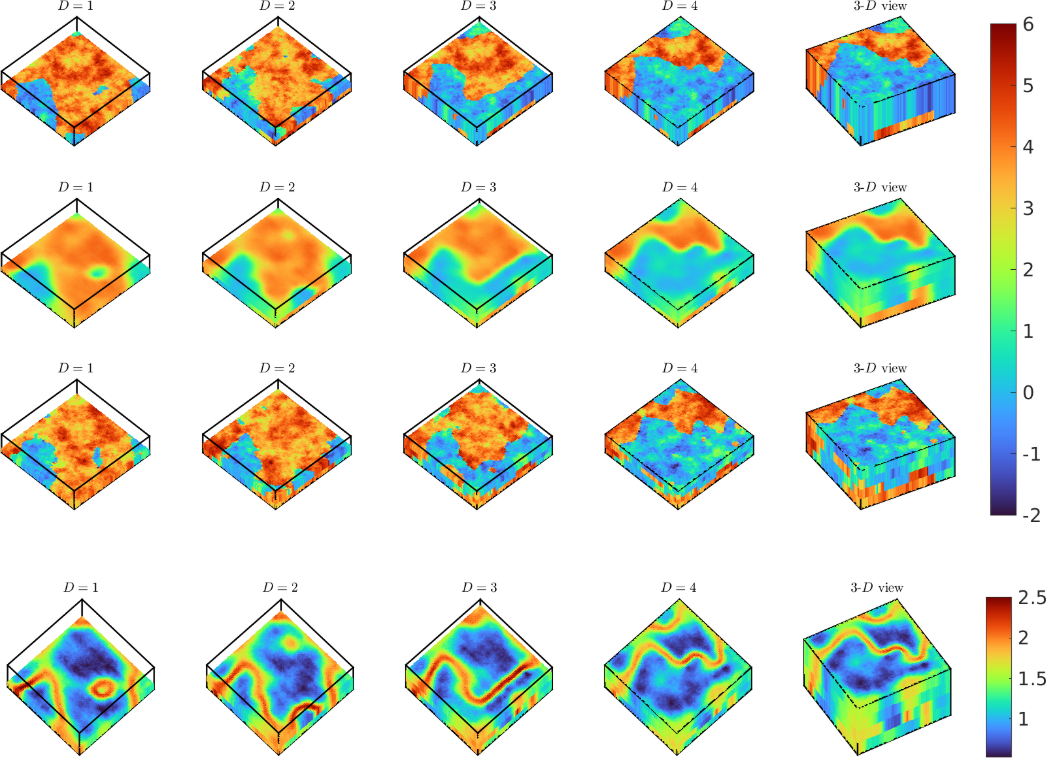}
  \caption{$3$-D problem and configuration-$1$: the first row corresponds to the actual log-permeability field for each depth $D$ and a $3$-D view. The second row corresponds to the mean for all the samples for each depth ($D$) and a $3$-D view. The third row shows a sample at each depth $D$ and a $3$-D view and finally the last shows the standard deviation of the predicted log-permeability based on the generated samples for the $3$-D problem.}
  \label{fig:mean_figure_3D1}
\end{figure}

\begin{figure}[H]
    \centering
    \includegraphics[scale=0.4]{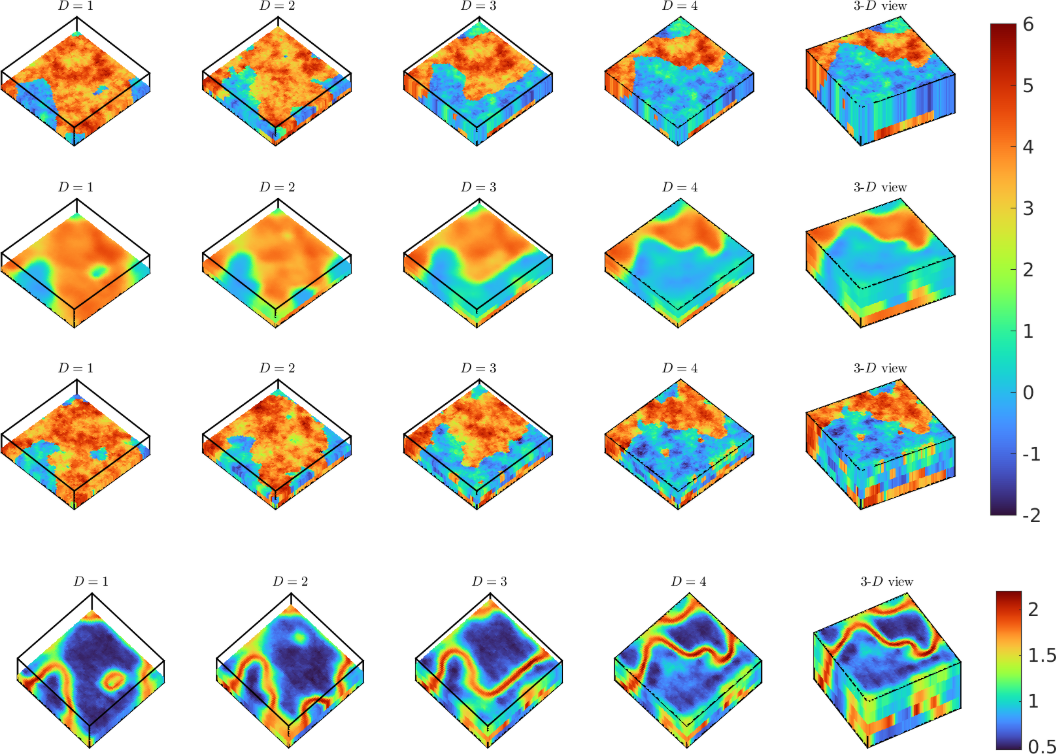}
  \caption{$3$-D problem and configuration-$2$: the first row corresponds to the actual log-permeability field for each depth $D$ and a $3$-D view. The second row corresponds to the mean for all the samples for each depth $D$ and a $3$-D view. The third row shows a sample at each depth $D$ and a $3$-D view and finally the last shows the standard deviation of the predicted log-permeability based on the generated samples for the $3$-D problem.}
  \label{fig:mean_figure_3D2}
\end{figure}

In this work, we train the $2$-D and $3$-D cINN surrogate model with $N = 6000$, $8000$, and $10000$ training data. Here, the synthetic training data consists of the log-permeability field and the corresponding pressure and saturation observations that are obtained using the forward (well-posed) model. The mini-batch and the learning rate remain the same when training for all three training data sets. The details regarding the training procedure and hyperparameters of the model were discussed earlier in Section~\ref{sec:train_model}. The mean, samples of the predicted log-permeability field, and the ground truth for the configuration-$1$ and configuration-$2$ test data, where the model trained with $10000$ training data and  $1$\% observation noise are shown in Fig.~\ref{fig:mean_figure}  for $2$-D case and Fig.~\ref{fig:mean_figure_3D1} and Fig.~\ref{fig:mean_figure_3D2} for $3$-D,  respectively. For the same setup, we show the standard deviation for both the configurations in Fig.~\ref{fig:std_2D} for $2$-D and Fig.~\ref{fig:mean_figure_3D1} and Fig.~\ref{fig:mean_figure_3D2} for $3$-D case. For both the cases, we observe that the predicted mean of all the samples illustrates a similar channel structure as that of the ground truth. Also, the individual samples have similar high-permeability channels and for low-permeability non-channel patterns compared to the actual ground truth sample. Note that we do not apply any thresholding step or any post-processing as on the predicted samples. 
\par
We show the 2-$D$ and 3-$D$ test mean NLL error during training with various training data in Fig.~\ref{fig:2_D_l2}(a)-(b) and Fig.~\ref{fig:3_D_l2}(a)-(b) for $1\%$ observation noise and configuration-$2$, respectively. Overall the cINN surrogate model converges  around $50$ epochs for the $2$-D case and $150$ epochs for the $3$-D case when the model is trained with $6000$, $8000$, and $10000$ training data. Also, we observe that the mean NLL loss decreases as we increase the number of training data, and this trend is also observed in configuration-$1$.
Figure~\ref{fig:2_D_l2}(c) -(d) shows the test relative $L_2$ error for configuration-$1$ and configuration-$2$, respectively. Similarly, we show the test relative $L_2$ error for the $3$-D case for both configurations in Fig.~\ref{fig:3_D_l2}(c)-(d). In each sub-figure, we show the performance of the inverse surrogate for various training data and observation noise. In all the cases, the test relative $L_2$ error for each test data, the predictive mean of the log-permeability is estimated with $S = 1000$ samples from the conditional density by sampling $1000$ realizations of noise $\mathcal{\bm{Z}} = \{\bm{z^{(i)}}\}_{i=1}^{S}$ as illustrated in Algorithm~\ref{algo_test}. In sub-figures, Fig.~\ref{fig:2_D_l2}(c)-(d) for the $2$-D case and Fig.~\ref{fig:3_D_l2}(c)-(d) for the $3$-D case, we observe that the test relative $L_2$ error decreases as we increase the number of training data from the $6000$ to $10000$ and we observe this trend for all values of   the observation noise.  
\begin{figure}[H]
    \begin{minipage}[b]{0.3\linewidth}
    \centering
    \includegraphics[width=0.95\linewidth]{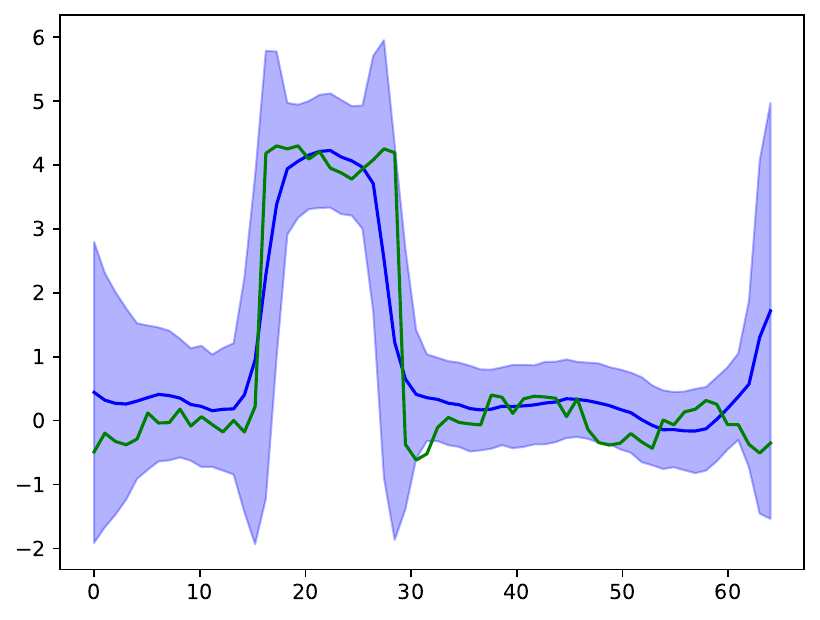} \\
    {(a)} 
  \end{minipage} 
      \begin{minipage}[b]{0.3\linewidth}
    \centering
    \includegraphics[width=0.95\linewidth]{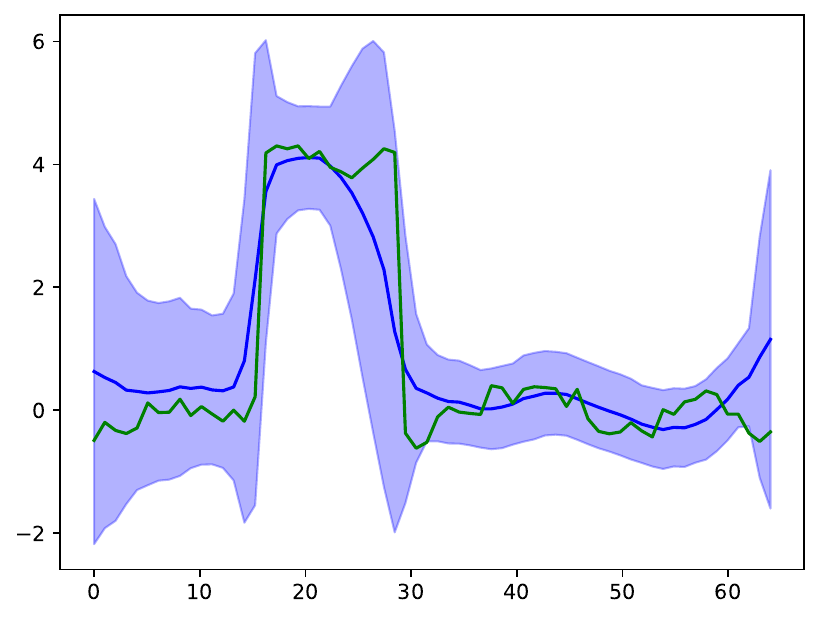}\\{(b)} 
  \end{minipage}
      \begin{minipage}[b]{0.3\linewidth}
    \centering
    \includegraphics[width=0.95\linewidth]{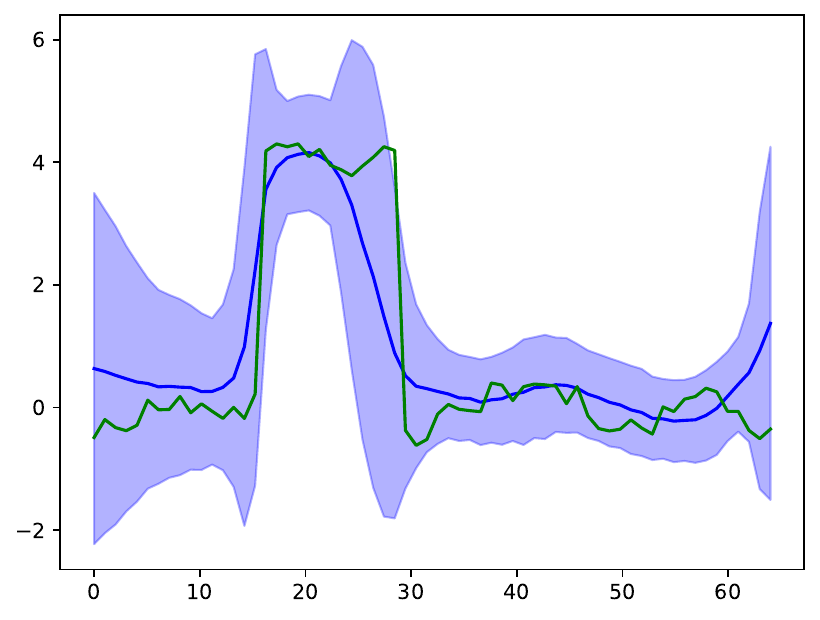}\\{(c)} 
  \end{minipage} 
    \newline
    \begin{minipage}[b]{0.3\linewidth}
    \centering
    \includegraphics[width=0.95\linewidth]{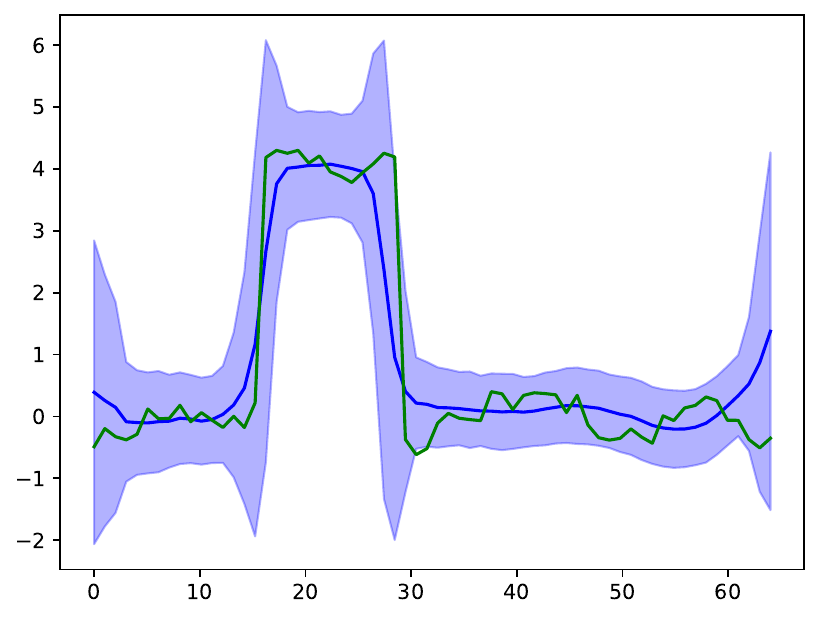}\\{(d)} 
  \end{minipage} 
    \begin{minipage}[b]{0.3\linewidth}
    \centering
    \includegraphics[width=0.95\linewidth]{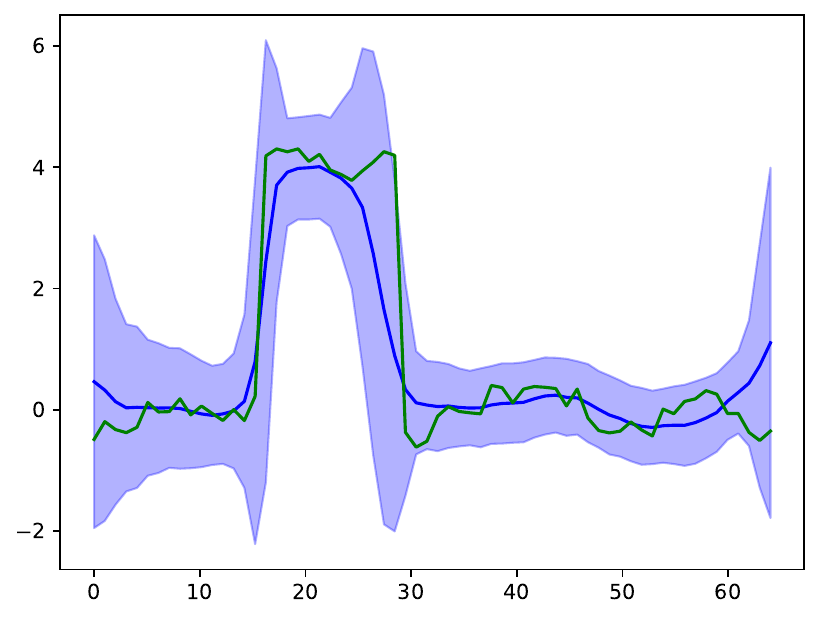}\\{(e)} 
  \end{minipage}
    \begin{minipage}[b]{0.3\linewidth}
    \centering
    \includegraphics[width=0.95\linewidth]{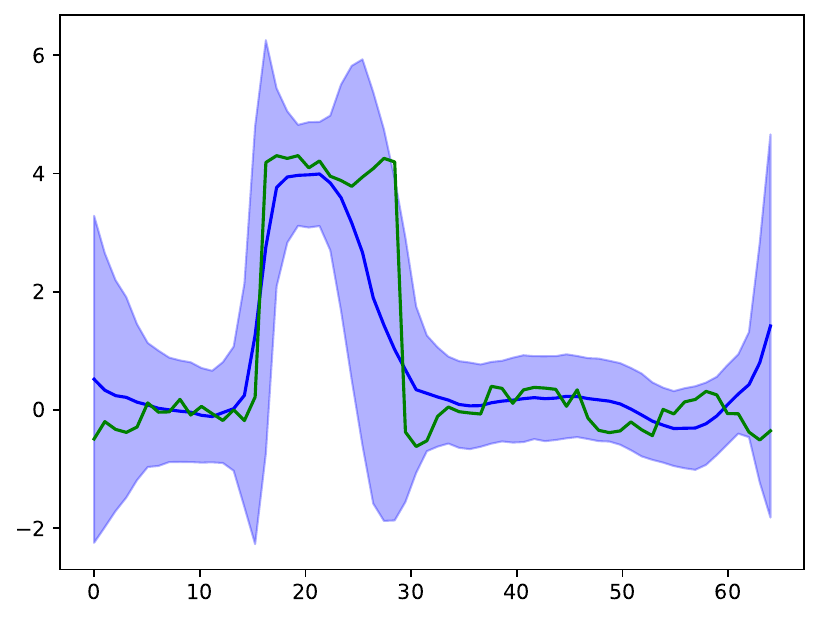}\\{(f)} 
  \end{minipage}
 \caption{One-dimensional cut along the diagonal of the predicted cINN samples and the ground truth log-permeability domain for the $2$-D problem. (a)-(c) configuration-$1$ for $1$\%, $3$\%, and $5$\% observation noise, respectively. (d)-(f) configuration-$2$ for $1$\%, $3$\%, and $5$\% observation noise, respectively. The green curve is the actual (ground truth) field; the blue curve is the mean log-permeability, and the blue shaded area is the $95$\% confidence interval.}
 \label{fig:error_fig}
\end{figure}
\begin{figure}[H]
    \begin{minipage}[b]{0.32\linewidth}
    \centering
    \includegraphics[width=0.95\linewidth]{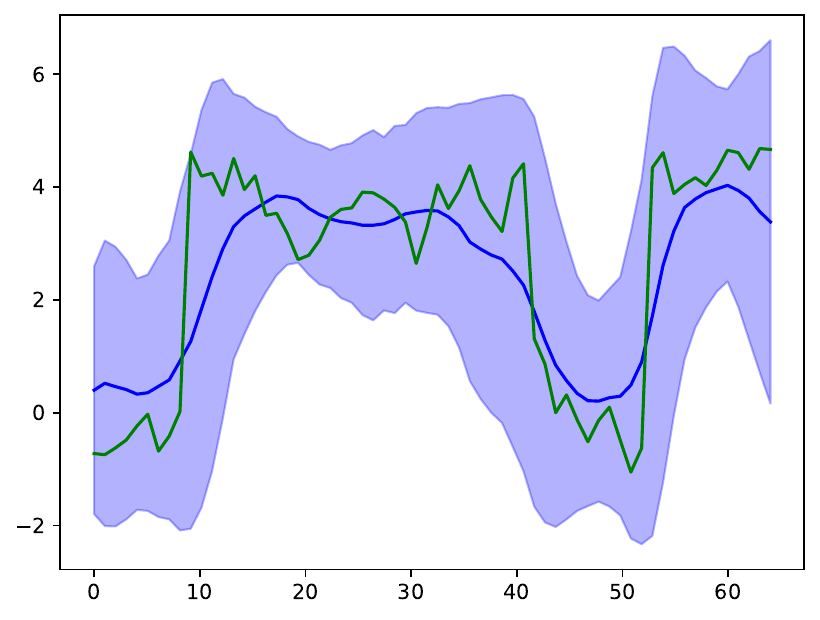} \\
    {(a)} 
  \end{minipage} 
      \begin{minipage}[b]{0.32\linewidth}
    \centering
    \includegraphics[width=0.95\linewidth]{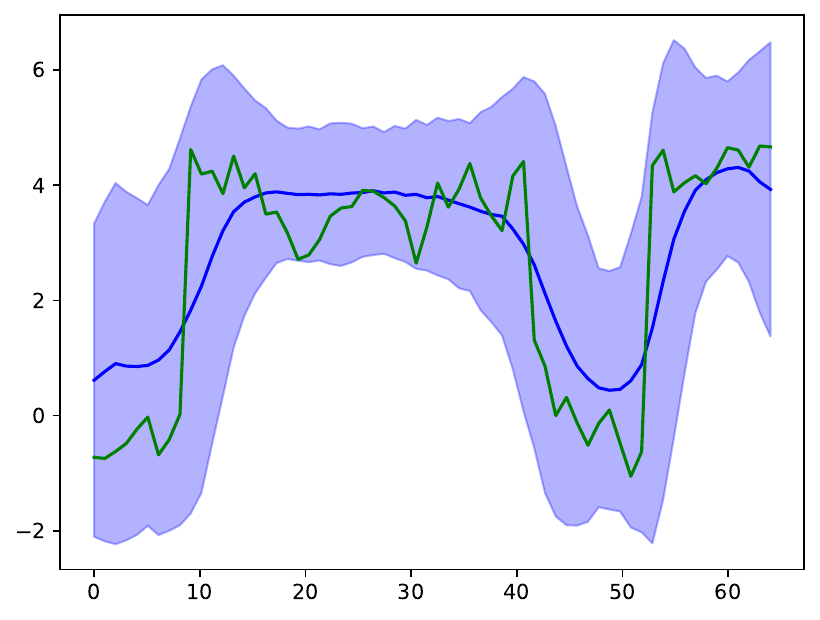}\\{(b)} 
  \end{minipage}
      \begin{minipage}[b]{0.32\linewidth}
    \centering
    \includegraphics[width=0.95\linewidth]{{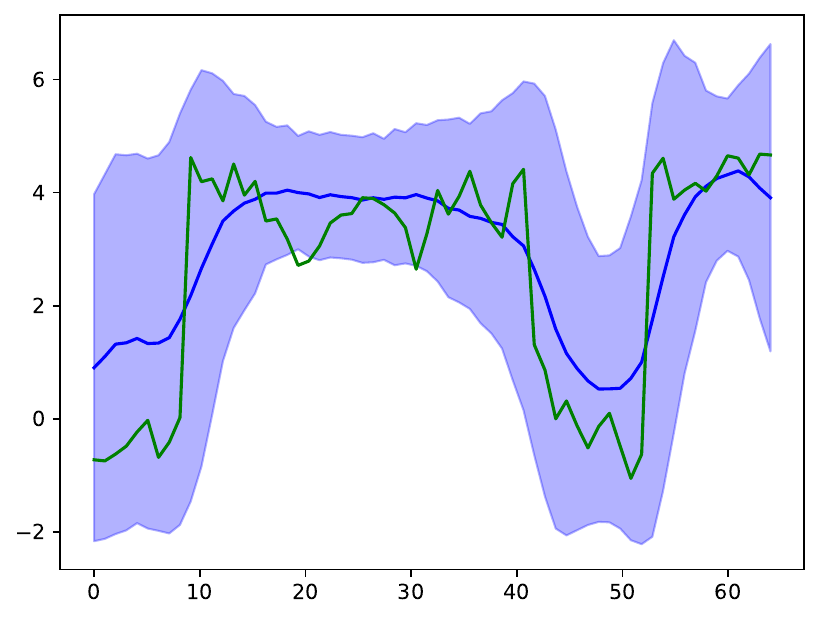}}\\{(c)} 
  \end{minipage} 
    \newline
    \begin{minipage}[b]{0.32\linewidth}
    \centering
    \includegraphics[width=0.95\linewidth]{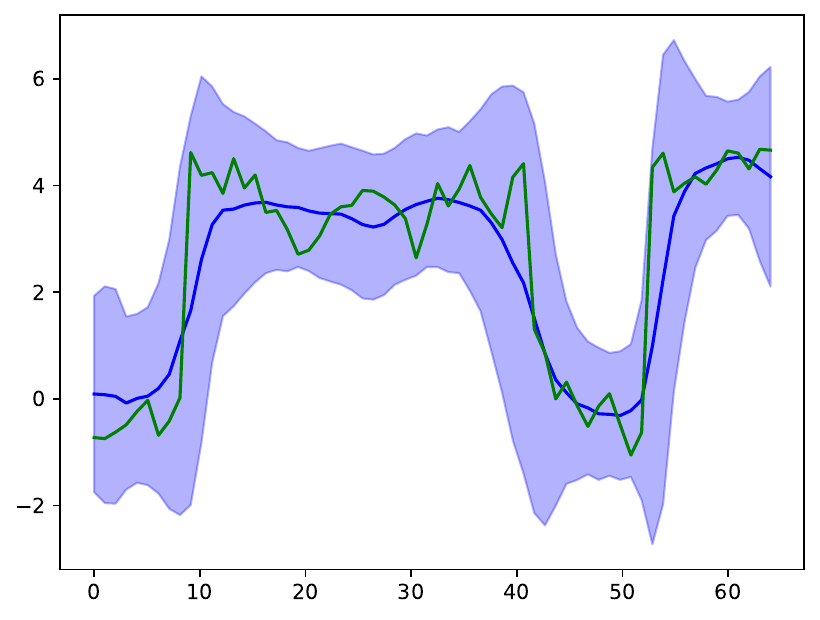}\\{(d)} 
  \end{minipage} 
    \begin{minipage}[b]{0.32\linewidth}
    \centering
    \includegraphics[width=0.95\linewidth]{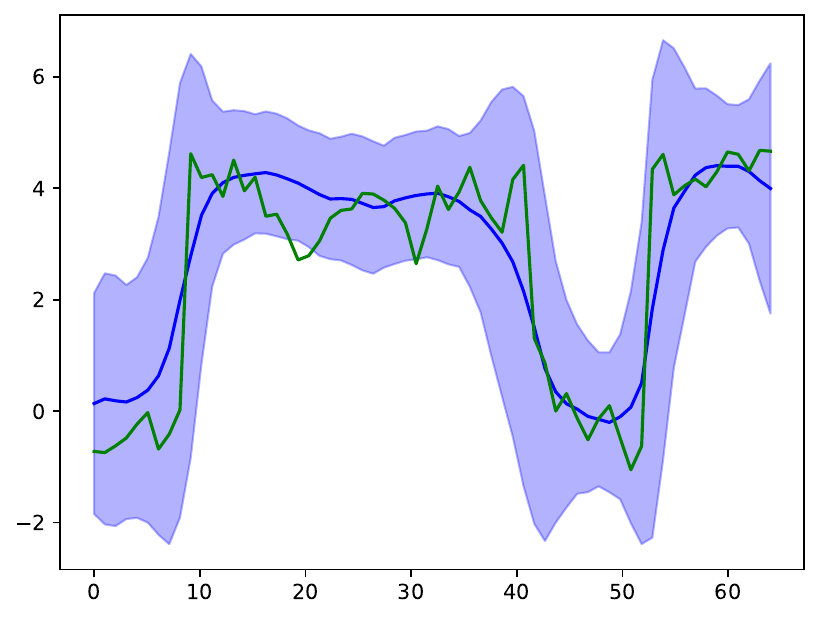}\\{(e)} 
  \end{minipage}
    \begin{minipage}[b]{0.32\linewidth}
    \centering
    \includegraphics[width=0.95\linewidth]{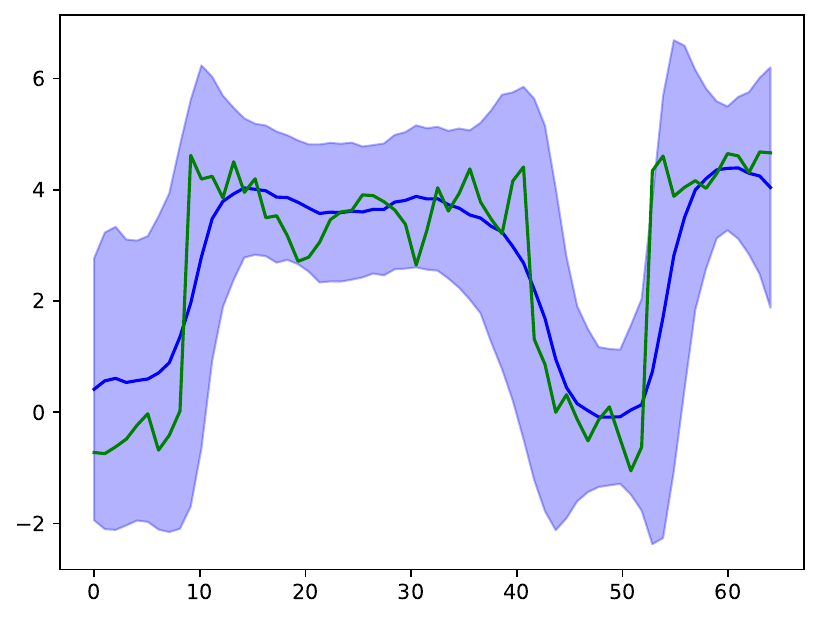}\\{(f)} 
  \end{minipage}
 \caption{One-dimensional cut along the diagonal of the predicted cINN samples and the ground truth log-permeability domain at $D = 4$  for the $3$-D problem. (a)-(c) configuration-$1$ for $1$\%, $3$\%, and $5$\%, respectively. (d)-(f) configuration-$2$ for $1$\%, $3$\%, and $5$\%, respectively. The green curve is the actual (ground truth) field; the blue curve is the mean log-permeability, and the blue shaded area is the $95$\% confidence interval.}
 \label{fig:error_fig_3D}
\end{figure}
\subsection{Effect of the number of observations}
In the present study, we consider two configurations by varying the number of observations in the temporal component. Specifically, we consider the saturation observations at the last time instant and three regular time instants. For the $2$-D, we consider a one-dimensional cut along the diagonal of the domain, as illustrated in Fig.~\ref{fig:error_fig}. For the $3$-D case, we first consider a two-dimensional cut at depth $D$ = 4 and then a one-dimensional cut along the diagonal of that $2$-D domain. 
\par
\begin{figure}[H]
  \begin{minipage}[b]{0.48\linewidth}
    \centering
    \includegraphics[scale=0.45]{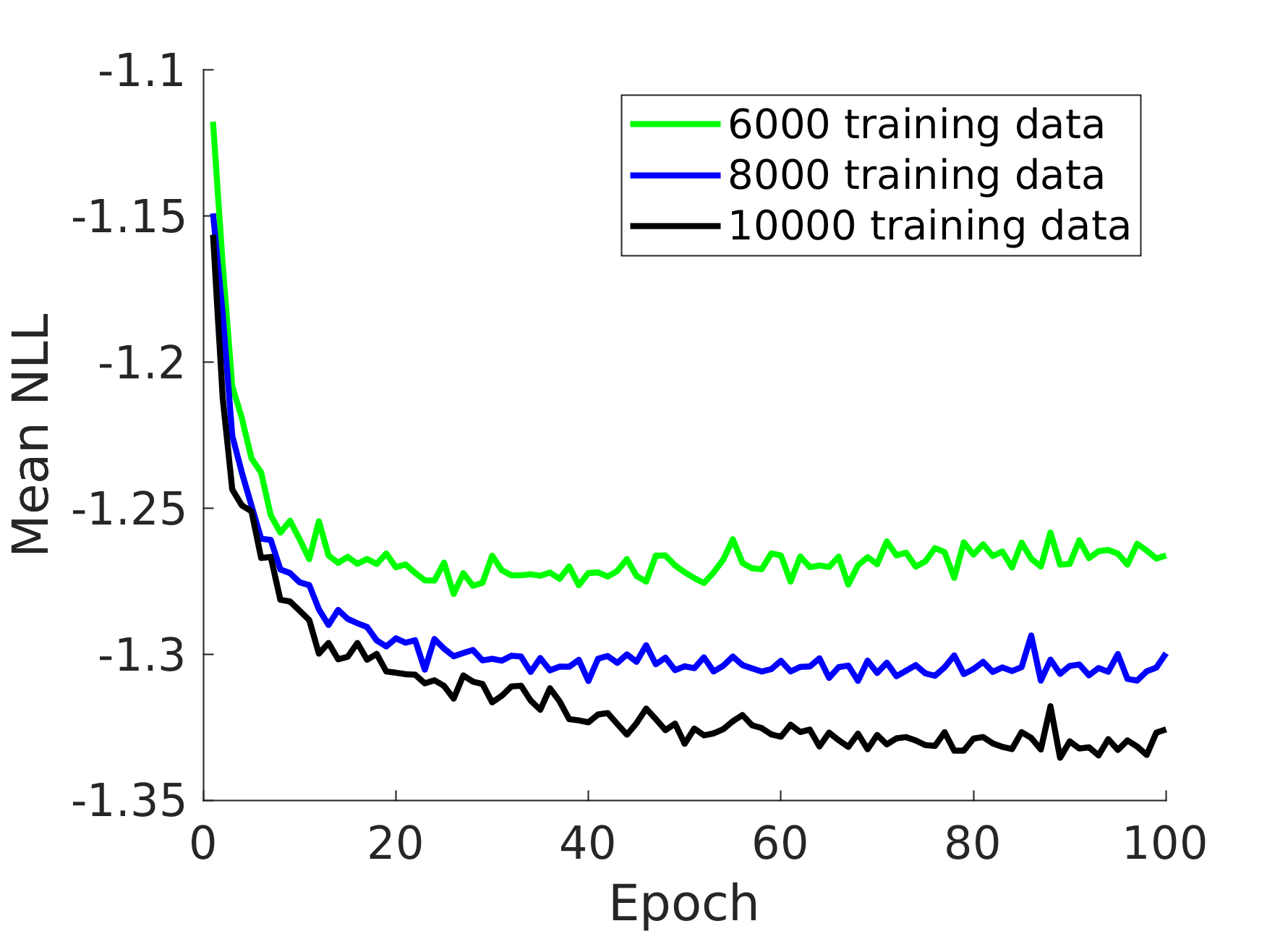}\\
    {(a)}  
  \end{minipage}
  \hspace{4ex}
  \begin{minipage}[b]{0.48\linewidth}
    \centering
    \includegraphics[scale=0.45]{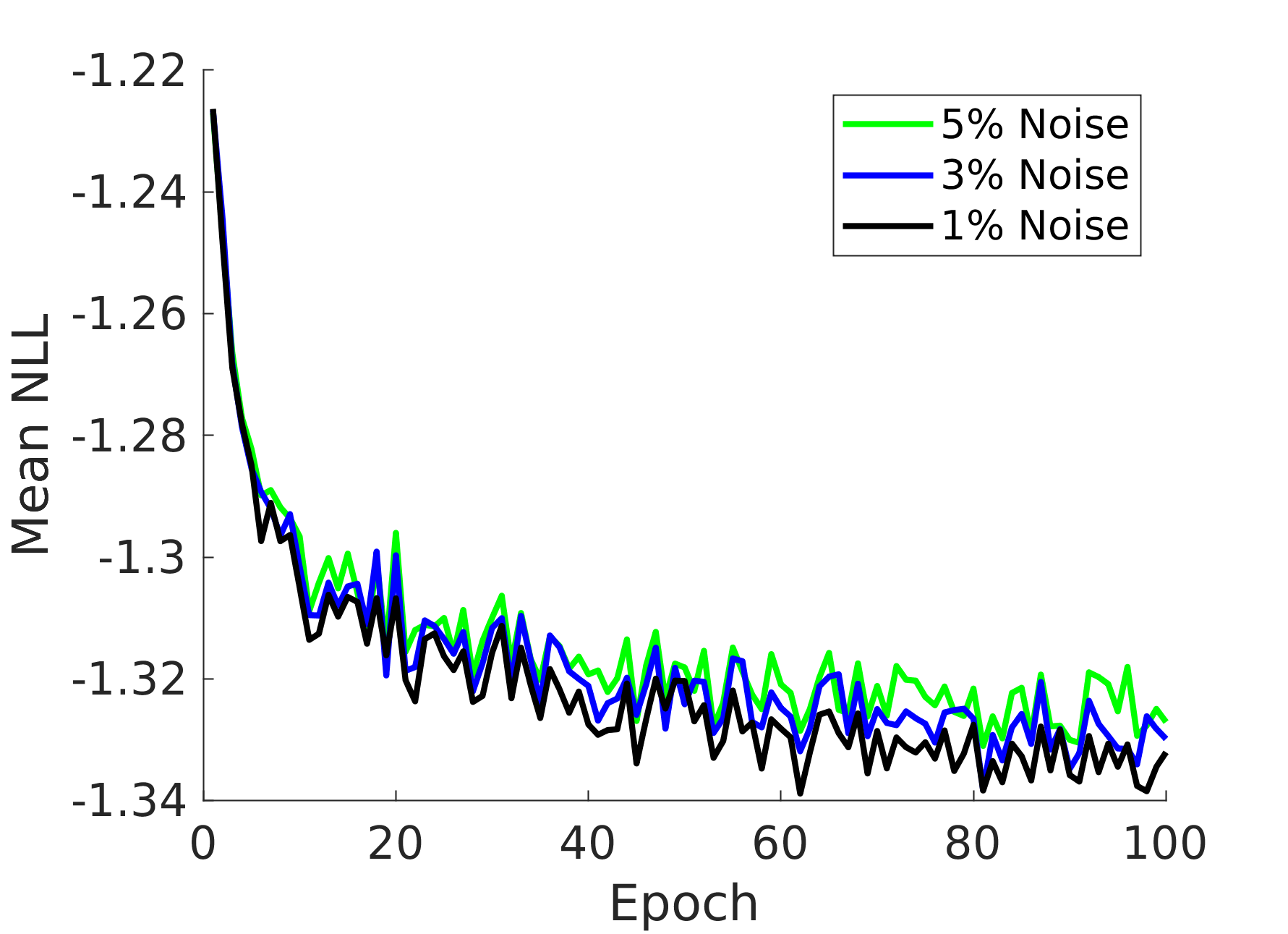} \\
    {(b)}  
  \end{minipage}
  \newline
   \begin{minipage}[b]{0.48\linewidth}
    \centering
    \includegraphics[scale=0.45]{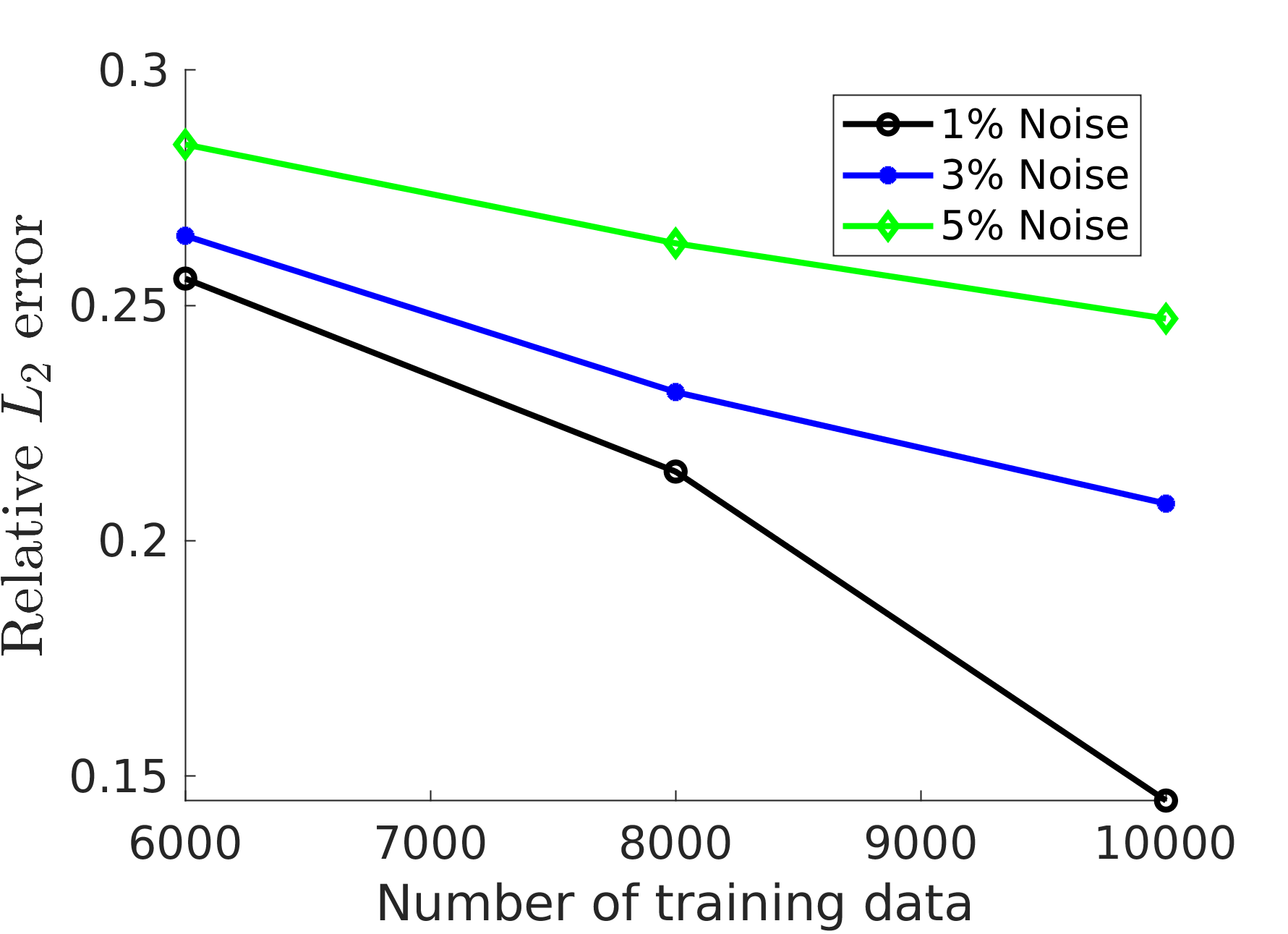}\\
    {(c)}  
  \end{minipage}
  \hspace{4ex}
  \begin{minipage}[b]{0.48\linewidth}
    \centering
    \includegraphics[scale=0.45]{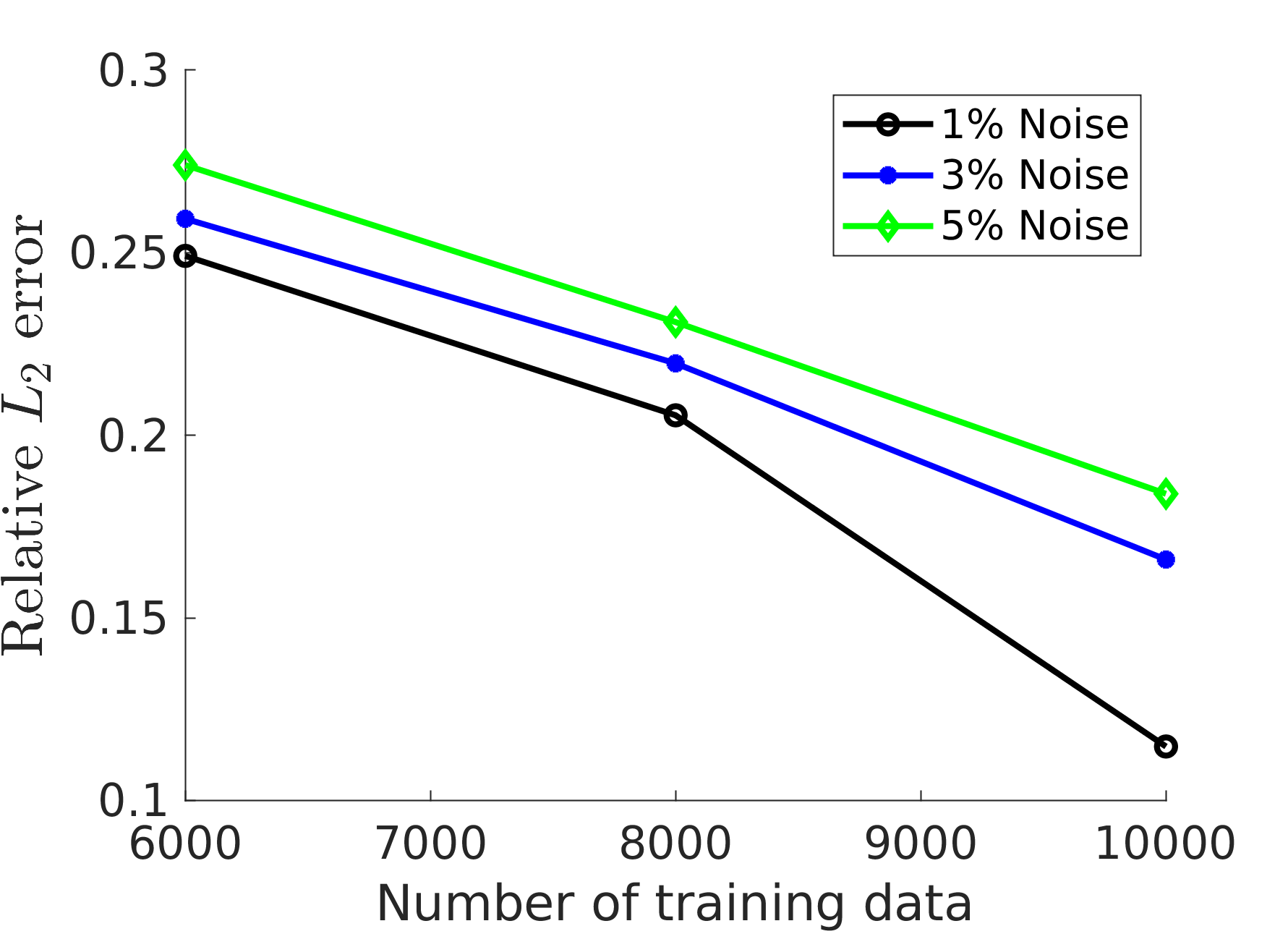} \\
    {(d)}  
  \end{minipage}
 \caption{$2$-D: (a) Test mean NLL loss for various training data with $1\%$ observation noise. The inverse surrogate model is trained with  configuration-$2$, (b) Test mean NLL loss for various levels of observation noise with $10000$ training data. The surrogate model is trained with configuration-$1$, (c) The relative $L_2$ error of the predicted mean of the log-permeability samples for configuration-$1$ with varying training data and different levels of observation noise and (d) The relative $L_2$ error of the predicted mean of the log-permeability samples for configuration-$2$ with varying training data and different observation noise.}
 \label{fig:2_D_l2}
\end{figure}

For both the $2$-D and $3$-D cases, we observe the following.  
First, we see that the width of the confidence interval of the predicted log-permeability gradually reduces as the number of observations increases. For example, if we compare configuration-$1$ and configuration-$2$, we observe a decrease in the confidence interval's width as we increase the number of observations. At the same time, the approximate mean posterior log-permeability gradually follows the trend of the ground truth as we increase the number of observations. Also, the confidence interval is high near the corner pixels of the log-permeability field as the inverse problem is very ill-conditioned, i.e., the number of data (observation data) is not sufficient to completely specify the input log-permeability field. Next, from Fig.~\ref{fig:2_D_l2}(c)-(d) and Fig.~\ref{fig:3_D_l2}(c)-(d), for a given number of training data and observation noise, the relative $L_2$ error decreases as the inverse surrogate is trained with more number of observations. Overall, the addition of the saturation data ($S_1, S_2$ and $S_3$) over a regular interval of time  helps in improving the accuracy of the inverse prediction. 
\subsection{Effect of the observation noise}
\begin{figure}[H]
  \begin{minipage}[b]{0.48\linewidth}
    \centering
    \includegraphics[scale=0.45]{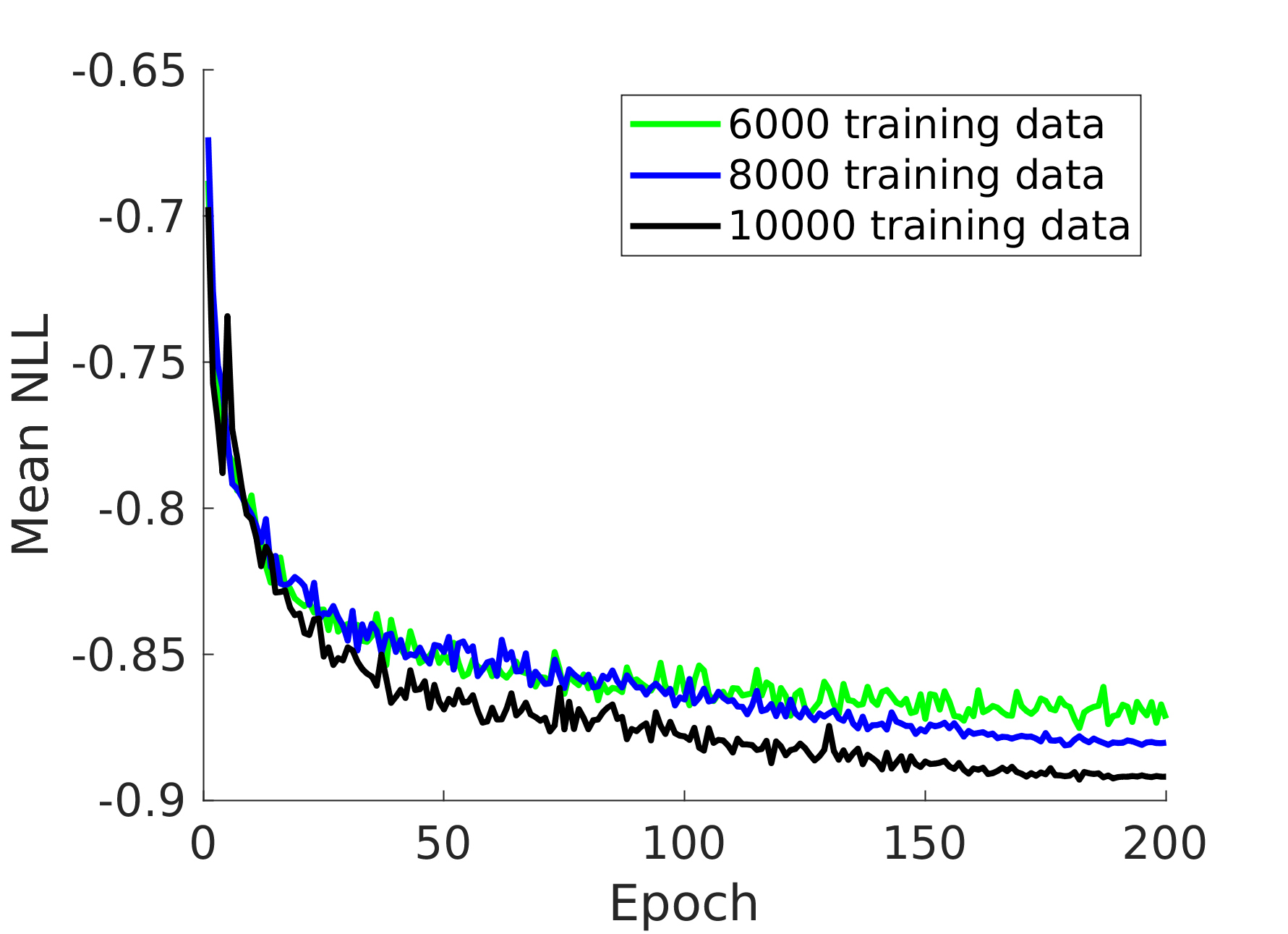}\\
    {(a)}  
  \end{minipage}
  \hspace{4ex}
  \begin{minipage}[b]{0.48\linewidth}
    \centering
    \includegraphics[scale=0.45]{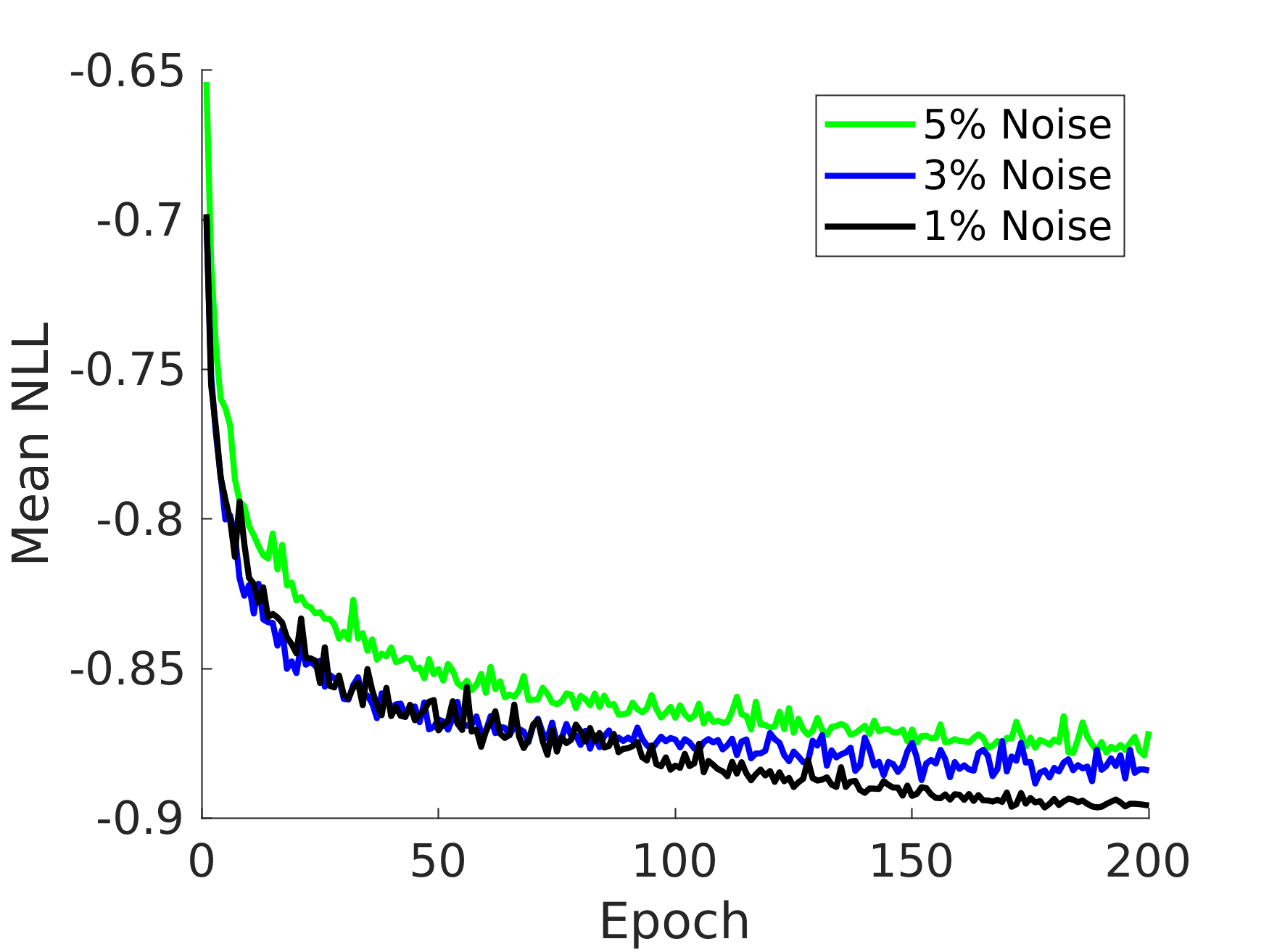} \\
    {(b)}  
  \end{minipage}
  \newline
   \begin{minipage}[b]{0.48\linewidth}
    \centering
    \includegraphics[scale=0.45]{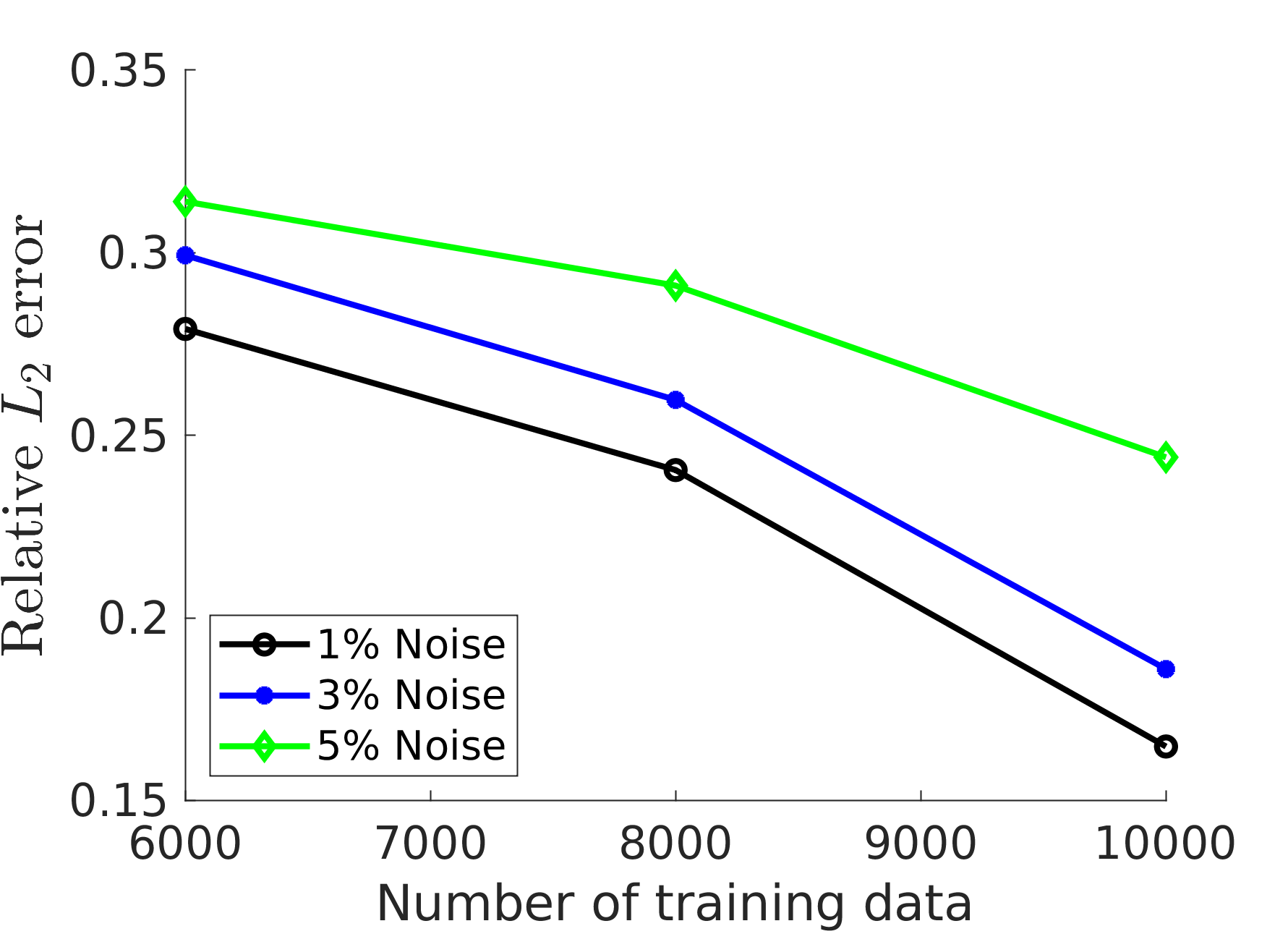}\\
    {(c)}  
  \end{minipage}
  \hspace{4ex}
  \begin{minipage}[b]{0.48\linewidth}
    \centering
    \includegraphics[scale=0.45]{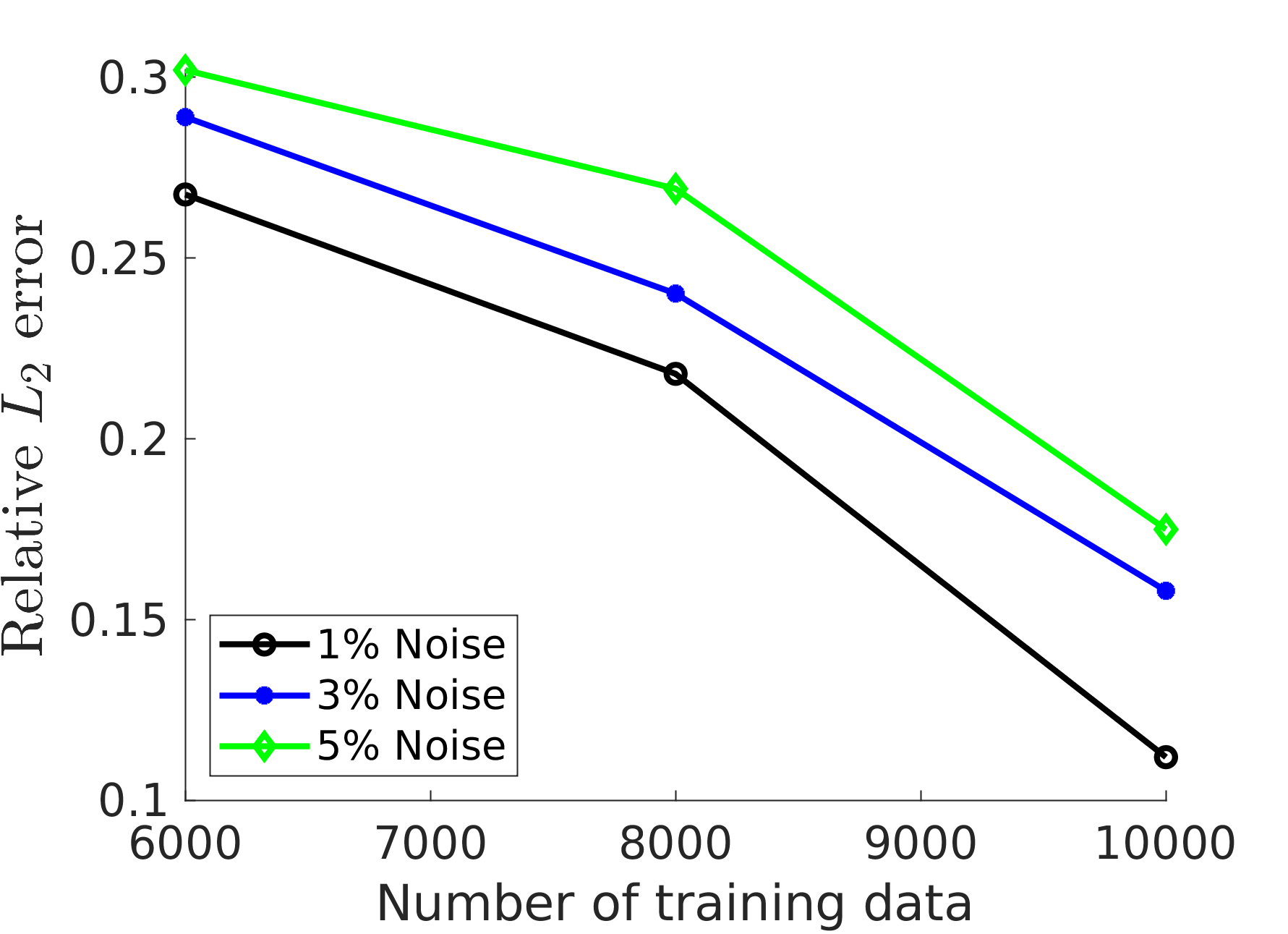} \\
    {(d)}  
  \end{minipage}
 \caption{$3$-D: (a) Test mean NLL loss for various training data with $1\%$ observation noise. The surrogate model is trained with  configuration-$2$, (b) Test mean NLL loss for varying levels of the observation noise with $10000$ training data. The surrogate model is trained with configuration-$1$, (c) The relative $L_2$ error of the predicted mean of the log-permeability samples for configuration-$1$ with varying training data and   observation noise and (d) The relative $L_2$ error of the predicted mean of the log-permeability samples for configuration-$2$ with varying training data and  observation noise.}
 \label{fig:3_D_l2}
\end{figure}

The effect of the pressure and saturation observation noise plays a vital role in evaluating the performance of the developed $2$-D and $3$-D inverse surrogate model. In this work, we train the inverse surrogate with $1\%$, $3\%$, and $5\%$ Gaussian noise for both configurations. As illustrated from the one-dimensional cut along the diagonal of the domain in Fig.~\ref{fig:error_fig} and Fig.~\ref{fig:error_fig_3D}, the inverse surrogate model was trained with $10000$ training data. For both the $2$-D and $3$-D cases, we observe that as the observation noise decreases from $5$\% to $1$\%, the width of the confidence interval reduces. Fig.~\ref{fig:2_D_l2} (b) for $2$-D and Fig.~\ref{fig:3_D_l2} (b) for $3$-D shows the test mean NLL for various levels of observation noise, and it can be seen that the inverse surrogate model performs well even for the highest observation noise and the test mean NLL decreases as the observation noise is reduced from $5$\% to $1$\%. We also report the relative $L_2$ for both  configurations with different observation noise levels. Here, we observe   for both  configurations that the observation noise plays the predominant effect on the prediction of the model with very few pressure and saturation observations. One can observe a similar trend as that of the test mean NLL, i.e., as the observation noise decreases, the $L_2$ values decrease. From the test mean NLL plot and the relative $L_2$ plot for various observation noise, we observe that the $2$-D performs slightly better than the $3$-D model due to the additional dimensionality of depth in the $3$-D model. 
\section{Conclusions}\label{sec:summary}
We outlined the development of a surrogate model that maps the sparse and noisy pressure and saturation observations to a high-dimensional non-Gaussian log-permeability field. In this work, we construct a two- and three-dimensional inverse surrogate models that will allow us to estimate the unknown input field given the incomplete and noisy observations. The inverse surrogate model is a multiscale conditional invertible neural network (cINN) that consists of an invertible network and a conditioning network. Both   networks are trained in an end-to-end fashion by maximizing the conditional log-likelihood evaluated through the change of variables. The inverse surrogate model was then demonstrated on an inverse multiphase flow problem with   various numbers of training data, observation data, and lastly, different levels of the observation noise. The predicted output samples of the non-Gaussian log-permeability field from the model trained with limited pairs of pressure and saturation observations and the log-permeability field were diverse with the   predictive mean being close to the ground truth.
\par
One could further extend the presented model to a dynamic setting (sequential inversion tasks) by introducing modern architectures such as long short-term memory~\cite{Geneva2020Modeling,mackay2018reversible, xingjian2015convolutional, kumar2020convcast}. Simultaneous identification of the permeability and time-dependent source terms is also a natural extension~\cite{Mo2019Deep}. 
\section*{Acknowledgements}
This work on invertable neural networks has been supported by ARPA-E award \# DE-AR0001204. Computing resources were provided by the AFOSR Office
of Scientific Research through the DURIP program and by the University of Notre Dame’s Center for Research Computing (CRC).
\appendix
\section{Effect of the number of affine coupling layers in each invertible block}
\label{sec:Appendix-A}
The design of the invertible blocks is one of the key components of the conditional invertible neural network. To determine the optimum number of affine coupling layers in each invertible block, we train the network with a different number of affine coupling layers while maintaining the same number of training data and configurations, e.g., $10000$ training data, configuration $2$ and 1\% observation noise (see Section~\ref{sec:Observations}). We begin with a simple configuration: $1$-$1$-$1$-$1$ ($L_1$-$L_2$-$L_3$-$L_4$), and then vary the number of affine coupling layers in each convolutional invertible block. As mentioned in Section~\ref{sec:methods}, we construct three convolutional invertible blocks at multiple scales. For example, in the $2$-D case, we construct the convolutional invertible blocks: $1$, $2$ and $3$ with $64 \times 64$, $32 \times 32$ and $16 \times 16$ feature maps, respectively.
\par
\begin{figure}[H]
  \begin{minipage}[b]{0.48\linewidth}
    \centering
    \includegraphics[scale=0.38]{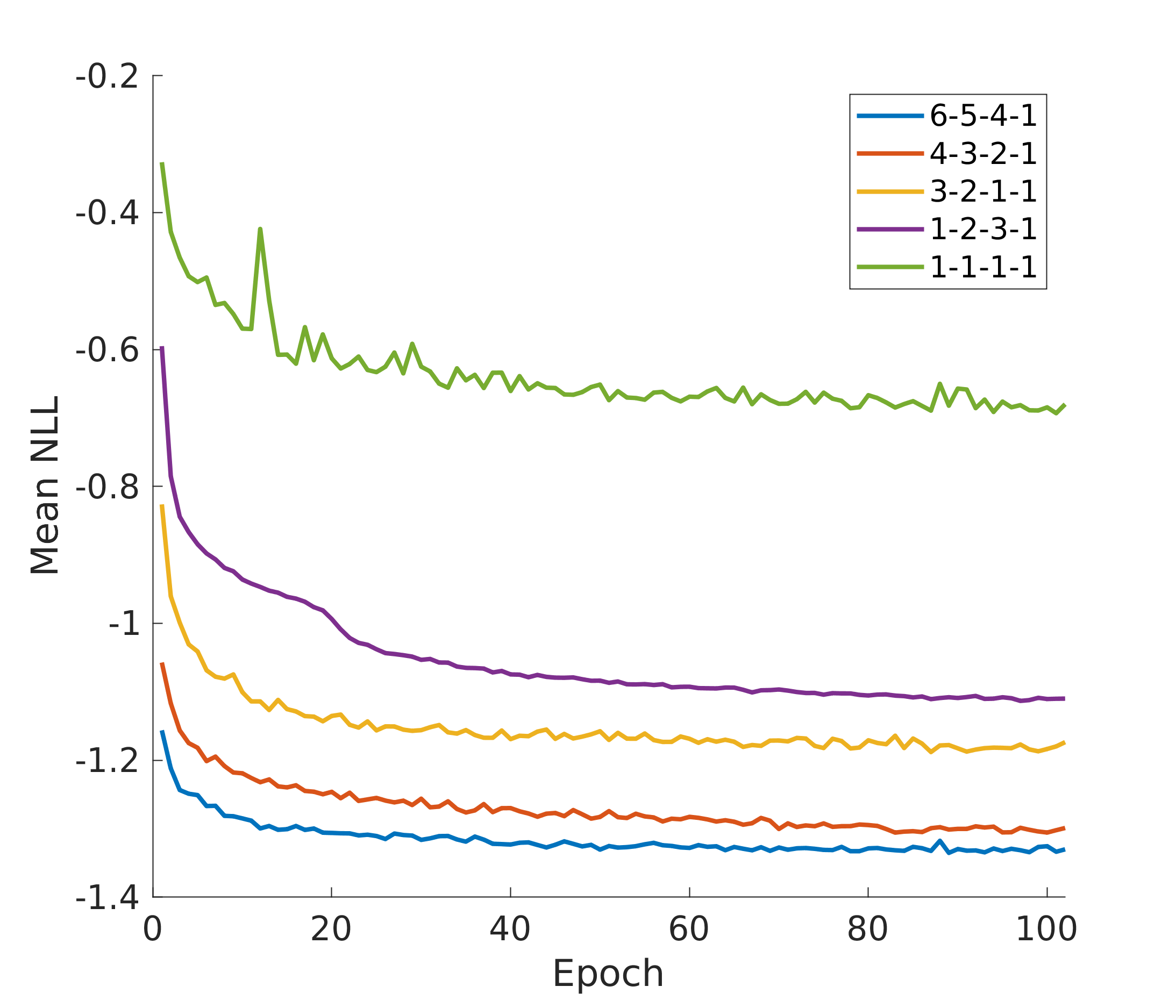}\\
    {(a)}  
  \end{minipage}
  \begin{minipage}[b]{0.48\linewidth}
    \centering
    \includegraphics[scale=0.38]{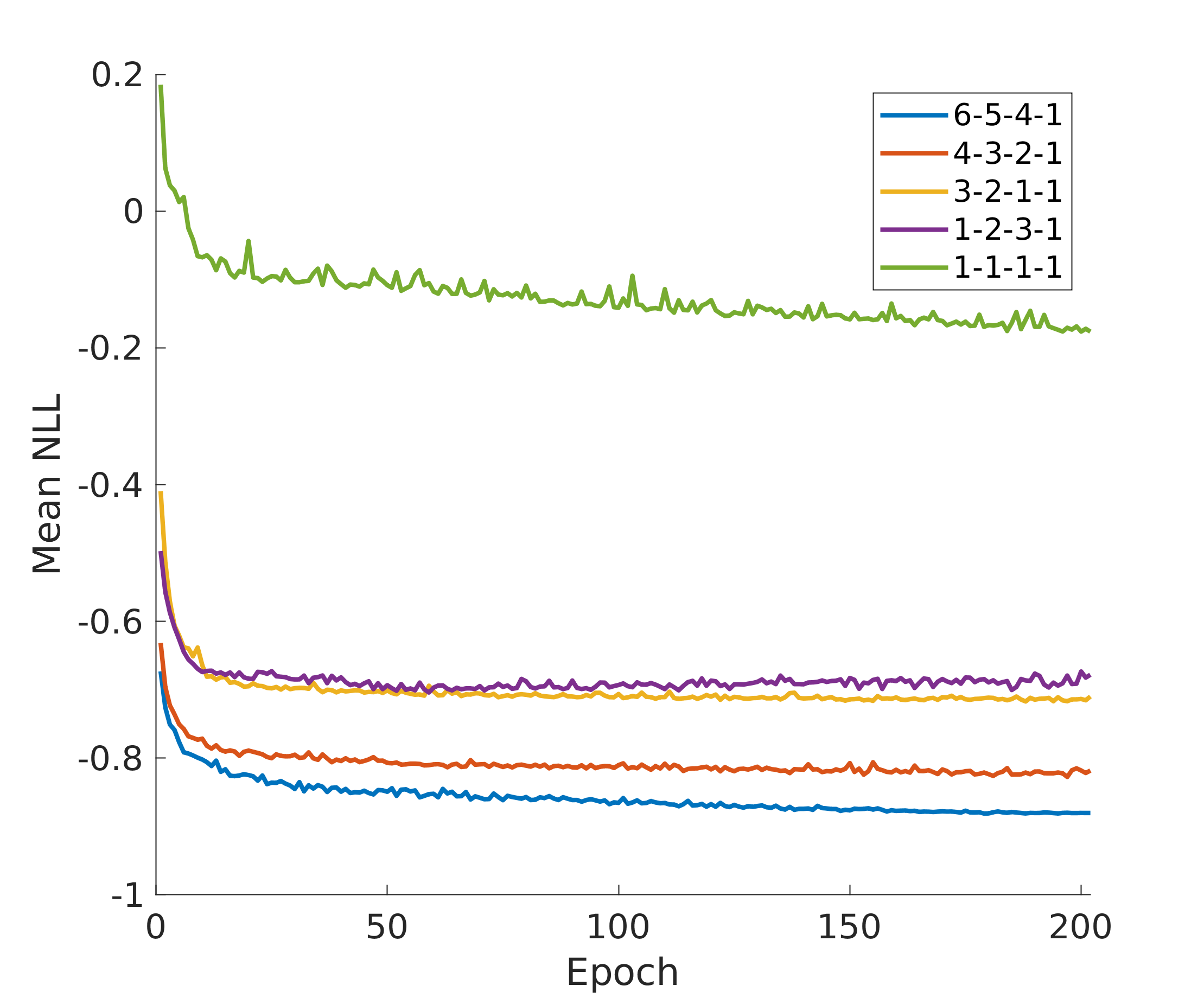} \\
    {(b)}  
  \end{minipage}
 \caption{The mean NLL for various invertible blocks for (a) $2$-D multiscale architecture and (b) $3$-D multiscale architecture.}
 \label{fig:diff_block}
\end{figure}

First, we increase the number of affine coupling layers in the convolutional invertible block$-1$ and train the model with the configuration: $3-2-1-1$. Second, we construct another model with increasing the number of affine coupling layers in the convolutional invertible block$-3$ and train the model with the configuration: $1-2-3-1$. The mean NLL for the above configurations is shown in Fig.~\ref{fig:diff_block}. We find that the configuration $3-2-1-1$ performs better than the configuration: $1-2-3-1$. This is mainly due to the fact that the convolutional invertible block$-1$ extracts higher-level information of the input field compared to the convolutional invertible block$-3$. Therefore, we increase the number of affine coupling layers with the following configurations: $4-3-2-1$ and  $6-5-4-1$. From Fig.~\ref{fig:diff_block}(a) and (b) for the $2$-D and $3$-D problem respectively, we observe that the mean NLL decreases as we increase the number of blocks from $3-2-1-1$ to $4-3-2-1$. Also, we observe a slight decrease in the mean NLL as we increase the number of blocks from $4-3-2-1$ to $6-5-4-1$. A further increase in the number of affine coupling layers leads to only a slight improvement in the model predictions but with a higher computational cost (higher number of parameters to train the model). Therefore, we construct the conditional invertible neural network with the following number of invertible blocks ($L_1-L_2-L_3-L_4$): $6-5-4-1$. The hyperparameters considered for this examples are tabulated in Table~\ref{tab:hyperparameters1}.   
\begin{table}[H]
\caption{The hyper-parameters parameters used for training various affine coupling layers in each invertible block.}
\begin{tabular}{ccccc}
                                      & \multicolumn{4}{c}{$2$-D problem}                                                                                                 \\ \cline{2-5} 
\multicolumn{1}{c|}{}                 & \multicolumn{1}{c|}{$1-1-1-1$} & \multicolumn{1}{c|}{$1-2-3-1$} & \multicolumn{1}{c|}{$3-2-1-1$} & \multicolumn{1}{c|}{$6-5-4-1$} \\ \hline
\multicolumn{1}{c|}{Learning rate}    & $1e-3$                         & $5e-3$                         & $1e-3$                         & $1e-4$                         \\
\multicolumn{1}{c|}{Weight decay}     & $1e-5$                         & $5e-5$                         & $5e-5$                         & $1e-5$                         \\
\multicolumn{1}{c|}{Number of epochs} & $100$                          & $100$                          & $100$                          & $100$                          \\ \hline
                                      & \multicolumn{4}{c}{$3$-D problem}                                                                                                 \\ \cline{2-5} 
\multicolumn{1}{c|}{}                 & \multicolumn{1}{c|}{$1-1-1-1$} & \multicolumn{1}{c|}{$1-2-3-1$} & \multicolumn{1}{c|}{$3-2-1-1$} & \multicolumn{1}{c|}{$6-5-4-1$} \\ \hline
\multicolumn{1}{c|}{Learning rate}    & $5e-4$                         & $1e-4$                         & $9e-5$                         & $5e-5$                         \\
\multicolumn{1}{c|}{Weight decay}     & $7e-6$                         & $7e-6$                         & $5e-6$                         & $5e-6$                         \\
\multicolumn{1}{c|}{Number of epochs} & $200$                          & $200$                          & $200$                          & $200$                          \\ \cline{1-1}
\end{tabular}
\label{tab:hyperparameters1}
\end{table}
\section{Effect of conditioning input scales}
\label{sec:Appendix-B}
The multi-scale nature of the conditional invertible network is critical in recovering the high-dimensional log-permeability field given  noisy and sparse observations. To demonstrate the effect of the conditioning input scales, we construct and train three models separately with the same number of training data and configurations, e.g., $10000$ training data, configuration $2$ and 1\% observation noise (see Section~\ref{sec:Observations}). The three models are constructed as follows. For the first model, we consider the conditioning input scale of $64 \times 64$, i.e., $\bm{c}_1$  and a fully-connected block $\bm{c}_4$ (see Section~\ref{sec:methods}) and their corresponding invertible blocks with $6$ and $1$ affine coupling layers, respectively, thereby limiting the conditioning input scales. The second model consists of two 
%convolutions 
conditioning input scales: $64 \times 64$ (conditional block $\bm{c}_1$)  and $32 \times 32$ (conditional block $\bm{c}_2$),  a fully-connected block $\bm{c}_4$ and their corresponding invertible blocks with $6$, $5$ and $1$ affine coupling layers, respectively. For the third model, we consider all the conditioning input scales (at multiple scales) as described in Section~\ref{sec:cinn}, i.e., $64 \times 64$ ($\bm{c}_1$), $32 \times 32$ ($\bm{c}_2$), $16 \times 16$ ($\bm{c}_3$), a fully-connected block $\bm{c}_4$ and their corresponding invertible blocks with $6$, $5$, $4$ and $1$ affine coupling layers, respectively. The hyperparameters considered for these examples are tabulated in Table~\ref{tab:hyperparameters2}.
\par
\begin{figure}[H]
  \begin{minipage}[b]{0.325\linewidth}
    \centering
    \includegraphics[width=0.95\linewidth]{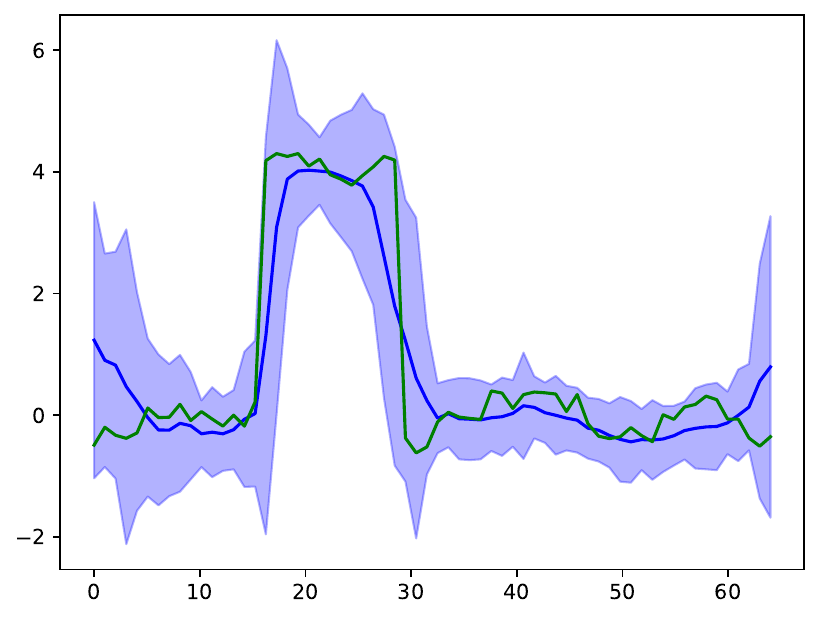}\\
    {(a)}  
  \end{minipage}
  \begin{minipage}[b]{0.325\linewidth}
    \centering
    \includegraphics[width=0.95\linewidth]{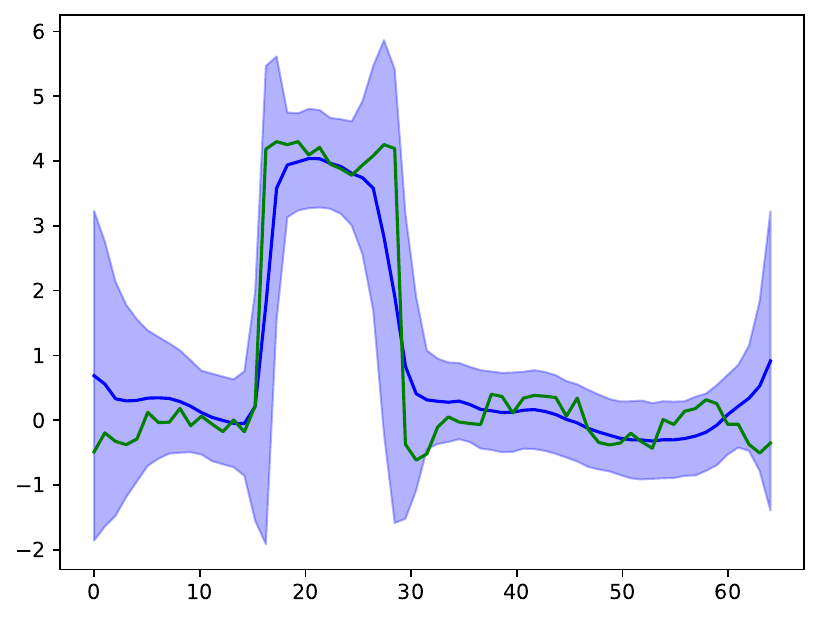} \\
    {(b)}  
  \end{minipage}
   \begin{minipage}[b]{0.325\linewidth}
    \centering
    \includegraphics[width=0.95\linewidth]{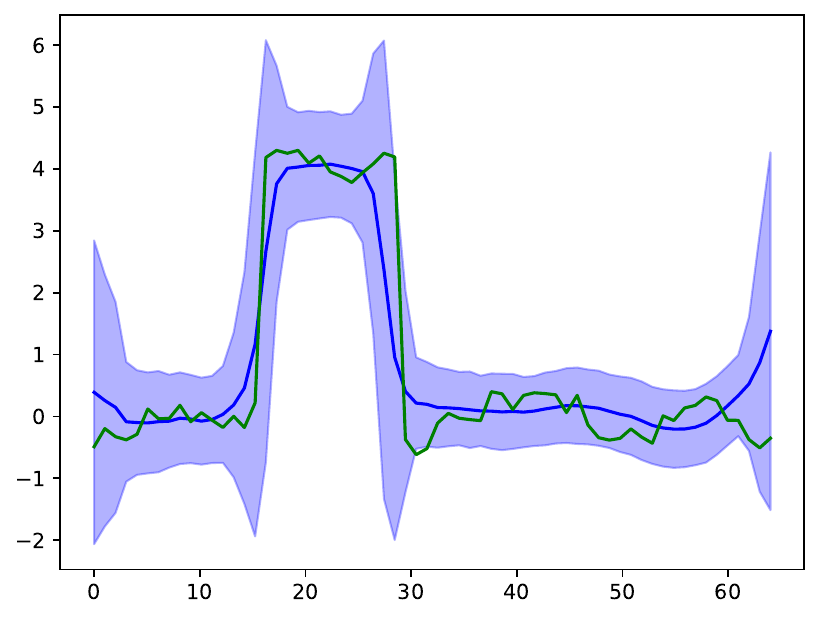}\\
    {(c)}  
  \end{minipage}
  \newline
    \centering
    \includegraphics[scale=0.5]{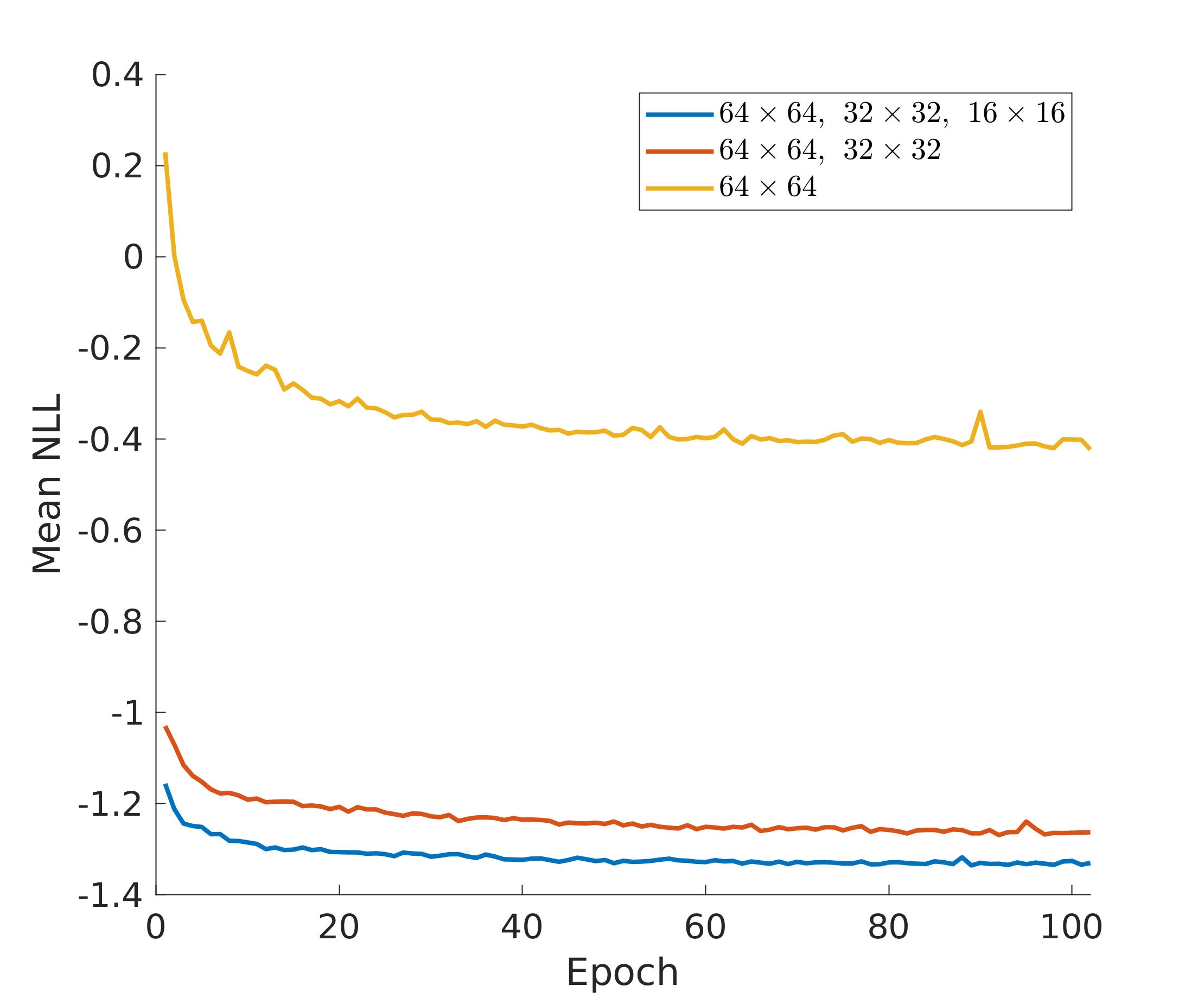} \\
    {(d)}  
 \caption{One-dimensional cut along the diagonal of the predicted cINN samples and the ground truth log-permeability domain for the $2$-D problem: (a) model-$1$, (b) model-$2$, and (c) model-$3$. The green curve is the actual (ground truth) field; the blue curve is the mean log-permeability, and the blue shaded area is the $95$\% confidence interval. (d) The mean NLL for various models for the $2$-D problem.}
 \label{fig:2_D_diff_scale}
\end{figure}

\begin{figure}[H]
  \begin{minipage}[b]{0.32\linewidth}
    \centering
    \includegraphics[width=0.95\linewidth]{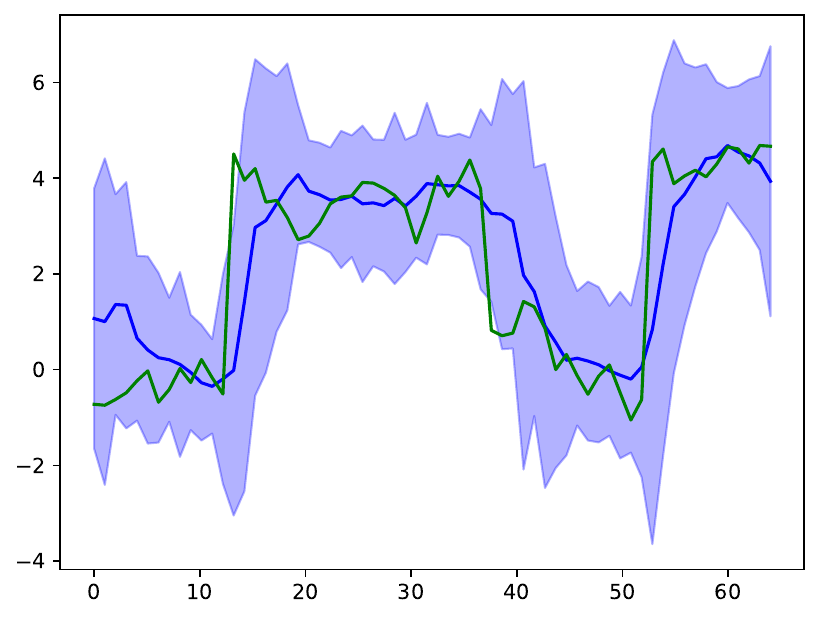}\\
    {(a)}  
  \end{minipage}
  \begin{minipage}[b]{0.32\linewidth}
    \centering
    \includegraphics[width=0.95\linewidth]{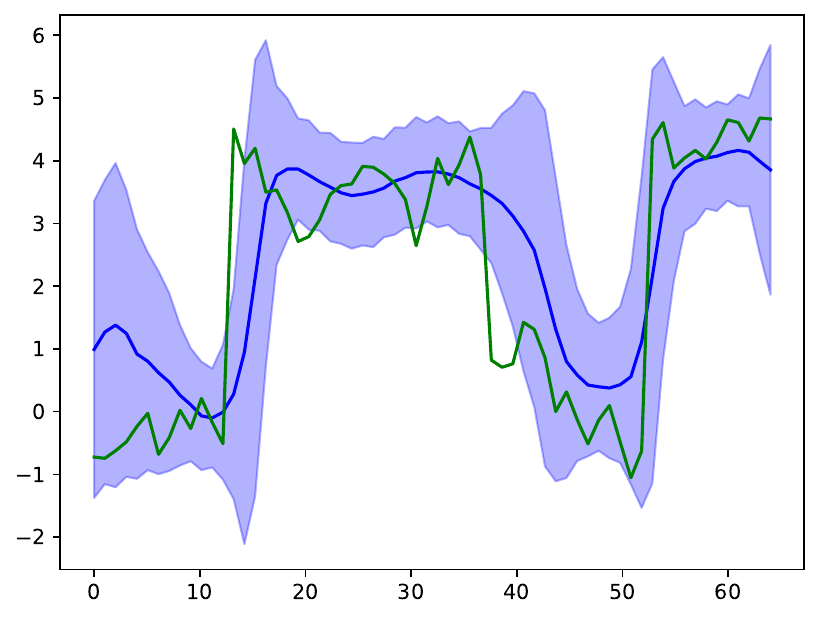} \\
    {(b)}  
  \end{minipage}
   \begin{minipage}[b]{0.32\linewidth}
    \centering
    \includegraphics[width=0.95\linewidth]{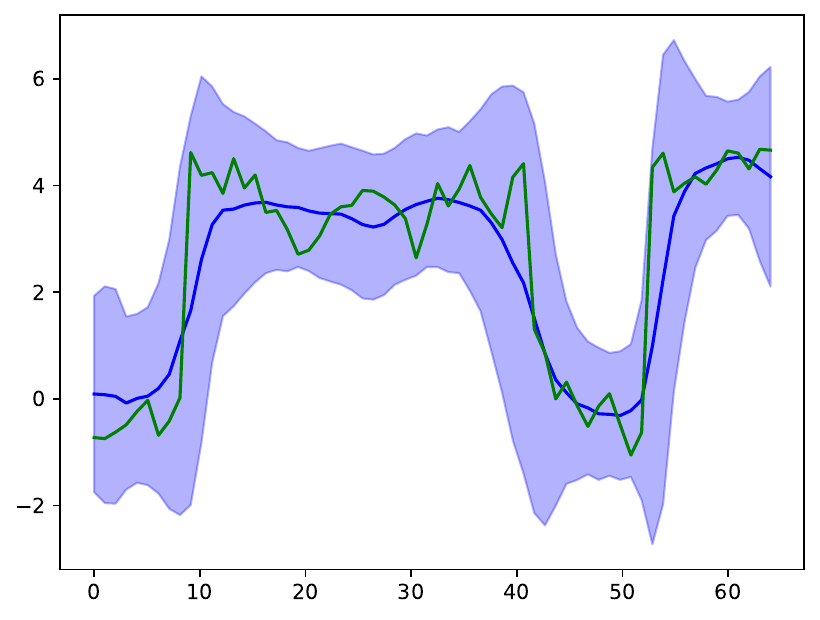}\\
    {(c)}  
  \end{minipage}
  \newline
    \centering
    \includegraphics[scale=0.5]{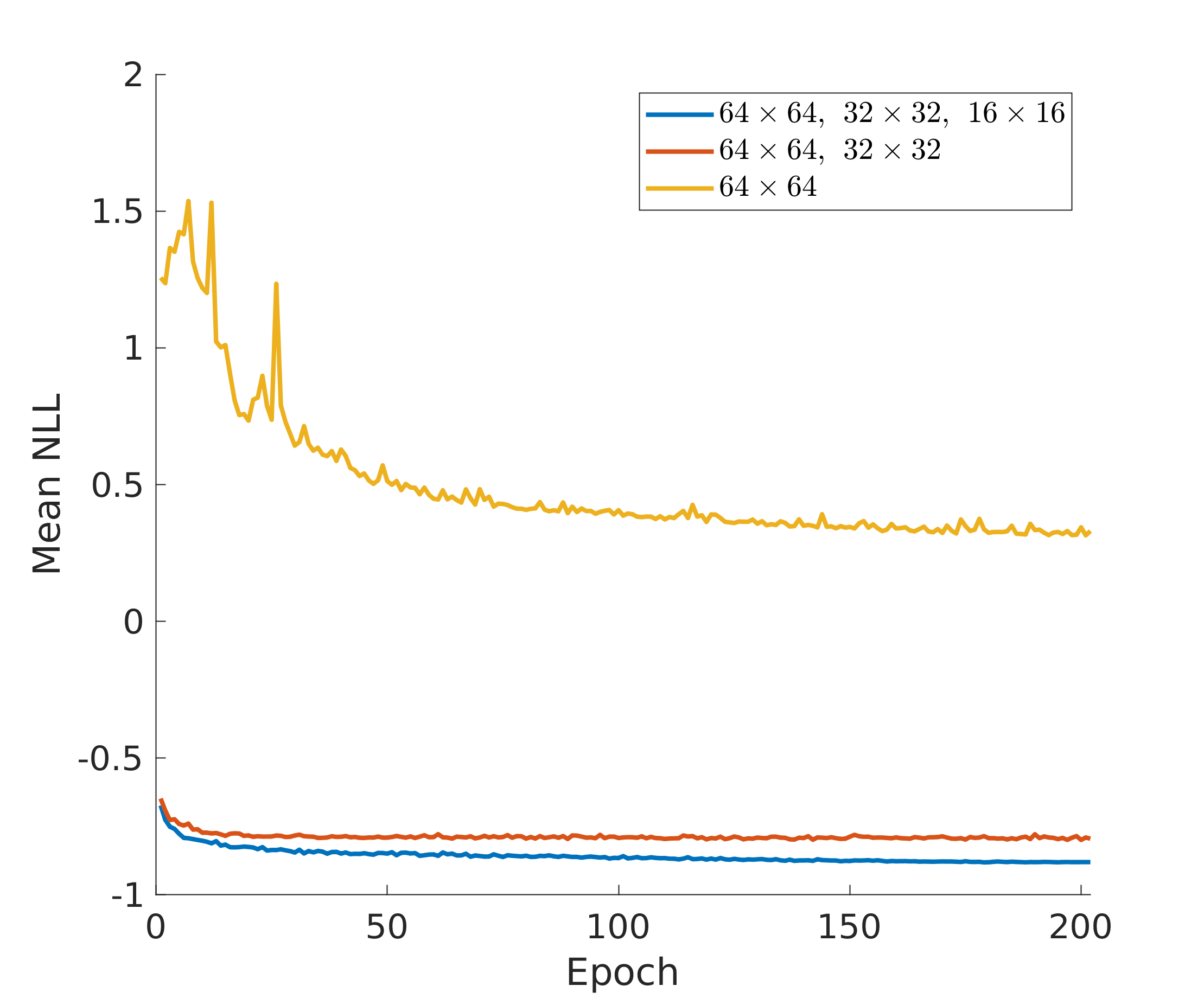} \\
    {(d)}  
 \caption{One-dimensional cut along the diagonal of the predicted cINN samples and the ground truth log-permeability domain at $D = 4$  for the $3$-D problem: (a) model-$1$, (b) model-$2$, and (c) model-$3$. The green curve is the actual (ground truth) field; the blue curve is the mean log-permeability, and the blue shaded area is the $95$\% confidence interval. (d) The mean NLL for various models for the $3$-D problem.}
 \label{fig:3_D_diff_scale}
\end{figure}

 
The mean NLL for the aforementioned models are shown in Fig.~\ref{fig:2_D_diff_scale} (d) and Fig.~\ref{fig:3_D_diff_scale} (d) for the $2$-D and $3$-D problems, respectively. We observe that the mean NLL decreases as we increase the number of conditioning input scales from $64 \times 64$ (model-$1$) to $64 \times 64$ and $32 \times 32$ (model-$2$). Also, we see a slight decrease in the mean NLL when we increase the number conditioning input scales from $64 \times 64$ and $32 \times 32$ (i.e., model-$2$) to $64 \times 64$, $32 \times 32$ and $16 \times 16$ (model-$3$). We also show the one-dimensional cut along the diagonal of the domain for the $2$-D problem in Fig.~\ref{fig:2_D_diff_scale} (a), (b) and (c) for models $1$, $2$ and $3$, respectively. Similarly, for the $3$-D case, we first consider a two-dimensional cut at depth $D = 4$ and then a one-dimensional cut along the diagonal of that $2$-D domain as shown in  Figs.~\ref{fig:3_D_diff_scale} (a), (b) and (c) for models $1$, $2$ and $3$, respectively. We find that as we increase the conditioning input scales from $64 \times 64$ (model-$1$) to $64 \times 64$ and $32 \times 32$ (model-$2$), the confidence interval’s width decreases.  There is a slight decrease in the confidence interval’s width when we increase the number of conditioning input scales from model-$2$ to model-$3$. Also, the approximate mean posterior log-permeability (indicated in blue curve) gradually follows the trend of the ground truth (indicated in green curve) as we increase the number of conditioning scales i.e., at multiple scales.
\begin{table}[]
\caption{The hyperparameters parameters used for training models $1$, $2$, and $3$.}
\centering
\begin{tabular}{c|ccc|ccc|}
                 & \multicolumn{3}{c|}{$2$-D problem}                                          & \multicolumn{3}{c|}{$3$-D problem}                                          \\ \cline{2-7} 
                 & \multicolumn{1}{c|}{Model-$1$} & \multicolumn{1}{c|}{Model-$2$} & Model-$3$ & \multicolumn{1}{c|}{Model-$1$} & \multicolumn{1}{c|}{Model-$2$} & Model-$3$ \\ \hline
Learning rate    & $9e-3$                         & $5e-4$                         & $1e-4$    & $5e-4$                         & $5e-4$                         & $5e-5$    \\
Weight decay     & $5e-5$                         & $5e-5$                         & $1e-5$    & $7e-6$                         & $5e-6$                         & $5e-6$    \\
Number of epochs & $100$                          & $100$                          & $100$     & $200$                          & $200$                          & $200$    
\end{tabular}
\label{tab:hyperparameters2}
\end{table}
\bibliographystyle{unsrt}
\bibliography{sample1.bib}

\begin{thebibliography}{10}

\bibitem{herbei2008gyres}
Radu Herbei, Ian~W McKeague, and Kevin~G Speer.
\newblock Gyres and jets: {Inversion} of tracer data for ocean circulation
  structure.
\newblock {\em Journal of Physical Oceanography}, 38(6):1180--1202, 2008.
\newblock
  \url{https://pdfs.semanticscholar.org/fe92/7ba3be89cab4beb6ab929ecec699fd38468e.pdf}.

\bibitem{aguilo2010inverse}
Miguel~A Aguilo, Wilkins Aquino, John~C Brigham, and Mostafa Fatemi.
\newblock An inverse problem approach for elasticity imaging through
  vibroacoustics.
\newblock {\em IEEE transactions on medical imaging}, 29(4):1012--1021, 2010.
\newblock \url{https://doi.org/10.1109/TMI.2009.2039225}.

\bibitem{russell1988introduction}
Brian~H Russell.
\newblock {\em Introduction to seismic inversion methods}.
\newblock Society of Exploration Geophysicists, 1988.
\newblock \url{https://doi.org/10.1190/1.9781560802303}.

\bibitem{haario2004markov}
Heikki Haario, Marko Laine, Markku Lehtinen, Eero Saksman, and Johanna
  Tamminen.
\newblock {Markov chain Monte Carlo} methods for high dimensional inversion in
  remote sensing.
\newblock {\em Journal of the Royal Statistical Society: series B (statistical
  methodology)}, 66(3):591--607, 2004.
\newblock \url{ https://doi.org/10.1111/j.1467-9868.2004.02053.x}.

\bibitem{vrugt2016markov}
Jasper~A Vrugt.
\newblock Markov {chain Monte Carlo} simulation using the {DREAM} software
  package: {Theory}, concepts, and {MATLAB} implementation.
\newblock {\em Environmental Modelling \& Software}, 75:273--316, 2016.
\newblock \url{https://doi.org/10.1016/j.envsoft.2015.08.013}.

\bibitem{laloy2012high}
Eric Laloy and Jasper~A Vrugt.
\newblock High-dimensional posterior exploration of hydrologic models using
  multiple-try {DREAM (ZS)} and high-performance computing.
\newblock {\em Water Resources Research}, 48(1), 2012.
\newblock \url{ https://doi.org/10.1029/2011WR010608}.

\bibitem{sun2009sequential}
Alexander~Y Sun, Alan~P Morris, and Sitakanta Mohanty.
\newblock Sequential updating of multimodal hydrogeologic parameter fields
  using localization and clustering techniques.
\newblock {\em Water Resources Research}, 45(7), 2009.
\newblock \url{https://doi.org/10.1029/2008WR007443}.

\bibitem{ju2018adaptive}
Lei Ju, Jiangjiang Zhang, Long Meng, Laosheng Wu, and Lingzao Zeng.
\newblock An adaptive {Gaussian} process-based iterative ensemble smoother for
  data assimilation.
\newblock {\em Advances in Water Resources}, 115:125--135, 2018.
\newblock \url{https://doi.org/10.1016/j.advwatres.2018.03.010}.

\bibitem{bilionis2013solution}
Ilias Bilionis and Nicholas Zabaras.
\newblock Solution of inverse problems with limited forward solver evaluations:
  {A Bayesian} perspective.
\newblock {\em Inverse Problems}, 30(1):015004, 2013.
\newblock \url{https://doi.org/10.1088/0266-5611/30/1/015004}.

\bibitem{rasmussen2003gaussian}
Carl~Edward Rasmussen.
\newblock Gaussian processes in machine learning.
\newblock In {\em Summer School on Machine Learning}, pages 63--71. Springer,
  2003.
\newblock \url{https://doi.org/10.1007/978-3-540-28650-9_4}.

\bibitem{marzouk2009dimensionality}
Youssef~M Marzouk and Habib~N Najm.
\newblock Dimensionality reduction and polynomial chaos acceleration of
  {Bayesian} inference in inverse problems.
\newblock {\em Journal of Computational Physics}, 228(6):1862--1902, 2009.
\newblock \url{https://doi.org/10.1016/j.jcp.2008.11.024}.

\bibitem{xiu2002wiener}
Dongbin Xiu and George~Em Karniadakis.
\newblock The {Wiener--Askey} polynomial chaos for stochastic differential
  equations.
\newblock {\em SIAM journal on scientific computing}, 24(2):619--644, 2002.
\newblock \url{https://doi.org/10.1137/S1064827501387826}.

\bibitem{laloy2017inversion}
Eric Laloy, Romain H{\'e}rault, John Lee, Diederik Jacques, and Niklas Linde.
\newblock Inversion using a new low-dimensional representation of complex
  binary geological media based on a deep neural network.
\newblock {\em Advances in Water Resources}, 110:387--405, 2017.
\newblock \url{https://doi.org/10.1016/j.advwatres.2017.09.029}.

\bibitem{canchumuni2019history}
Smith~WA Canchumuni, Alexandre~A Emerick, and Marco Aurelio~C Pacheco.
\newblock History matching geological facies models based on ensemble smoother
  and deep generative models.
\newblock {\em Journal of Petroleum Science and Engineering}, 177:941--958,
  2019.
\newblock \url{https://doi.org/10.1016/j.petrol.2019.02.037}.

\bibitem{mo2020integration}
Shaoxing Mo, Nicholas Zabaras, Xiaoqing Shi, and Jichun Wu.
\newblock Integration of adversarial autoencoders with residual dense
  convolutional networks for estimation of non-{Gaussian} hydraulic
  conductivities.
\newblock {\em Water Resources Research}, 56(2):e2019WR026082, 2020.
\newblock \url{https://doi.org/10.1002/2017WR022148}.

\bibitem{hamilton2018deep}
Sarah~Jane Hamilton and Andreas Hauptmann.
\newblock Deep d-bar: Real-time electrical impedance tomography imaging with
  deep neural networks.
\newblock {\em IEEE transactions on medical imaging}, 37(10):2367--2377, 2018.
\newblock \url{https://doi.org/10.1109/TMI.2018.2828303}.

\bibitem{jin2017deep}
Kyong~Hwan Jin, Michael~T McCann, Emmanuel Froustey, and Michael Unser.
\newblock Deep convolutional neural network for inverse problems in imaging.
\newblock {\em IEEE Transactions on Image Processing}, 26(9):4509--4522, 2017.
\newblock \url{https://doi.org/10.1109/TIP.2017.2713099}.

\bibitem{adler2017solving}
Jonas Adler and Ozan {\"O}ktem.
\newblock Solving ill-posed inverse problems using iterative deep neural
  networks.
\newblock {\em Inverse Problems}, 33(12):124007, 2017.
\newblock \url{https://doi.org/10.1088/1361-6420/aa9581}.

\bibitem{fan2020solving}
Yuwei Fan and Lexing Ying.
\newblock Solving electrical impedance tomography with deep learning.
\newblock {\em Journal of Computational Physics}, 404:109119, 2020.
\newblock \url{https://doi.org/10.1016/j.jcp.2019.109119}.

\bibitem{li2019novel}
Xiuyan Li, Yong Zhou, Jianming Wang, Qi~Wang, Yang Lu, Xiaojie Duan, Yukuan
  Sun, Jingwan Zhang, and Zongyu Liu.
\newblock A novel deep neural network method for electrical impedance
  tomography.
\newblock {\em Transactions of the Institute of Measurement and Control},
  41(14):4035--4049, 2019.
\newblock \url{https://doi.org/10.1177/0142331219845037}.

\bibitem{whang2020compressed}
Jay Whang, Qi~Lei, and Alexandros~G Dimakis.
\newblock Compressed sensing with invertible generative models and dependent
  noise.
\newblock {\em arXiv preprint arXiv:2003.08089}, 2020.
\newblock \url{https://arxiv.org/abs/2003.08089}.

\bibitem{mardani2017deep}
Morteza Mardani, Enhao Gong, Joseph~Y Cheng, Shreyas Vasanawala, Greg
  Zaharchuk, Marcus Alley, Neil Thakur, Song Han, William Dally, John~M Pauly,
  et~al.
\newblock Deep generative adversarial networks for compressed sensing automates
  mri.
\newblock {\em arXiv preprint arXiv:1706.00051}, 2017.
\newblock \url{https://arxiv.org/abs/1706.00051}.

\bibitem{Mo2019Deep}
Shaoxing Mo, Nicholas Zabaras, Xiaoqing Shi, and Jichun Wu.
\newblock Deep autoregressive neural networks for high-dimensional inverse
  problems in groundwater contaminant source identification.
\newblock {\em Water Resources Research}, 55(5):3856--3881, 2019.
\newblock \url{https://doi.org/10.1029/2018WR024638}.

\bibitem{laloy2018training}
Eric Laloy, Romain H{\'e}rault, Diederik Jacques, and Niklas Linde.
\newblock Training-image based geostatistical inversion using a spatial
  generative adversarial neural network.
\newblock {\em Water Resources Research}, 54(1):381--406, 2018.
\newblock \url{ https://doi.org/10.1002/2017WR022148}.

\bibitem{goodfellow2014generative}
Ian Goodfellow, Jean Pouget-Abadie, Mehdi Mirza, Bing Xu, David Warde-Farley,
  Sherjil Ozair, Aaron Courville, and Yoshua Bengio.
\newblock Generative adversarial nets.
\newblock In {\em Advances in neural information processing systems}, pages
  2672--2680, 2014.
\newblock
  \url{https://proceedings.neurips.cc/paper/2014/file/5ca3e9b122f61f8f06494c97b1afccf3-Paper.pdf}.

\bibitem{kingma2013auto}
Diederik~P Kingma and Max Welling.
\newblock Auto-encoding variational bayes.
\newblock {\em arXiv preprint arXiv:1312.6114}, 2013.
\newblock \url{https://arxiv.org/abs/1312.6114}.

\bibitem{Densenet}
Gao Huang, Zhuang Liu, Laurens Van Der~Maaten, and Kilian~Q Weinberger.
\newblock Densely connected convolutional networks.
\newblock In {\em Proceedings of the IEEE conference on computer vision and
  pattern recognition}, pages 4700--4708, 2017.
\newblock \url{ https://doi.org/10.1109/cvpr.2017.243}.

\bibitem{res}
Kaiming He, Xiangyu Zhang, Shaoqing Ren, and Jian Sun.
\newblock Deep residual learning for image recognition.
\newblock In {\em Proceedings of the IEEE conference on computer vision and
  pattern recognition}, pages 770--778, 2016.
\newblock \url{ https://doi.org/10.1109/CVPR.2016.90}.

\bibitem{Alom}
Md~Zahangir Alom, Tarek~M Taha, Christopher Yakopcic, Stefan Westberg, Paheding
  Sidike, Mst Nasrin, Brian Van, Abdul Awwal, and Vijayan~K Asari.
\newblock The history began from {AlexNet}: A comprehensive survey on deep
  learning approaches.
\newblock {\em arXiv preprint arXiv:1803.01164}, 2018.
\newblock \url{https://arxiv.org/abs/1803.01164}.

\bibitem{Ian}
Ian Goodfellow, Yoshua Bengio, and Aaron Courville.
\newblock Deep learning.
\newblock 2016.
\newblock \url{http://www.deeplearningbook.org}.

\bibitem{DNN_main}
Yann LeCun, Yoshua Bengio, and Geoffrey Hinton.
\newblock Deep learning.
\newblock {\em nature}, 521(7553):436, 2015.
\newblock \url{ https://doi.org/10.1038/nature14539}.

\bibitem{Yinhao}
Yinhao Zhu and Nicholas Zabaras.
\newblock Bayesian deep convolutional encoder–decoder networks for surrogate
  modeling and uncertainty quantification.
\newblock {\em Journal of Computational Physics}, 366:415 -- 447, 2018.
\newblock \url{https://doi.org/10.1016/j.jcp.2018.04.018}.

\bibitem{Bilionis}
Rohit~K. Tripathy and Ilias Bilionis.
\newblock Deep {UQ}: Learning deep neural network surrogate models for high
  dimensional uncertainty quantification.
\newblock {\em Journal of Computational Physics}, 375:565--588, December 2018.
\newblock \url{https://doi.org/10.1016/j.jcp.2018.08.036}.

\bibitem{Mo}
Shaoxing Mo, Yinhao Zhu, Nicholas Zabaras, Xiaoqing Shi, and Jichun Wu.
\newblock Deep convolutional encoder-decoder networks for uncertainty
  quantification of dynamic multiphase flow in heterogeneous media.
\newblock {\em Water Resources Research}, 55(1):703--728, January 2019.
\newblock \url{https://doi.org/10.1029/2018wr023528}.

\bibitem{zhu2019physics}
Yinhao Zhu, Nicholas Zabaras, Phaedon-Stelios Koutsourelakis, and Paris
  Perdikaris.
\newblock Physics-constrained deep learning for high-dimensional surrogate
  modeling and uncertainty quantification without labeled data.
\newblock {\em Journal of Computational Physics}, 394:56--81, 2019.
\newblock \url{https://doi.org/10.1016/j.jcp.2019.05.024}.

\bibitem{Nick}
Nicholas Geneva and Nicholas Zabaras.
\newblock Quantifying model form uncertainty in {Reynolds}-averaged turbulence
  models with {Bayesian} deep neural networks.
\newblock {\em Journal of Computational Physics}, 383:125--147, 2019.
\newblock \url{https://doi.org/10.1016/j.jcp.2019.01.021}.

\bibitem{TUM}
Nils Thuerey, Konstantin Weissenow, Harshit Mehrotra, Nischal Mainali, Lukas
  Prantl, and Xiangyu Hu.
\newblock A study of deep learning methods for {Reynolds-Averaged
  Navier-Stokes} simulations.
\newblock {\em arXiv preprint arXiv:1810.08217}, 2018.
\newblock \url{https://arxiv.org/abs/1810.08217}.

\bibitem{krizhevsky2012imagenet}
Alex Krizhevsky, Ilya Sutskever, and Geoffrey~E Hinton.
\newblock Imagenet classification with deep convolutional neural networks.
\newblock In {\em Advances in neural information processing systems}, pages
  1097--1105, 2012.
\newblock \url{https://doi.org/10.1145/3065386}.

\bibitem{xie2018tempogan}
You Xie, Erik Franz, Mengyu Chu, and Nils Thuerey.
\newblock Tempogan: {A} temporally coherent, volumetric gan for
  super-resolution fluid flow.
\newblock {\em ACM Transactions on Graphics (TOG)}, 37(4):1--15, 2018.
\newblock \url{https://doi.org/10.1145/3197517.3201304}.

\bibitem{chan2017parametrization}
Shing Chan and Ahmed~H Elsheikh.
\newblock Parametrization and generation of geological models with generative
  adversarial networks.
\newblock {\em arXiv preprint arXiv:1708.01810}, 2017.
\newblock \url{https://arxiv.org/abs/1708.01810}.

\bibitem{mosser2018stochastic}
Lukas Mosser, Olivier Dubrule, and Martin~J Blunt.
\newblock Stochastic reconstruction of an {Oolitic} limestone by generative
  adversarial networks.
\newblock {\em Transport in Porous Media}, 125(1):81--103, 2018.
\newblock \url{https://doi.org/10.1007/s11242-018-1039-9}.

\bibitem{mosser2017reconstruction}
Lukas Mosser, Olivier Dubrule, and Martin~J Blunt.
\newblock Reconstruction of three-dimensional porous media using generative
  adversarial neural networks.
\newblock {\em Physical Review E}, 96(4):043309, 2017.
\newblock \url{https://doi.org/10.1103/PhysRevE.96.043309}.

\bibitem{sonderby2016ladder}
Casper~Kaae S{\o}nderby, Tapani Raiko, Lars Maal{\o}e, S{\o}ren~Kaae
  S{\o}nderby, and Ole Winther.
\newblock Ladder variational autoencoders.
\newblock In {\em Advances in neural information processing systems}, pages
  3738--3746, 2016.
\newblock
  \url{https://proceedings.neurips.cc/paper/2016/file/6ae07dcb33ec3b7c814df797cbda0f87-Paper.pdf}.

\bibitem{hsu2018scalable}
Wei-Ning Hsu and James Glass.
\newblock Scalable factorized hierarchical variational autoencoder training.
\newblock {\em arXiv preprint arXiv:1804.03201}, 2018.
\newblock \url{https://arxiv.org/abs/1804.03201}.

\bibitem{dinh2016density}
Laurent Dinh, Jascha Sohl-Dickstein, and Samy Bengio.
\newblock Density estimation using real {NVP}.
\newblock {\em arXiv preprint arXiv:1605.08803}, 2016.
\newblock \url{https://arxiv.org/abs/1605.08803}.

\bibitem{kingma2018glow}
Durk~P Kingma and Prafulla Dhariwal.
\newblock Glow: {Generative} flow with invertible 1x1 convolutions.
\newblock In {\em Advances in Neural Information Processing Systems}, pages
  10215--10224, 2018.
\newblock
  \url{https://papers.nips.cc/paper/2018/file/d139db6a236200b21cc7f752979132d0-Paper.pdf}.

\bibitem{dinh2014nice}
Laurent Dinh, David Krueger, and Yoshua Bengio.
\newblock Nice: {Non-linear} independent components estimation.
\newblock {\em arXiv preprint arXiv:1410.8516}, 2014.
\newblock \url{https://arxiv.org/abs/1410.8516}.

\bibitem{vo2014new}
Hai~X Vo and Louis~J Durlofsky.
\newblock A new differentiable parameterization based on principal component
  analysis for the low-dimensional representation of complex geological models.
\newblock {\em Mathematical Geosciences}, 46(7):775--813, 2014.
\newblock \url{https://doi.org/10.1007/s11004-014-9541-2}.

\bibitem{sarma2008kernel}
Pallav Sarma, Louis~J Durlofsky, and Khalid Aziz.
\newblock Kernel principal component analysis for efficient, differentiable
  parameterization of multipoint geostatistics.
\newblock {\em Mathematical Geosciences}, 40(1):3--32, 2008.
\newblock \url{https://doi.org/10.1007/s11004-007-9131-7}.

\bibitem{zhang2018iterative}
Jiangjiang Zhang, Guang Lin, Weixuan Li, Laosheng Wu, and Lingzao Zeng.
\newblock An iterative local updating ensemble smoother for estimation and
  uncertainty assessment of hydrologic model parameters with multimodal
  distributions.
\newblock {\em Water Resources Research}, 54(3):1716--1733, 2018.
\newblock \url{https://doi.org/10.1002/2017WR020906}.

\bibitem{ardizzone2019guided}
Lynton Ardizzone, Carsten L{\"u}th, Jakob Kruse, Carsten Rother, and Ullrich
  K{\"o}the.
\newblock Guided image generation with conditional invertible neural networks.
\newblock {\em arXiv preprint arXiv:1907.02392}, 2019.
\newblock \url{https://arxiv.org/abs/1907.02392}.

\bibitem{prenger2019waveglow}
Ryan Prenger, Rafael Valle, and Bryan Catanzaro.
\newblock Waveglow: {A} flow-based generative network for speech synthesis.
\newblock In {\em ICASSP 2019-2019 IEEE International Conference on Acoustics,
  Speech and Signal Processing (ICASSP)}, pages 3617--3621. IEEE, 2019.
\newblock \url{https://doi.org/10.1109/ICASSP.2019.8683143}.

\bibitem{geneva2020multi}
Nicholas Geneva and Nicholas Zabaras.
\newblock Multi-fidelity generative deep learning turbulent flows.
\newblock {\em Foundations of Data Science}, page In Press., 2020.
\newblock \url{https://arxiv.org/abs/2006.04731}.

\bibitem{aarnes2007introduction}
J{\o}rg~E Aarnes, Tore Gimse, and Knut-Andreas Lie.
\newblock An introduction to the numerics of flow in porous media using matlab.
\newblock In {\em Geometric modelling, numerical simulation, and optimization},
  pages 265--306. Springer, 2007.
\newblock \url{https://doi.org/10.1007/978-3-540-68783-2_9}.

\bibitem{software}
SINTEF.
\newblock Sintef mrst project web page, 2015.
\newblock \url{http://www.sintef.no/Projectweb/MRST/}.

\bibitem{lie2019introduction}
Knut-Andreas Lie.
\newblock {\em An introduction to reservoir simulation using MATLAB/GNU Octave:
  User guide for the MATLAB Reservoir Simulation Toolbox (MRST)}.
\newblock Cambridge University Press, 2019.
\newblock \url{https://doi.org/10.1017/9781108591416.023}.

\bibitem{rudin1992nonlinear}
Leonid~I Rudin, Stanley Osher, and Emad Fatemi.
\newblock Nonlinear total variation based noise removal algorithms.
\newblock {\em Physica D: nonlinear phenomena}, 60(1-4):259--268, 1992.
\newblock \url{https://doi.org/10.1016/0167-2789(92)90242-F}.

\bibitem{isola2017image}
Phillip Isola, Jun-Yan Zhu, Tinghui Zhou, and Alexei~A Efros.
\newblock Image-to-image translation with conditional adversarial networks.
\newblock In {\em Proceedings of the IEEE conference on computer vision and
  pattern recognition}, pages 1125--1134, 2017.
\newblock \url{https://doi.org/10.1109/CVPR.2017.632}.

\bibitem{mirza2014conditional}
Mehdi Mirza and Simon Osindero.
\newblock Conditional generative adversarial nets.
\newblock {\em arXiv preprint arXiv:1411.1784}, 2014.
\newblock \url{https://arxiv.org/abs/1411.1784}.

\bibitem{sohn2015learning}
Kihyuk Sohn, Honglak Lee, and Xinchen Yan.
\newblock Learning structured output representation using deep conditional
  generative models.
\newblock In {\em Advances in neural information processing systems}, pages
  3483--3491, 2015.
\newblock
  \url{https://proceedings.neurips.cc/paper/2015/file/8d55a249e6baa5c06772297520da2051-Paper.pdf}.

\bibitem{ardizzone2018analyzing}
Lynton Ardizzone, Jakob Kruse, Sebastian Wirkert, Daniel Rahner, Eric~W
  Pellegrini, Ralf~S Klessen, Lena Maier-Hein, Carsten Rother, and Ullrich
  K{\"o}the.
\newblock Analyzing inverse problems with invertible neural networks.
\newblock {\em arXiv preprint arXiv:1808.04730}, 2018.
\newblock \url{https://arxiv.org/abs/1808.04730}.

\bibitem{srivastava2014dropout}
Nitish Srivastava, Geoffrey Hinton, Alex Krizhevsky, Ilya Sutskever, and Ruslan
  Salakhutdinov.
\newblock Dropout: a simple way to prevent neural networks from overfitting.
\newblock {\em The journal of machine learning research}, 15(1):1929--1958,
  2014.
\newblock
  \url{https://www.jmlr.org/papers/volume15/srivastava14a/srivastava14a.pdf}.

\bibitem{glorot2010understanding}
Xavier Glorot and Yoshua Bengio.
\newblock Understanding the difficulty of training deep feedforward neural
  networks.
\newblock In {\em Proceedings of the thirteenth international conference on
  artificial intelligence and statistics}, pages 249--256, 2010.
\newblock
  \url{https://citeseerx.ist.psu.edu/viewdoc/download?doi=10.1.1.207.2059&rep=rep1&type=pdf}.

\bibitem{tripathy2018deep}
Rohit~K Tripathy and Ilias Bilionis.
\newblock Deep uq: {Learning} deep neural network surrogate models for high
  dimensional uncertainty quantification.
\newblock {\em Journal of computational physics}, 375:565--588, 2018.
\newblock \url{https://arxiv.org/abs/1802.00850}.

\bibitem{kingma2014adam}
Diederik~P Kingma and Jimmy Ba.
\newblock Adam: {A} method for stochastic optimization.
\newblock {\em arXiv preprint arXiv:1412.6980}, 2014.
\newblock \url{https://arxiv.org/abs/1412.6980}.

\bibitem{paszke2017automatic}
Adam Paszke, Sam Gross, Soumith Chintala, Gregory Chanan, Edward Yang, Zachary
  DeVito, Zeming Lin, Alban Desmaison, Luca Antiga, and Adam Lerer.
\newblock Automatic differentiation in pytorch.
\newblock 2017.
\newblock \url{https://openreview.net/pdf?id=BJJsrmfCZ}.

\bibitem{Geneva2020Modeling}
Nicholas Geneva and Nicholas Zabaras.
\newblock Modeling the dynamics of pde systems with physics-constrained deep
  auto-regressive networks.
\newblock {\em Journal of Computational Physics}, 403:109056, 2020.
\newblock \url{https://doi.org/10.1016/j.jcp.2019.109056}.

\bibitem{mackay2018reversible}
Matthew MacKay, Paul Vicol, Jimmy Ba, and Roger~B Grosse.
\newblock Reversible recurrent neural networks.
\newblock In {\em Advances in Neural Information Processing Systems}, pages
  9029--9040, 2018.
\newblock
  \url{https://papers.nips.cc/paper/2018/file/4ff6fa96179cdc2838e8d8ce64cd10a7-Paper.pdf}.

\bibitem{xingjian2015convolutional}
SHI Xingjian, Zhourong Chen, Hao Wang, Dit-Yan Yeung, Wai-Kin Wong, and
  Wang-chun Woo.
\newblock Convolutional {LSTM} network: {A} machine learning approach for
  precipitation nowcasting.
\newblock In {\em Advances in neural information processing systems}, pages
  802--810, 2015.
\newblock
  \url{https://papers.nips.cc/paper/2015/file/07563a3fe3bbe7e3ba84431ad9d055af-Paper.pdf}.

\bibitem{kumar2020convcast}
Ashutosh Kumar, Tanvir Islam, Yoshihide Sekimoto, Chris Mattmann, and Brian
  Wilson.
\newblock Convcast: An embedded convolutional {LSTM} based architecture for
  precipitation nowcasting using satellite data.
\newblock {\em {PLoS} one}, 15(3):e0230114, 2020.
\newblock \url{https://doi.org/10.1371/journal.pone.0230114}.

\end{thebibliography}
\end{document}